\documentclass{aa}  
\usepackage[version=4]{mhchem}
\usepackage{url}
\newcommand{\water}{\ce{H_2O } }
\newcommand{\waterp}{\ce{H_2O }$^+$ }
\newcommand{\cii}{[C\,{\sc ii}] }
\newcommand{\barolo}{{\sc $^{\rm 3D}$Barolo }}
\usepackage{mathtools}
\usepackage{comment}
\usepackage{graphicx}
\usepackage[dvipsnames]{xcolor}
\usepackage{natbib}
\usepackage[breaklinks=true]{hyperref}
\hypersetup{colorlinks=true, linkcolor=red,citecolor=blue,filecolor=cyan,urlcolor=magenta}
\bibpunct{(}{)}{;}{a}{}{,} 
\usepackage{amsmath}
\usepackage{graphicx}
\usepackage{txfonts}
\usepackage{sidecap}
\usepackage{float}

\begin{document}

    \title{Detailed lens modeling and kinematics of the submillimeter galaxy G09v1.97}

     \subtitle{An analysis of CO, \ce{H_2O}, \ce{H_2O}$^+$, and dust continuum emission}

   \author{K. Kade
          \inst{1}
          \and
          C. Yang 
          \inst{1}
          \and
          M. Yttergren 
          \inst{1}
          \and
          K. K. Knudsen 
          \inst{1}
          \and
          S. K\"onig 
          \inst{1} 
          \and 
          A. Amvrosiadis
          \inst{2}
          \and 
          S. Dye 
          \inst{3}
          \and \\
          J. Nightingale
          \inst{4}  
          \and 
          L. Zhang 
          \inst{5}
          \and 
          Z. Zhang
          \inst{5}
          \and 
          A. Cooray
          \inst{6}
          \and 
          P. Cox
          \inst{7}
          \and 
          R. Gavazzi
          \inst{7, 8}
          \and 
          E. Ibar
          \inst{9, 13}
          \and \\
          M.~J.~Micha{\l}owski
          \inst{10}
          \and
          P. van der Werf
          \inst{11}
          \and
          R. Xue
          \inst{12}
          }

\institute{
Department of Space, Earth \& Environment, Chalmers University of Technology, SE-412 96 Gothenburg, Sweden \\ \email{kiana.kade@chalmers.se} 
\and 
Department of Physics, Institute for Computational Cosmology, Durham University, South Road, Durham, DH 13LE, UK 
\and
School of Physics \& Astronomy, University of Nottingham, University Park, Nottingham, NG7 2RD, UK 
\and 
School of Mathematics, Statistics and Physics, Newcastle University, Newcastle upon Tyne, NE1 7RU, UK 
\and
School of Astronomy and Space Science, Nanjing University, Nanjing 210093, China 
\and
Department of Physics \& Astronomy, University of California Irvine, Irvine CA 92697, USA 
\and 
Sorbonne Universit{\'e}, UPMC Universit{\'e} Paris 6 and CNRS, UMR 7095, Institut d'Astrophysique de Paris, 98bis boulevard Arago, 75014 Paris, France 
\and 
Laboratoire d’Astrophysique de Marseille, Aix-Marseille Univ, CNRS, CNES, Marseille, France 
\and 
Instituto de F\'isica y Astronom\'ia, Universidad de Valpara\'iso, Avda. Gran Breta\~na 1111, Valpara\'iso, Chile 
\and
Astronomical Observatory Institute, Faculty of Physics, Adam Mickiewicz University, ul.~S{\l}oneczna 36, 60-286 Pozna{\'n}, Poland 
\and 
Leiden Observatory, Leiden University, PO Box 9513, 2300 RA Leiden, The Netherlands 
\and
National Radio Astronomy Observatory, Charlottesville, VA, US 
\and 
Millennium Nucleus for Galaxies (MINGAL)
}

   \date{Received July 11, 2024; accepted January 15, 2026}

  \abstract
   {The formation mechanisms of intensely starbursting galaxies at high redshift remain unknown. One possible mechanism for producing these intense rates of star formation is through mergers and interactions, but detecting these at high redshift remains a challenge. Observations of high-redshift gravitationally lensed galaxies provide a method for studying the interstellar medium and environment of these extreme starbursts in detail.}
   {We aim to use high angular resolution observations of dust continuum, CO(6--5), H$_2$O($2_{11} - 2_{02}$), and H$_2$O$^{+}$($2_{02}\text{--}1_{11}$) emission to constrain the ongoing processes in the $z = 3.63$ gravitationally lensed submillimeter galaxy H-ATLAS J083051.0+013224 (G09v1.97). }
   {We used the sophisticated lens modeling software {\sc PyAutoLens} to perform both parametric and non-parametric source modeling. We created a de-magnified source plane CO(6--5) emission line cube and performed kinematic modeling using {\sc $^{\rm 3D}$Barolo}. Additionally, we investigated the properties of the continuum and molecular line emission in the source plane. }
   {We find that the regions of CO(6--5) and H$_2$O($2_{11} - 2_{02}$) emission match closely in the source plane but that the dust continuum emission is more compact. We find that our lens modeling results do not require more than one source, contrary to what has been found in previous studies. Instead, we find that G09v1.97 resembles a rotating disk with $\rm V_{max}/\bar{\sigma} = 2.8 \pm 0.4$ with evidence for residual emission indicative of non-circular motions such as outflows, tidal tails, or an additional background galaxy.}
   {We suggest that the origin of the non-circular motions may be associated with a bi-conical outflow, a tidal tail from an interaction, or indicate the possible presence of an additional galaxy. We calculate the dynamical mass, gas mass, star-formation rate, and depletion time for G09v1.97 and find a high star-formation rate and low gas depletion time. In combination, this suggests that G09v1.97 has recently undergone an interaction, triggering intense star formation, and is in the process of settling into a disk.}

   \keywords{Galaxies: high-redshift --  galaxies: evolution -- galaxies: starburst -- galaxies: ISM: kinematics and dynamics}

   \maketitle

\section{Introduction}

Some of the brightest and most strongly star-forming galaxies at high-redshift are heavily dust-obscured galaxies \citep[submillimeter galaxies, SMGs;][]{Smail97, Barger98, Hughes98, Dudzeviciute20} and dusty star-forming galaxies \citep[DSFGs, see reviews from][ and references therein]{Blain02, Casey14}. These galaxies have prodigious star-formation rates (SFRs), prompting questions about the underlying formation mechanisms and the drivers behind such intense star formation. SMGs can reach total infrared luminosities of $\sim$\,$10^{13}$\,L$_{\odot}$ and in certain cases maintain `maximum starbursts' of $>1000$\,M$_{\odot}$\,yr$^{-1}$ \citep[e.g.,][]{Barger14, Simpson15, Riechers13a}. Consequently, these galaxies represent a crucial stage in the growth of massive galaxies, making them key to understanding the evolution of massive, high-redshift galaxies.

One process thought to lead to such intense SFRs and infrared luminosities is mergers and/or interactions between high-redshift galaxies, where the molecular gas fuels the nuclear regions, triggering intense star formation \citep[e.g.,][]{Mihos1994, Hopkins06, Engel10}. This is consistent with both simulations and observations of the high-redshift universe, which have shown the importance and high prevalence of mergers and interactions \citep[e.g.,][]{Keres05, Hopkins08, Genzel10}. However, some simulations suggest that these dusty starburst galaxies may form through more secular processes \citep[e.g., ][]{Dekel09, Dave10}. This is supported by observations of massive high redshift galaxies that show no signs of merger or interaction activity, only well-ordered rotation \citep[e.g.,][]{Tacconi13, Hodge16, JA18, Smit18, Lelli2021, Fernanda23}. Thus, although this phase represents a critical stage in galaxy growth, the formation mechanisms of these galaxies remain uncertain.

\begin{table*}[t]
    \centering
    \caption{Details of observations$^{h}$.}
    \begin{tabular}{l c c c c c c} \hline \hline
        Line & Spectral Window$^a$ & Date of Obs.$^b$ & $\nu_{\mathrm{spw, central}}^c$ & Channel Width$^d$ & Synthesized Beam$^e$ & RMS$^f$\\
         & & [yyyy mm dd] & [GHz] & [$\mathrm{km\,s^{-1}} $] & [$'' \times ''$] & [mJy/beam]\\ \hline 
        CO(6--5) & 0 & 2019-07-14 & 149.28 & 20 & $0.11 \times 0.085$ & 0.19 \\
        H$_2$O(2$_{11}$--2$_{02}$) & 2 & 2019-07-14 & 162.36 & 20 & $0.11 \times 0.081$ & 0.50 \\
        \llap{${^g}$}\waterp($2_{02}\text{--}1_{11}$) & 1 & combined & 160.28 & 29 & $0.34 \times 0.31$ & 0.13 \\
        \hline
    \end{tabular}
    \tablefoot{
        \tablefoottext{a}{Spectral window and observed lines therein.}
        \tablefoottext{b}{Date of ALMA observation.}
        \tablefoottext{c}{Central frequency of the spectral window containing the specific line.}
        \tablefoottext{d}{Channel width of the calibrated and imaged data for the specific line.}
        \tablefoottext{e}{Synthesized beam of the spectral window using natural weighting.}
        \tablefoottext{f}{Per-channel RMS of the spectral window.} 
        \tablefoottext{g}{The combined \waterp lines are the hyperfine structure H$_2$O$^+$(2$_{02}$--1$_{11}$)\,$_{5/2\text{--}3/2}$ lines at 160.28\,GHz.}
        \tablefoottext{h}{The observational phase center for the observations detailed in the table was (J2000) RA 08:30:51.156, DEC +01:32:24.35. These are the same coordinates as the observations used in \citet{Yang19}.}
        }
    \label{tab:obs_details}
\end{table*}

One of the primary reasons for this lack of understanding is the difficulty inherent in observing high-redshift galaxies. Resolving the environment of high-redshift galaxies remains challenging, even for telescopes like the Atacama Large Millimeter Array (ALMA). SMGs have been shown to have an average redshift of $\approx 2.5$ for those discovered at $870\,\mu$m \citep[e.g.,][]{Chapman05, Danielson17}, where an angular resolution of $1''$ corresponds to $\approx8.2$\,kpc. Even at an angular resolution of $0.1''$, the physical scales probed within the galaxy would be at $\approx800$\,pc scales. At this resolution, it is possible to begin to resolve the regions of denser and warmer gas \citep[e.g.,][]{Tacconi08, Riechers13, Spilker15, Swinbank15, Dye15, Rybak20, Rybak23}, but reaching beyond this limit is challenging. Gravitational lensing circumvents the need for extremely high angular resolution observations. This phenomenon can, in effect, improve the angular resolution of observations by $1/\sqrt{\mu}$ where $\mu$ is the lensing magnification factor (under the assumption that the magnification is static across the image), meaning that details within the interstellar medium (ISM) and environment can, in some cases, be resolved in the source plane. 

To gain a better understanding of the processes governing SMGs, DSFGs, and other high-redshift galaxies with intense star formation rates, it is paramount to understand in detail the internal ongoing processes in the ISM and the influence of the surrounding environment (i.e., mergers/interactions). In recent years, studies of the ISM of gravitationally lensed galaxies have shed considerable light on these processes \citep[e.g.,][]{Amvrosiadis24, canameras2018, canameras2021, Dye22, Kade23, Kade24, Rybak15, Rybak20, Rybak23, Spilker22, Yang19, Yang23}. 

The strongly lensed SMG H-ATLAS~J083051.0+013224 (hereafter, G09v1.97) at $z = 3.63$ is an ideal laboratory for studying the properties of high-redshift sub-millimeter galaxies. First reported in \citet{Bussmann13}, G09v1.97 is exceptionally bright in the infrared with a lensing magnification factor of $\mu \sim 6-10$ and has been studied in detail in a number of previous works \citep[e.g.,][]{Bussmann13, Yang19, Maresca22}. G09v1.97 has an intrinsic infrared luminosity of $\sim 10^{13}$\,L$_{\odot}$, and an SFR of $>1000$\,M$_{\odot}$\,yr$^{-1}$, among the most intense starbursts known to date. \citet{Yang19} performed an in-depth study of G09v1.97 using ALMA data at an angular resolution of $0''.2 - 0''.4$ and, based on a parametric lensing model, concluded that the source was composed of two merging and/or interacting galaxies. 

The current paper utilizes very high angular-resolution ALMA observations ($\approx 0''.1$) of G09v1.97, achieving scales of a few hundred parsecs in the source plane, to study the ISM using the emission from the dust, CO(6--5), \water($2_{11} - 2_{02}$), hereafter H$_2$O and \waterp($2_{02}\text{--}1_{11}$), hereafter H$_2$O$^{+}$. We present a non-parametric source plane reconstruction of the dust continuum emission and detected molecular line emission. Additionally, we perform kinematic modeling and discuss the implications of our findings. In Section \ref{sec:observation_details}, we describe the observations and data reduction steps. The results of the continuum and line analysis, lens modeling, and kinematic modeling are reported in Section \ref{sec:results}. Section \ref{sec:discussion} provides a discussion of these results. Finally, our conclusions are presented in Section \ref{sec:conclusions}. Throughout this paper, we adopt a flat $\Lambda$CDM cosmology with ${H_{0}}$ = 70\,km\,$\mathrm{s^{-1}\,Mpc^{-1}}$, $\mathrm{\Omega_{\Lambda}}$ = 0.7, and $\mathrm{\Omega_{m}}$ = 0.3. At $z = 3.63$, $0''.1$ corresponds to $\approx 740$\,pc in the image plane. 

\begin{figure}
    \centering
    \includegraphics[width = 1.0\linewidth]{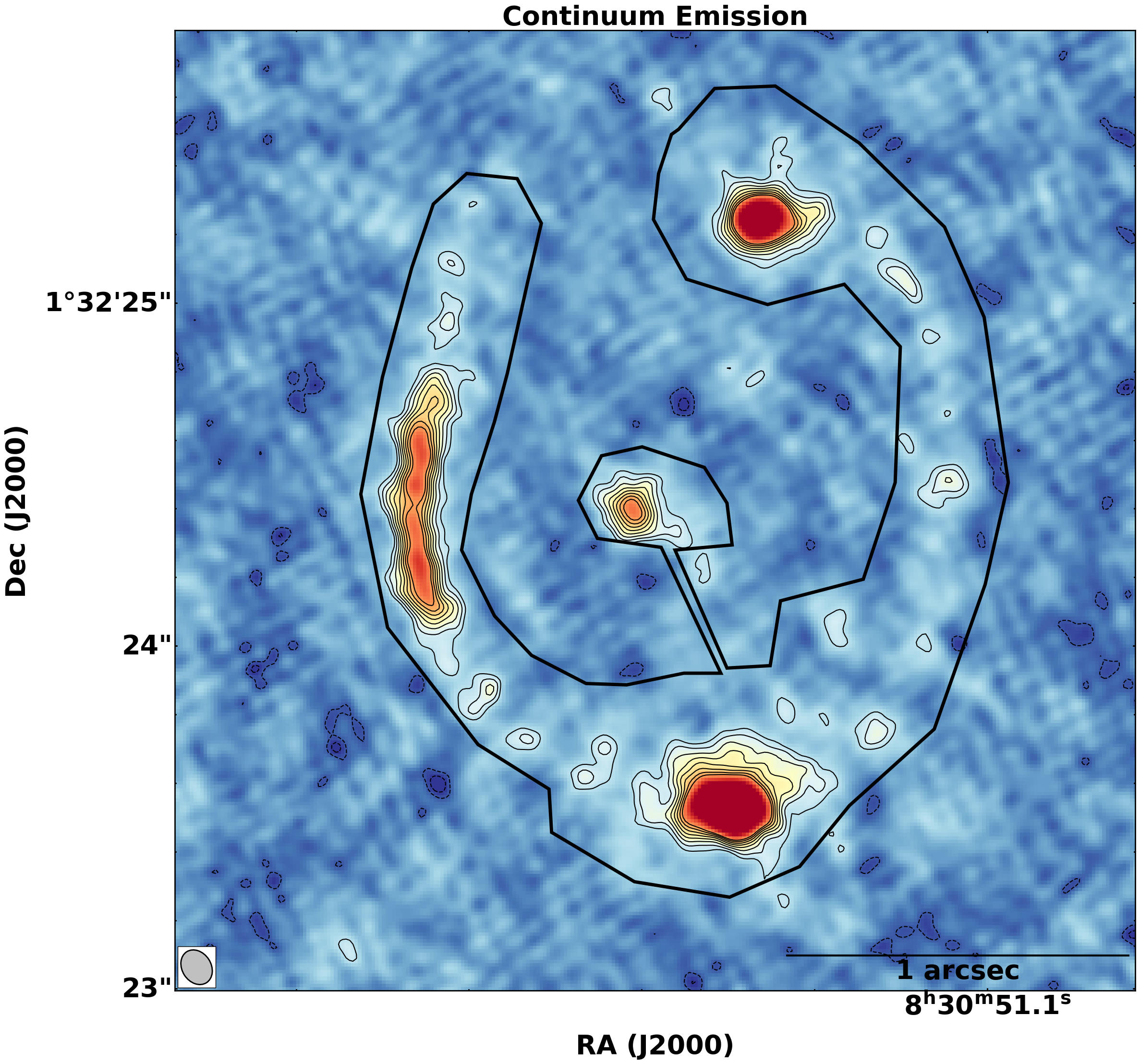}
    \caption{Continuum image of G09v1.97, imaged using a natural weighting scheme. The contours are shown at $-3, -2, 3, 4, 5, 6, 7, 8, 9, 10\sigma$ levels ($\sigma$=0.02\,mJy/beam). The black region shows the region from which the continuum emission was extracted. The synthesized beam is shown in the bottom left of the image and corresponds to $0.11'' \times 0.084''$.}
    \label{fig:continuum}
\end{figure}

\section{Observations and data reduction} \label{sec:observation_details}

G09v1.97 was observed in band 4 with ALMA as part of project ID 2018.1.01710.S (P.I. Yang). All calibrations and image processing were performed in the Common Astronomy Software Application package \citep[CASA;][]{CASA}. The band 4 observations were taken on 14-July-2019 using 41 antennas. The observations were performed using baselines between 138.5\,m, and 8547.6\,m, resulting in an angular resolution of $\approx 0''.1$ (using natural weighting). The on-source integration time was 47.2 minutes, with a total overhead time of 62.19 minutes. The overhead time includes the pointing, focusing, phase, flux density, and bandpass calibrations. J0825+0309 was used as the phase calibrator, J0725-0054 was used as the bandpass calibrator, and J0725-0054 was used as the flux calibrator. We report a conservative estimate of 10\% to represent the uncertainty in the absolute flux calibration (note that the fiducial value for band 4 is $\sim 5\%$) \footnote{https://almascience.eso.org/documents-and-tools/cycle10/alma- technical-handbook}. The weather conditions during the observations were good, with a mean precipitable water vapor of 0.8\,mm and a phase RMS of 0.239 degrees. 

The data were calibrated using the ALMA interferometric pipeline \citep{Hunter23}, including calibration of the phase, bandpass, flux, and gain. The pipeline-reduced data were checked to ensure quality. The images produced by the pipeline data reduction of the calibrated data were inspected visually to ensure data quality and identify line-free channels to be used in continuum subtraction. Continuum subtraction was performed using the CASA task {\sc uvcontsub} using a polynomial fit of order 1. Imaging of the calibrated data was performed in CASA using the {\sc tclean} task, cleaning in the region of expected emission using the mask encompassing the region of emission for the deep continuum and CO(6--5) emission images reported in \citet{Yang19}. Continuum images were created using the line-free channels and natural weighting and were cleaned down to a $1\sigma$ level, where $\sigma=0.02$\,mJy/beam, at $v_{\rm sky} = 154.5$\,GHz, corresponding to continuum emission at 1.94\,mm. Similarly, spectral cubes were created from the continuum subtracted measurement set using natural weighting and were cleaned down to a $1\sigma$ level. The native spectral resolution of the observations was $\approx 7.8$\,MHz. To ensure a sufficient per-channel signal-to-noise ratio (SNR) we used a channel width of 20\,km/s when cleaning the spectral cubes. We provide details of beam size, channel width, and image RMS in Table \ref{tab:obs_details}. 

\begin{figure}[htbp] 
    \centering
    \includegraphics[width = 0.8\linewidth]{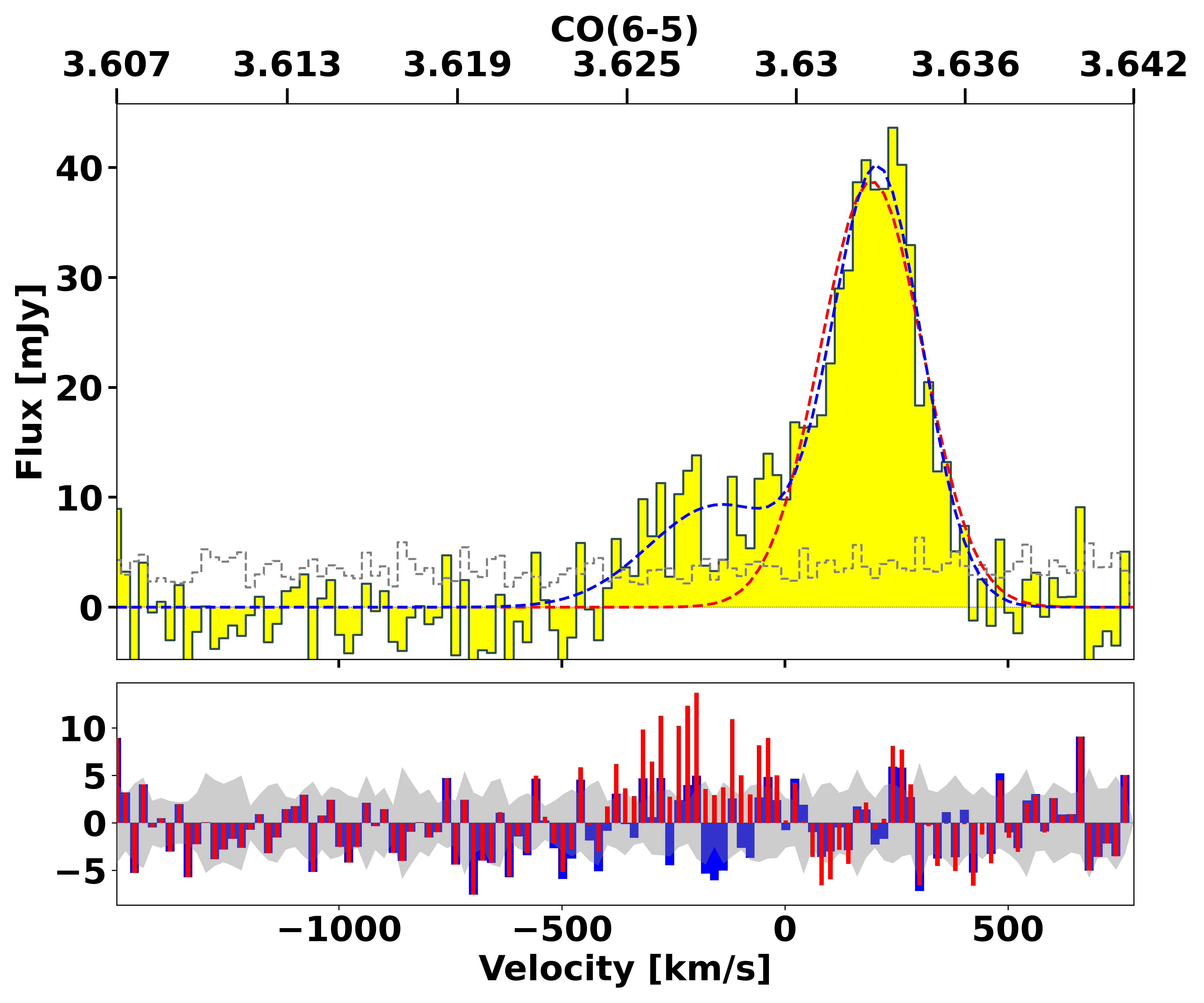}
    \includegraphics[width = 0.8\linewidth]{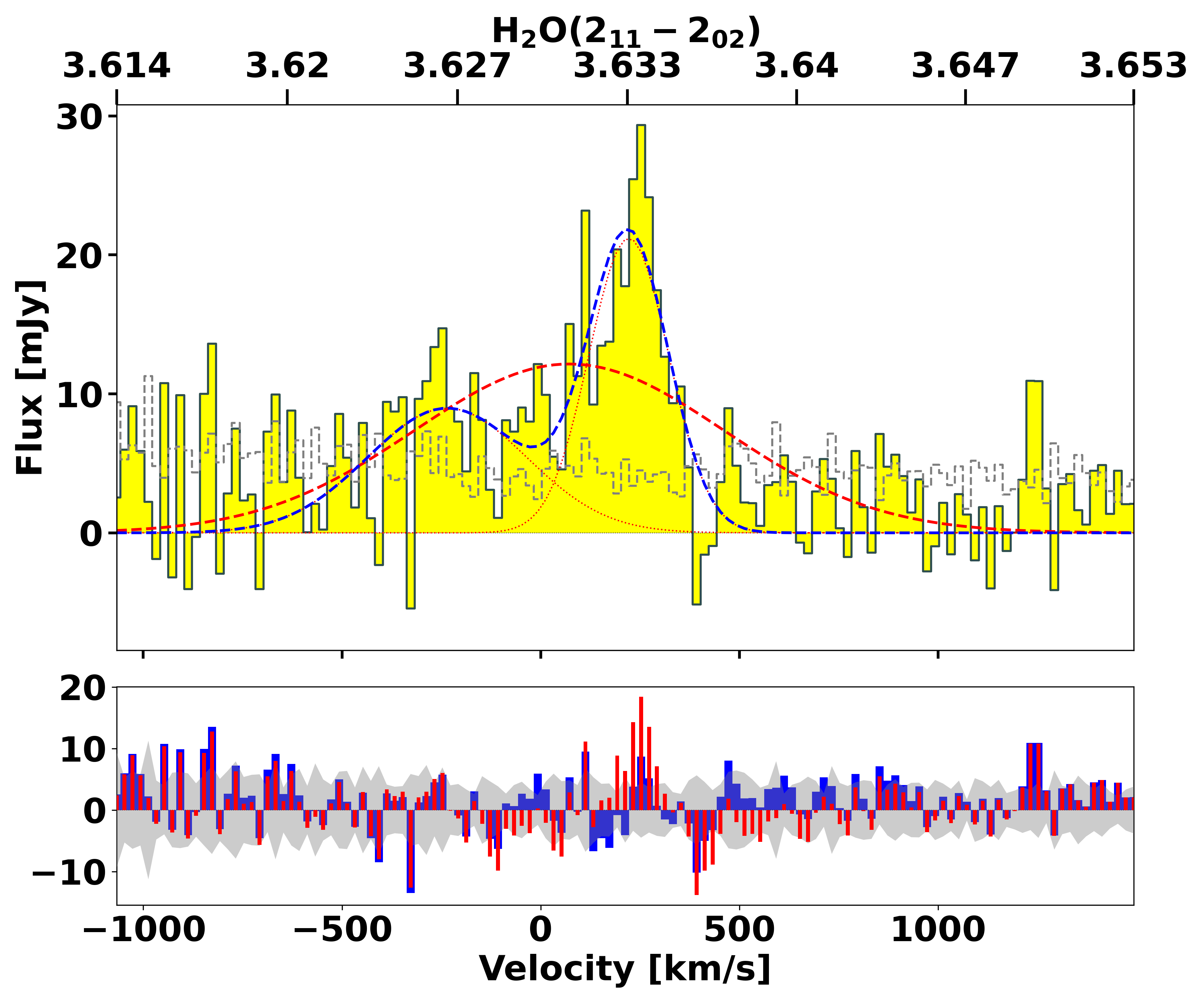}
    \includegraphics[width = 0.8\linewidth]{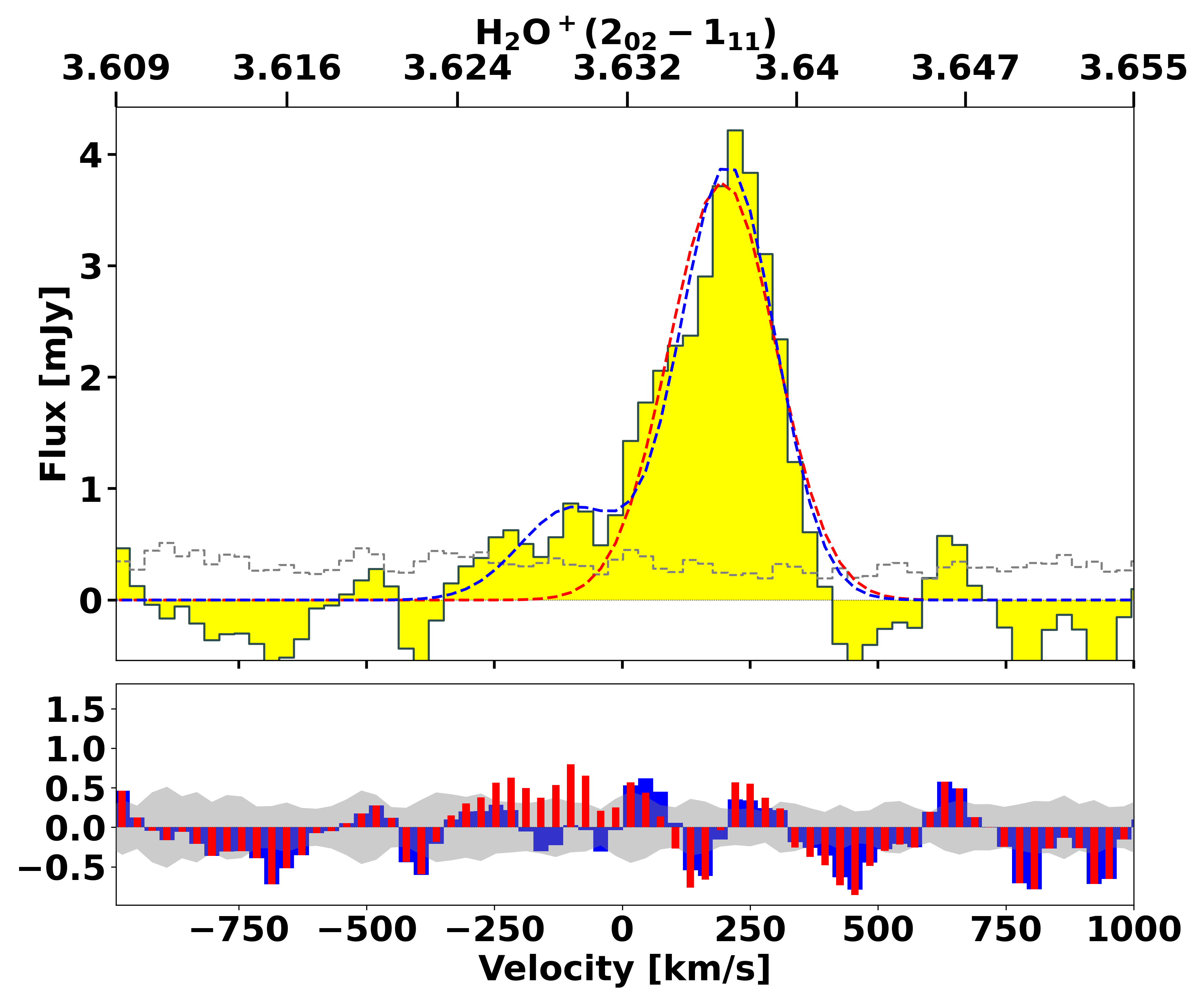}
    \caption{Spectra of the detected molecular line species in G09v1.97; top left: CO(6--5) spectrum, top right: \water spectrum, bottom: \waterp spectrum. Note that the \waterp emission was not detected in the high-resolution data, and thus, the spectrum was taken from the combined data (as described in Sections \ref{sec:observation_details} and \ref{subsec:lineemission_results}), resulting in a different spectral resolution than for the CO(6--5) and \water emission. The spectrum for each molecule was extracted from a region corresponding to the region shown in Fig. \ref{fig:continuum}. For each molecule, the spectrum is shown in the top panel, and the residuals from the Gaussian fits are shown in the lower panel. Single Gaussian profiles are shown in red, and double Gaussian profiles are shown in blue; similarly, in the lower panel, the residuals from fitting using a single Gaussian profile are shown in red and in blue for fits using two Gaussian profiles. It is clear that double Gaussian profiles fit the spectra better for all detected molecular line species. The dashed gray line in the top panel and the gray region in the lower panel represent the per-channel RMS. The top axis in the top panel for each spectrum shows the corresponding redshift. Note that the red residuals in the bottom panel are shown as slightly narrower (i.e., appear to have a smaller channel width) than the blue, purely for clarity purposes.}
    \label{fig:img_plane_spectra}
\end{figure}

To be able to study the fainter \waterp emission line, we combined the data presented above with lower-spatial-resolution observations (C36-5 configuration with baseline range 0.02--1.5\,km) used in \citet{Yang19} (project ID: 2015.1.01320.S, P.I.: Omont). As the frequency setups for the two observational data sets differed, we searched for a common range of frequencies and then re-gridded the individual data sets to the common frequency range and a common spectral resolution of 15.625\,MHz. After that, the individual data sets were combined to produce combined clean images. As a last step, and to correct for the intrinsic beam size differences in the combined data sets, we used the {\sc tp2vistweak} step from the Total Power to Visibilities tool \citep[TP2VIS,][]{koda11, koda19}. Images from the combined data were cleaned and imaged using the CASA task {\sc tclean} using Briggs weighting with a robust factor of 0.5 to keep a reasonable balance between SNR and angular resolution, while also maintaining an angular resolution as close to the higher resolution data as possible. We note that, due to the significantly lower angular resolution of the combined dataset, it is only used to study the fainter \waterp emission.

\section{Results} \label{sec:results}
\subsection{Continuum emission} \label{subsec:continuum_results}

We detect strong continuum emission towards G09v1.97 at a $\approx 22\sigma$ level. We extracted the continuum from the region of expected emission from \citet{Yang19}, shown in Fig. \ref{fig:continuum}. We find a 1.940\,mm continuum flux of $7.00 \pm 0.32$\,mJy. This value is lower than that found by \citet{Yang19} of $8.8 \pm 0.5$\,mJy, suggesting the possibility that some continuum emission has been resolved out in the high-resolution observations utilized in this paper. However, the extent of this discrepancy ($20\pm6$\%) is not large, and therefore, we do not consider it to have a significant impact on later results. 

\subsection{Line emission} \label{subsec:lineemission_results}

We detect CO(6--5) (rest frequency 691.4731\,GHz) and \water($2_{11} - 2_{02}$) (rest frequency 752.0331\,GHz) in the high-resolution data. We do not detect \waterp($2_{02}\text{--}1_{11})_{5/2\text{--}3/2}$ (rest frequency 742.1531\,GHz) in the high-resolution data and therefore use the combined dataset (as described in Section \ref{sec:observation_details}) for analysis of the \waterp line. 

We performed a regional spectral extraction within the expected region of emission described in \citet{Yang19}. We used the same region of extraction for all species to enable consistent comparisons between them; this region is shown in black in Fig. \ref{fig:continuum}. We do not detect any significant emission outside the region. We calculated the RMS for each species using a simple sampling method in which the RMS is sampled from distinct, emission-free regions of the cubes. We created moment-0 and moment-1 maps from the same region used to extract the spectra. For clarity, we removed flux at lower than $3\sigma$ levels in the moment-1 maps. The spectra for the \water and CO(6--5), as well as the moment-0 and moment-1 maps, were created from the naturally weighted data to match images produced during the lens modeling, as described in Section \ref{subsec:lens_modeling}. The \waterp emission spectrum, moment-0, and moment-1 maps were created from Briggs-weighted images using a robust factor of 0.5; see Section \ref{sec:observation_details}. 

Integrated flux densities and line luminosities were calculated following \citet{Solomon97}:
\begin{equation}
    L_{\rm line} = (1.04 \times 10^{-3})\,I_{\rm obs}\, \nu_{\rm rest}\,D^{2}_{L}\,(1+\textit{z})^{-1} [L_{\odot}],
\end{equation}
where $L_{\rm line}$ [L$_{\odot}$] gives the luminosity of the emission line, $I_{\rm obs}$ is the velocity integrated flux density defined as $I_{\mathrm{obs}} = S_{\mathrm{line}} \Delta V$ [Jy\,km/s], where $S_{\mathrm{line}}$ [mJy] is the observed flux density and $\Delta V$ [km/s] is the full-width half-maximum of the emission line, $\nu_{\rm rest}$ [GHz] is the rest-frame frequency of the line, $D_{\rm L}$ is the luminosity distance [Mpc], and $z$ is the redshift. Line properties for the emission lines detected towards G09v1.97 are provided in Table \ref{tab:Line_properties}. 

\begin{table*}[h]
    \centering
    \caption{Line properties from our observations. Note that none of the properties below are corrected for lensing magnification.}
    \begin{tabular}{l c c c c c c} \hline \hline
        Line & $\nu_{\rm rest}^{a}$ & $S_{v, \rm peak}^{b}$ & FWHM$_{\rm peak}^{c}$ & $I_{\mathrm{mol}}^{d}$ & $L_{\rm mol}^{e}$  \\  
         & [GHz] & [mJy] & [Jy km $\rm s^{-1}$] & [km\,$\rm s^{-1}$] & [$10^{9}\,L_{\odot}$] \\ \hline
        
        CO(6--5) & 691.47 & $8.3 \pm 1.3$  & $340 \pm 103$ & $12.7 \pm 1.3$ & $2.1 \pm 0.21$ \\
         & & $38.8 \pm 1.7$ & $235 \pm 16$ & & \\
        
        \water & 752.03 & $9.0 \pm 1.5$  & $470 \pm 124$ & $9.8 \pm 1.7$ & $1.8 \pm 0.3$ \\
        & & $21.2 \pm 2.1$ & $235 \pm 31$ & & \\
        
        \waterp & 742.15 & $3.8 \pm 0.3$ & $241 \pm 26$ & $1.1 \pm 0.2$ & $0.20 \pm 0.03$ \\
        & & $0.53 \pm 0.34$ & $235 \pm 182$ & & \\ \hline
    \end{tabular}
        \tablefoot{
        \tablefoottext{a}{Rest-frequency of the observed line.}
        \tablefoottext{b}{Specific intensity of the two peaks comprising the double Gaussian fit.}
        \tablefoottext{c}{FWHM of the two components comprising the double Gaussian fit.}
        \tablefoottext{d}{Intensity of the entire line.}
        \tablefoottext{e}{Luminosity of the entire line.}
        }
    \label{tab:Line_properties}
\end{table*}

\begin{figure*}[h]
    \centering
    \includegraphics[width = 0.8\linewidth]{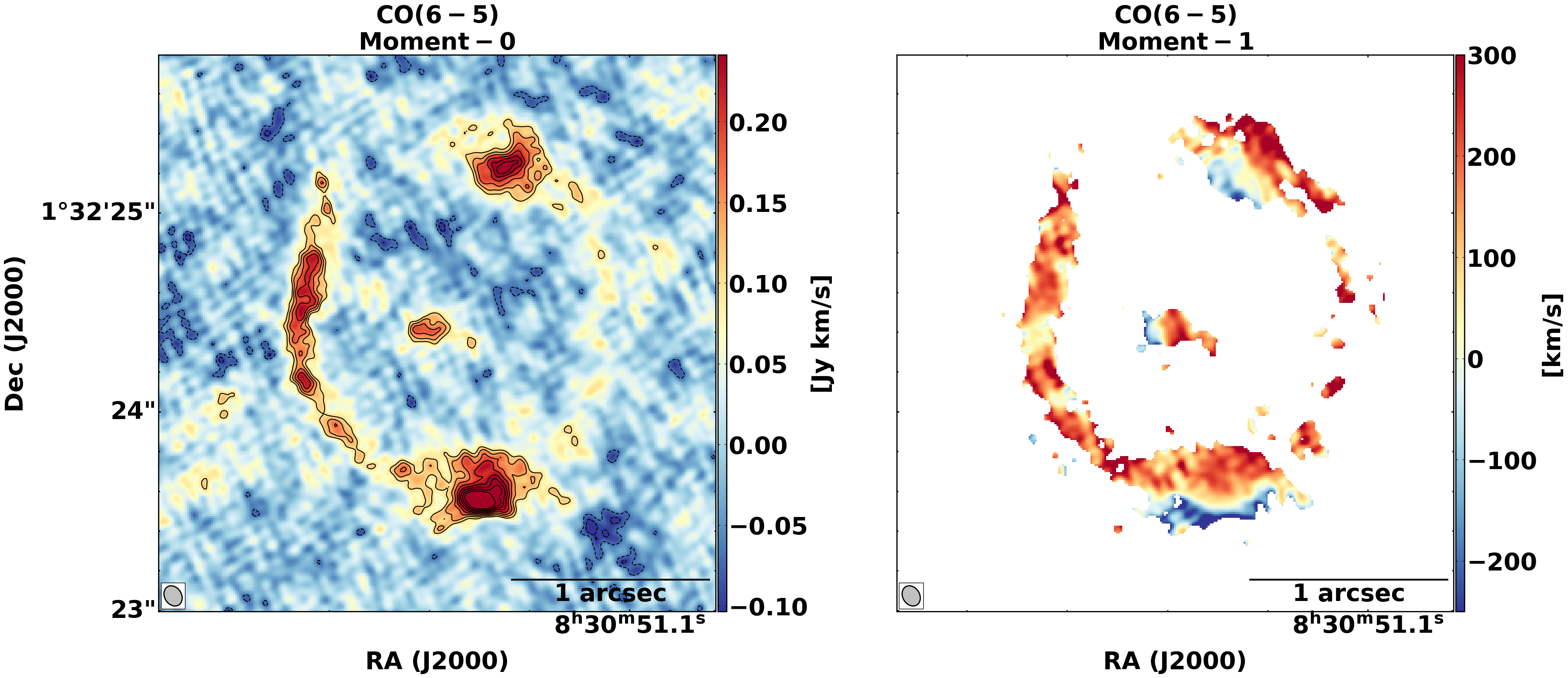}
    \includegraphics[width = 0.8\linewidth]{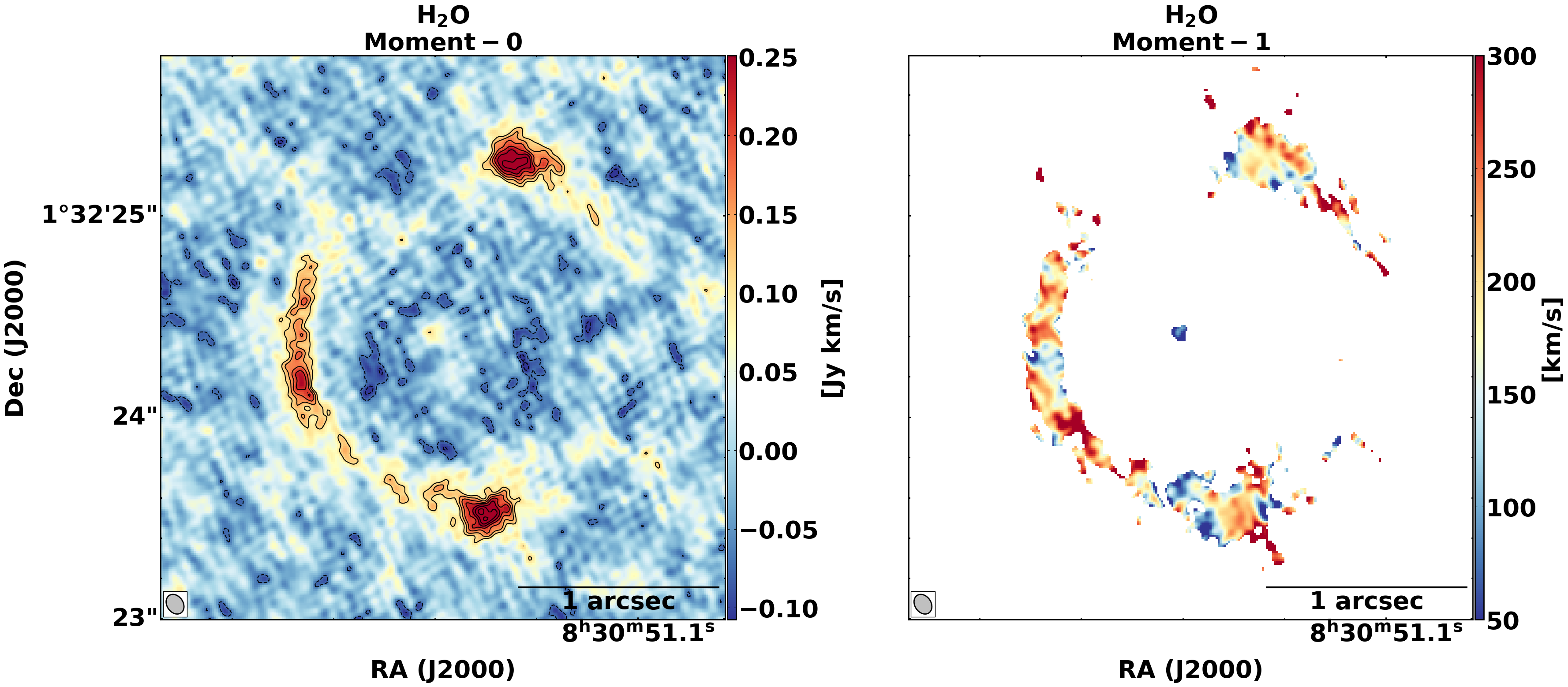}
    \includegraphics[width = 0.8\linewidth]{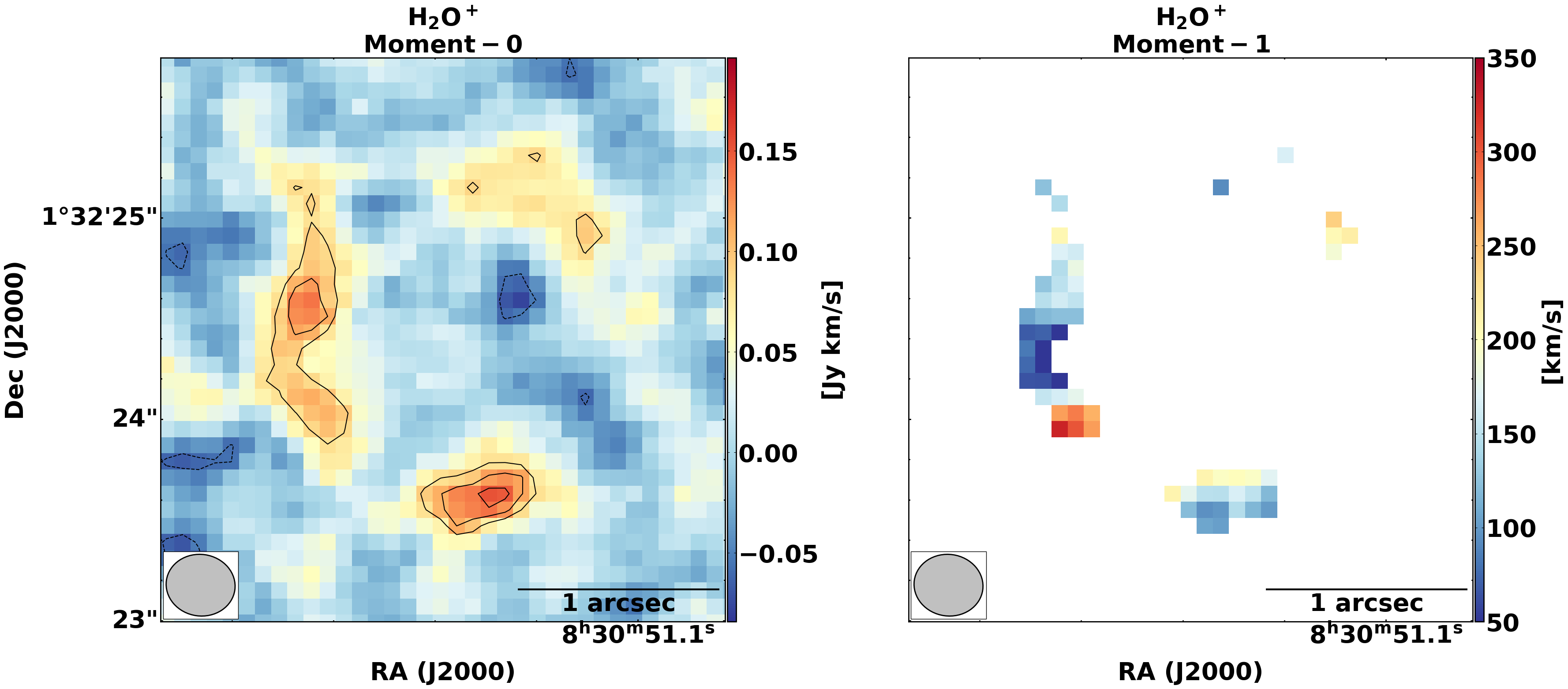}
    \caption{Moment-0 and moment-1 maps of the molecular line emission detected towards G09v1.97, top: CO(6--5), middle: H$_2$O, bottom: \ce{H_2O}$^+$. The contours are shown at $-3, -2, 3, 4, 5, 6, 7, 8, 9, 10\sigma$ levels for the CO(6--5) and \water emission and at $-3, -2, 3, 4, 5\sigma$ levels for the \waterp emission. The synthesized beam is shown in the bottom left of every image, beam sizes can be found in Table \ref{tab:obs_details} for each line. Note that the angular resolution is significantly lower for the \waterp emission as it was imaged from the combined dataset as described in Section \ref{sec:observation_details}. A clear velocity gradient is visible in the moment-1 map of the CO(6--5) emission. }
    \label{fig:img_plane_mom0_mom1}
\end{figure*}

\subsubsection{CO(6--5) emission} \label{subsubsec:co_line_results}

The spectrum of the CO(6--5) emission exhibits the same characteristic profiles as reported in \citet{Yang19}, with peaks in both the red ($>0$\,km/s) and blue ($<0$\,km/s) regions of the spectrum. The spectrum was fit with a double Gaussian profile to account for the asymmetric line shape. The ratio of the fitted peaks of the two emission lines from the Gaussian fitting is approximately $\mathrm{peak_{red} / peak_{blue}} \approx 4.5$ for CO(6--5). The flux of the CO(6--5) emission line was compared between the high-resolution ALMA dataset and the combined dataset, with no obvious difference observed (see Fig. \ref{fig:co_combined_highres_spec_comparison}). This suggests that, despite the very high resolution of the new CO(6--5) observations, no significant emission was resolved out. 

A clear velocity gradient is seen in the moment-1 maps of the CO(6--5) data in the brightest regions of emission in the north and south of the image (see Fig. \ref{fig:img_plane_mom0_mom1}). This is further investigated later in the paper following lens modeling; see Section \ref{subsec:kin_modeling}. We present the spectrum in Fig. \ref{fig:img_plane_spectra} and the moment maps in Fig. \ref{fig:img_plane_mom0_mom1}.

\subsubsection{H$_2$O emission} \label{subsubsec:h2o_line_results}
The spectrum of the \water emission also exhibits the same characteristic shape as previously reported \citep{Yang19} and closely matches the line profile of the CO(6--5) emission with the two peaks in the red and blue regions of the spectrum. We compare the normalized CO and \water spectra in Fig. \ref{fig:img_plane_spec_comparison} to illustrate the similarity of the emission line profiles. Similar to the CO(6--5) emission, we fit the spectrum with a double Gaussian profile to account for the asymmetric line shape. In comparison to the CO(6--5) spectrum, the peak in the blue part of the spectrum has significantly stronger flux. The ratio of the fitted peaks of the two emission lines from the Gaussian fitting is $\mathrm{peak_{red} / peak_{blue}} \approx 2.3$ for \water. This is $\sim 1.5\times$ smaller than the same ratio for the CO(6--5) emission, suggesting that the peak of the \water emission in the red portion of the spectrum is relatively lower compared to its counterpart in the blue portion of the spectrum. This is further discussed in Section \ref{sec:G09_in_source_plane}.   

By contrast to the CO(6--5) emission, the velocity gradient in the image-plane moment-1 map of the \water emission is much less prominent. This is likely due to the relatively lower SNR of the \water emission, in line with results reported in \citet{Yang19}, where similar velocity structures have been seen from a lower resolution data of CO(6--5) and the \water data. Given the SNR of the CO data, we focus on the CO(6--5) emission when discussing the kinematics of G09v1.97.  We present the spectrum in Fig. \ref{fig:img_plane_spectra} and the moment maps in Fig. \ref{fig:img_plane_mom0_mom1}.

\subsubsection{H$_2$O$^+$ emission} \label{subsubsec:h2op_line_results}
\begin{figure}[!htbp]
    \centering
    \includegraphics[width = 1.0\linewidth]{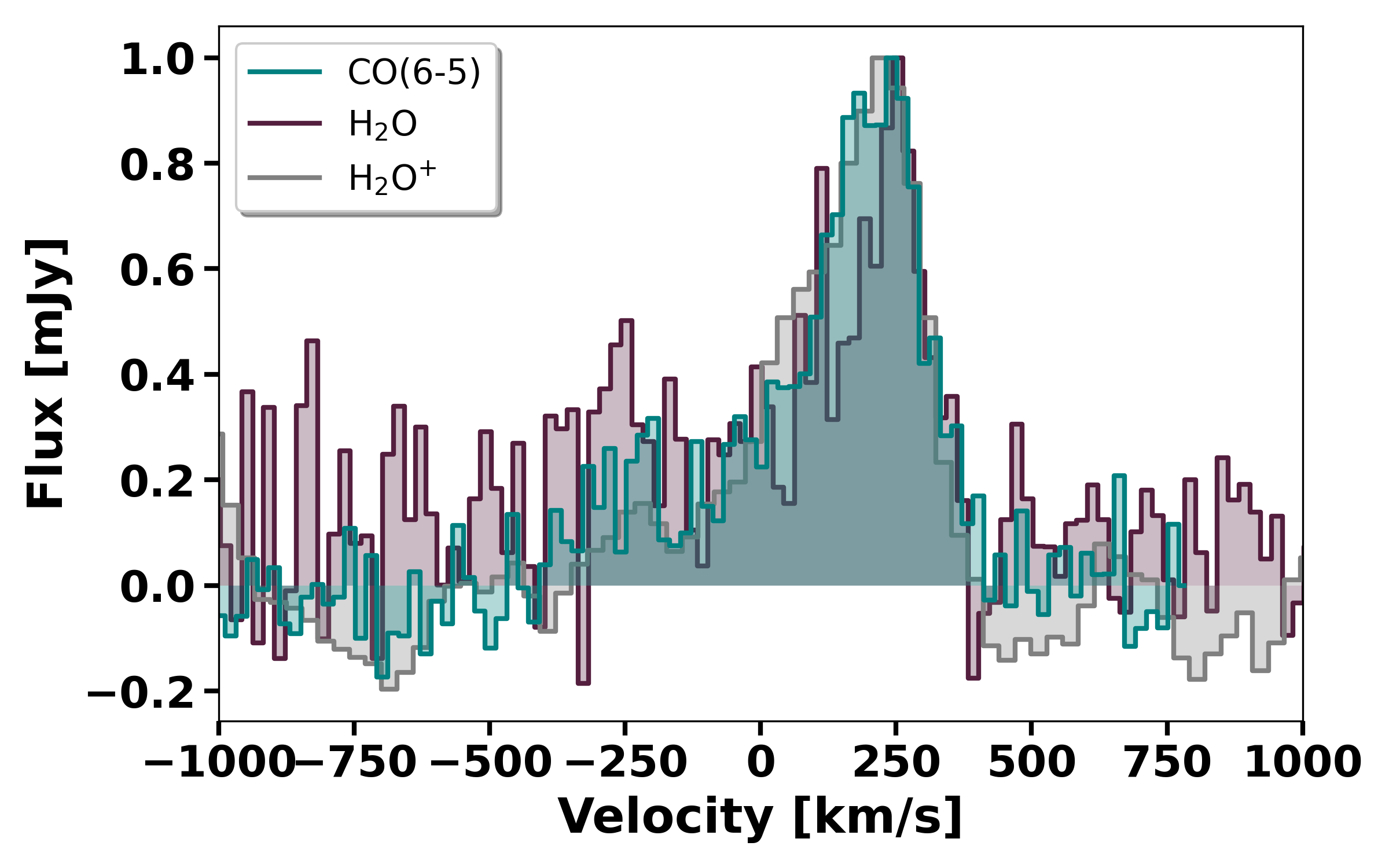}
    \caption{Comparison of the normalized CO(6--5), \ce{H_2O}, and \waterp emission. Note that the noise appears significantly higher in the \waterp emission due to the line's relatively lower SNR ratio. The line profiles of all three emission lines are very similar.}
    \label{fig:img_plane_spec_comparison}
\end{figure}

The spectrum of the \waterp emission is at sufficiently low SNR so as to be undetectable in the new, high-resolution observations. Instead, we used the combined data to improve on both the angular resolution and the SNR of the \waterp emission. As in \citet{Yang19}, we find a similar line profile to that of the CO(6--5) and \water emission with characteristic peaks in the red and blue regions of the spectrum, indicating that their spatial origins are comparable (Fig. \ref{fig:img_plane_spec_comparison}). We show the normalized spectral comparison in Fig \ref{fig:img_plane_spec_comparison}. As with the CO(6--5) and \water emission, we fit the spectrum using a double Gaussian profile to account for the asymmetric line shape. The ratio of the fitted peaks of the two emission lines from the Gaussian fitting is $\mathrm{peak_{red} / peak_{blue}} \approx 7$ for H$_2$O$^{+}$. This is $\sim3\times$ larger than the same ratio for the CO(6--5) emission and $\sim 5\times$ larger than the same ratio for the H$_2$O. However, it should be noted that the SNR ratio of the blue component in the \waterp spectrum is largely below the noise level of the spectrum, and therefore, it is difficult to draw firm conclusions about the nature of this emission and the spectral shape.

Similar to the \water emission, we do not detect a clear velocity gradient in the image-plane moment-1 map of the \waterp emission. This is not surprising given the low SNR of the \waterp emission and the low angular resolution of the images. Since the line profile of \waterp is similar to that of the CO(6--5), we expect an overall comparable kinematic structure. We present the spectrum in Fig. \ref{fig:img_plane_spectra} and the moment maps in Fig. \ref{fig:img_plane_mom0_mom1}.

\subsection{Lens modeling} \label{subsec:lens_modeling}
Advanced lens modeling software, such as the publicly available Python-based code {\sc PyAutoLens} \citep{Nightingale16, Nightingale18, Nightingale21}, allows for robust source-plane reconstructions of lensed galaxies. {\sc PyAutoLens} has been used in a number of recent studies of high-redshift galaxies to determine source-plane properties of the studied galaxies \citep[][]{Maresca22, Lei23, Amvrosiadis24}. In this work, we utilized {\sc PyAutoLens} to reconstruct the source plane emission of G09v1.97. {\sc PyAutoLens} is capable of handling interferometric data and performing reconstructions directly on the visibilities, thereby minimizing the effect of correlated noise in interferometric images. Additionally, {\sc PyAutoLens} can create source plane pixelizations from interferometric data, which eliminates the need for assumptions about the background galaxy's morphology. This approach is particularly advantageous for G09v1.97 as it does not require a priori assumptions to be made about whether the galaxy is a rotating disk or a merger. In its current form, intensity values from {\sc PyAutoLens} fits performed on interferometric visibilities are in arbitrary units. While this does not impact our analysis, it is discussed in later sections when relevant. Note also that images produced by {\sc PyAutoLens} are dirty images using approximately a natural weighting scheme.

\begin{table*}[]
    \centering
    \caption{Best-fit lens parameters, optimized from parametric models of the dust continuum emission.}
    \begin{tabular}{l c c c c c c} \hline \hline
        Lens Name & $z^{a}$ & $x_{\mathrm{off}}^{b}$ & $y_{\mathrm{off}}^{c}$ & q$^{d}$ & PA$^{e}$ & Einstein Radius \\
         & & [''] & [''] & & [degrees] & [''] \\ \hline
        \vspace{2.0mm}
        
        Lens SE & 0.626 & $-0.21^{+0.02}_{-0.02}$ & $-0.08^{+0.02}_{-0.02}$ & $0.51 ^{+0.04} _{-0.04}$ & $18.97 ^{+4.03} _{-3.78}$ & $0.44^{+0.01}_{-0.01}$  \\
        
        \vspace{1.0mm}
        
        Lens NW & 1.002 & $0.14^{+0.02}_{-0.02}$ & $0.38^{+0.02}_{-0.02}$ & $0.77 ^{+0.05} _{-0.06}$ & $-73.53 ^{+7.12} _{-6.08}$ & $0.48^{+0.01}_{-0.01}$  \\
     \hline 
    \end{tabular}
    \tablefoot{
        \tablefoottext{a}{Spectroscopic redshift of the lens.}
        \tablefoottext{b}{x-position of the lens defined in offset from the observational phase center.}
        \tablefoottext{c}{y-position of the lens defined in offset from the observational phase center.}
        \tablefoottext{d}{Minor-to-major axis ratio of the lens.}
        \tablefoottext{e}{Position angle of the lens, defined by {\sc PyAutoLens} as counter-clockwise from the x-axis.}}
    \label{tab:bestfit_lens_model}
\end{table*}

\begin{figure*}[h]
    \centering
    \includegraphics[width = 0.19\linewidth]{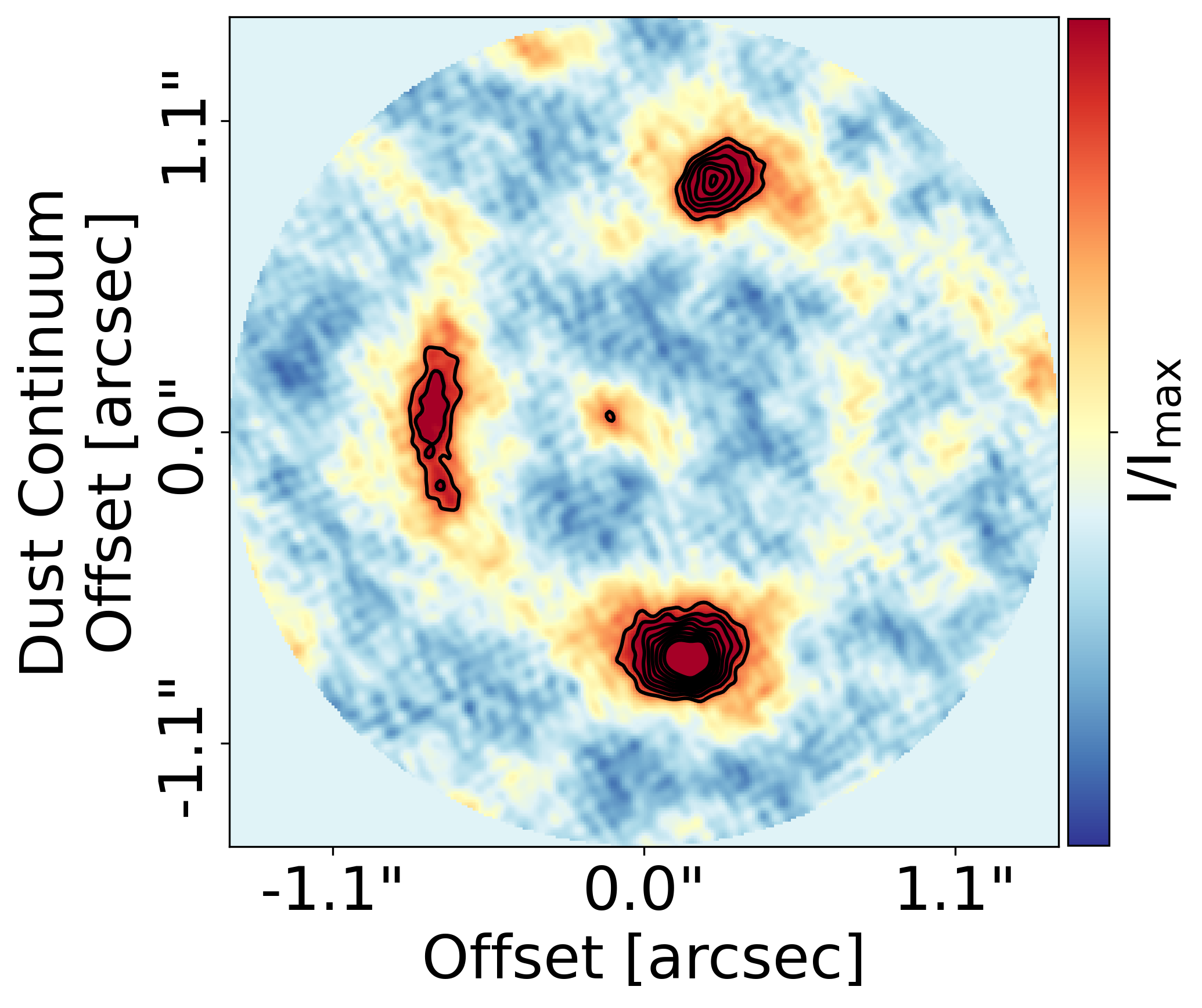}
    \includegraphics[width = 0.19\linewidth]{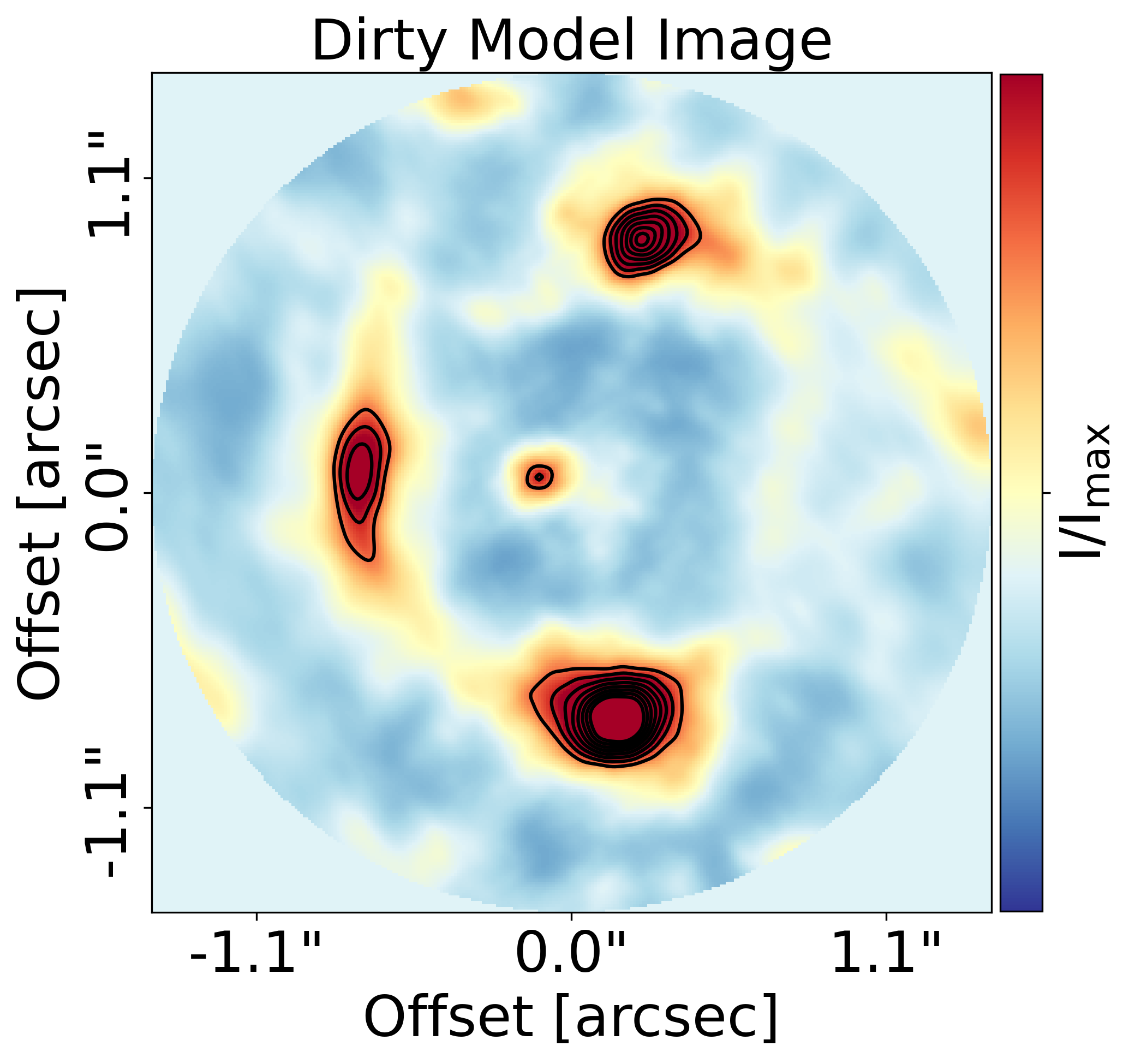}
    \includegraphics[width = 0.19\linewidth]{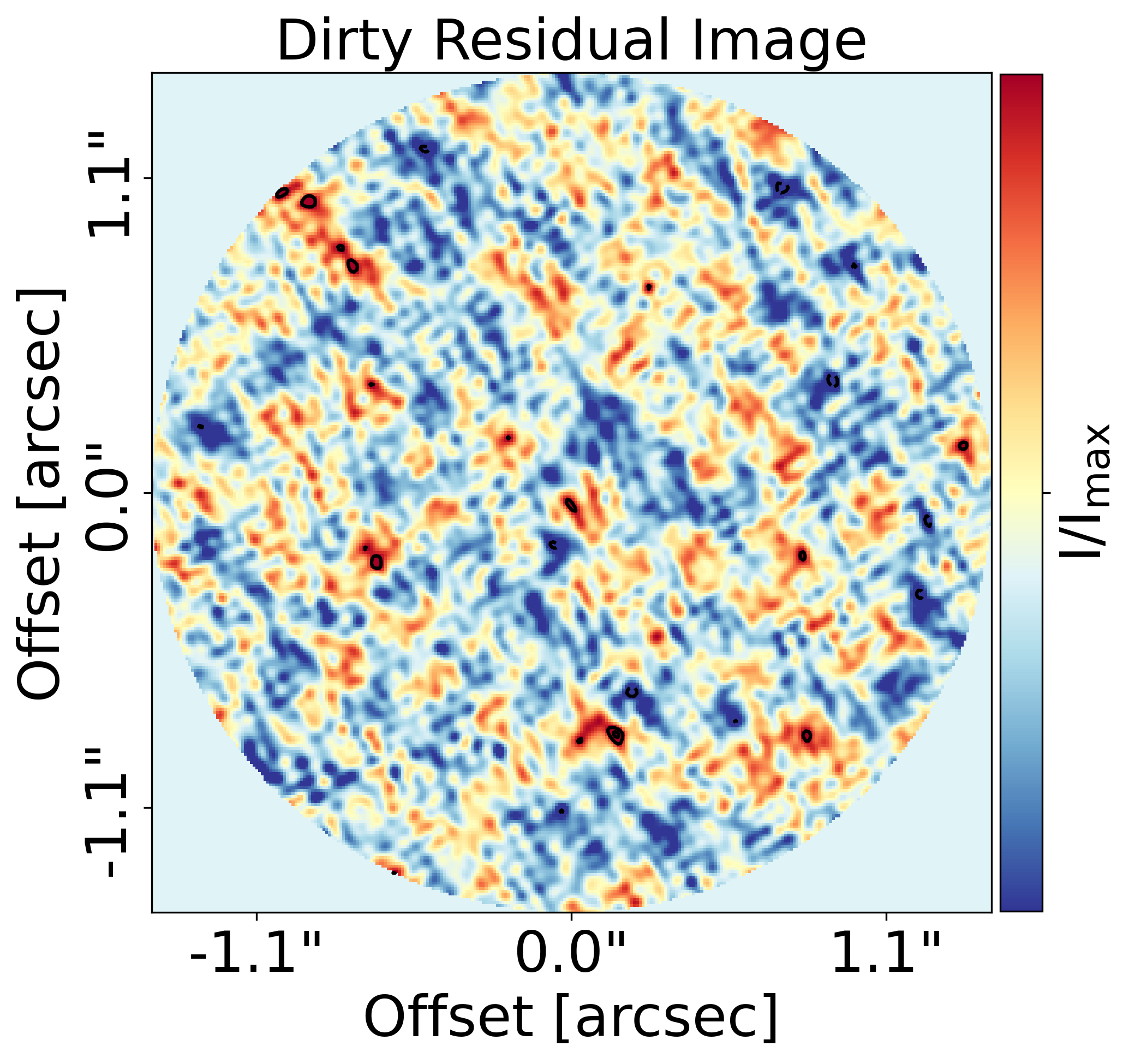}
    \includegraphics[width = 0.19\linewidth]{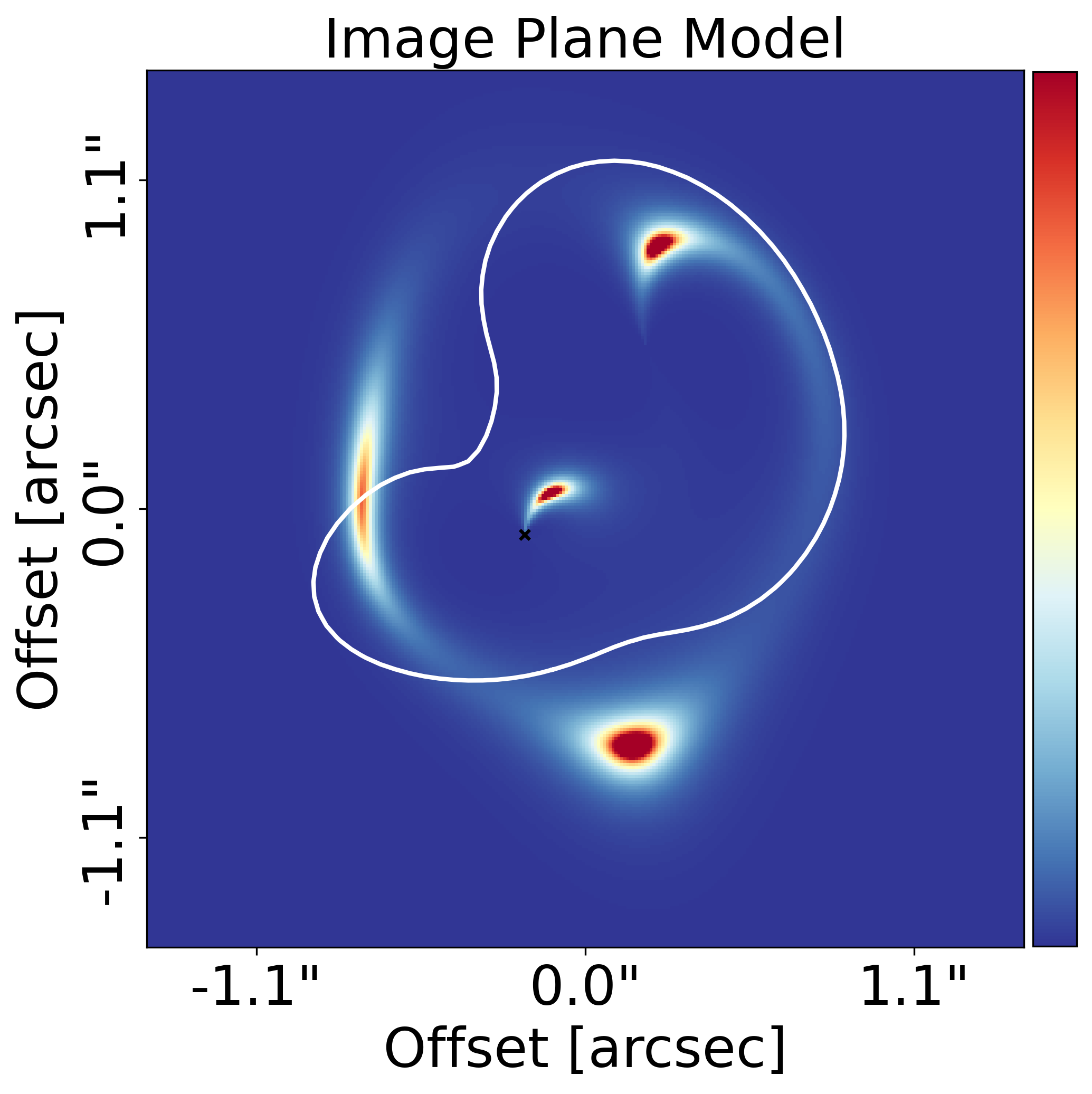}
    \includegraphics[width = 0.19\linewidth]{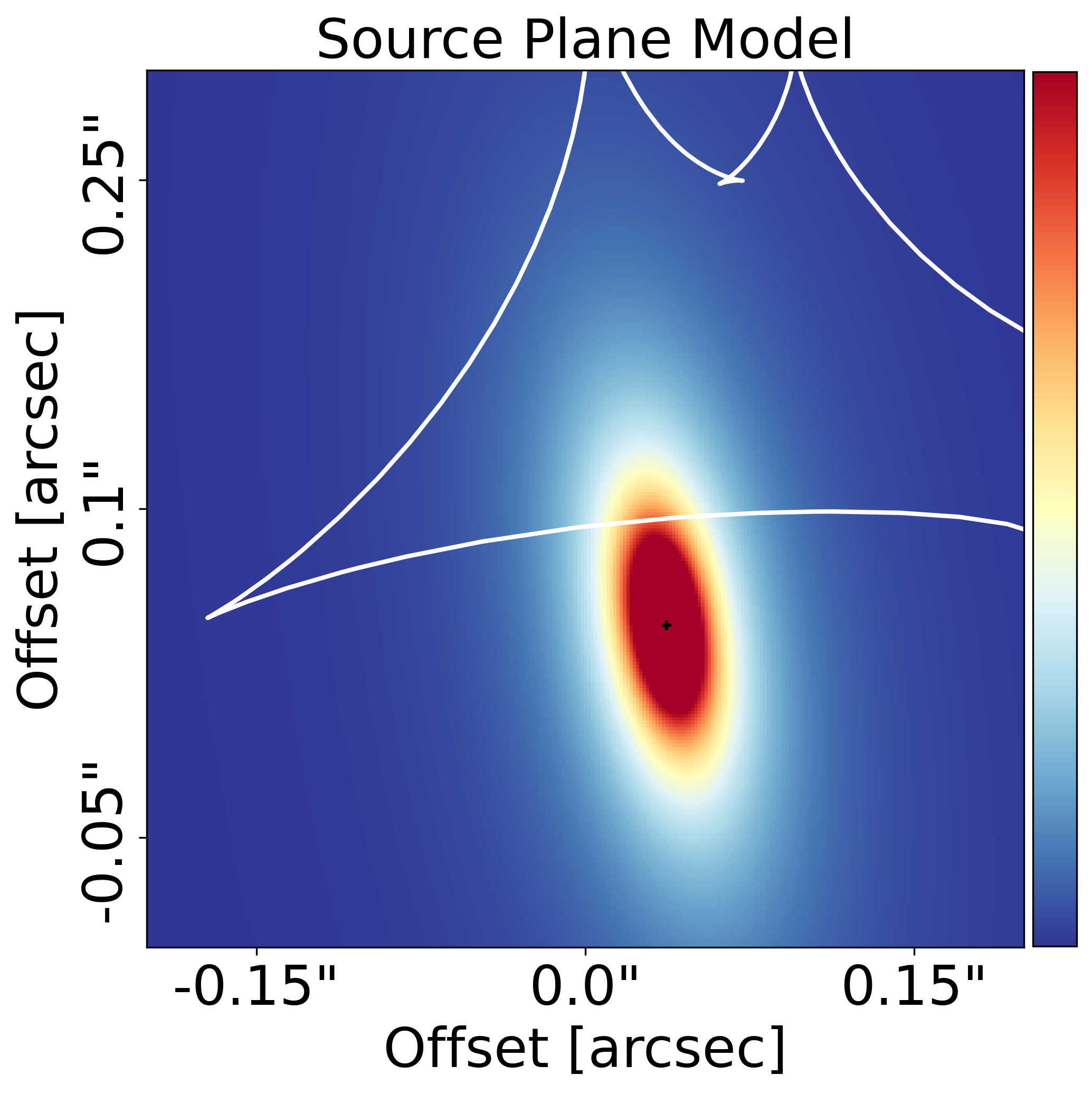}

    \includegraphics[width = 0.19\linewidth]{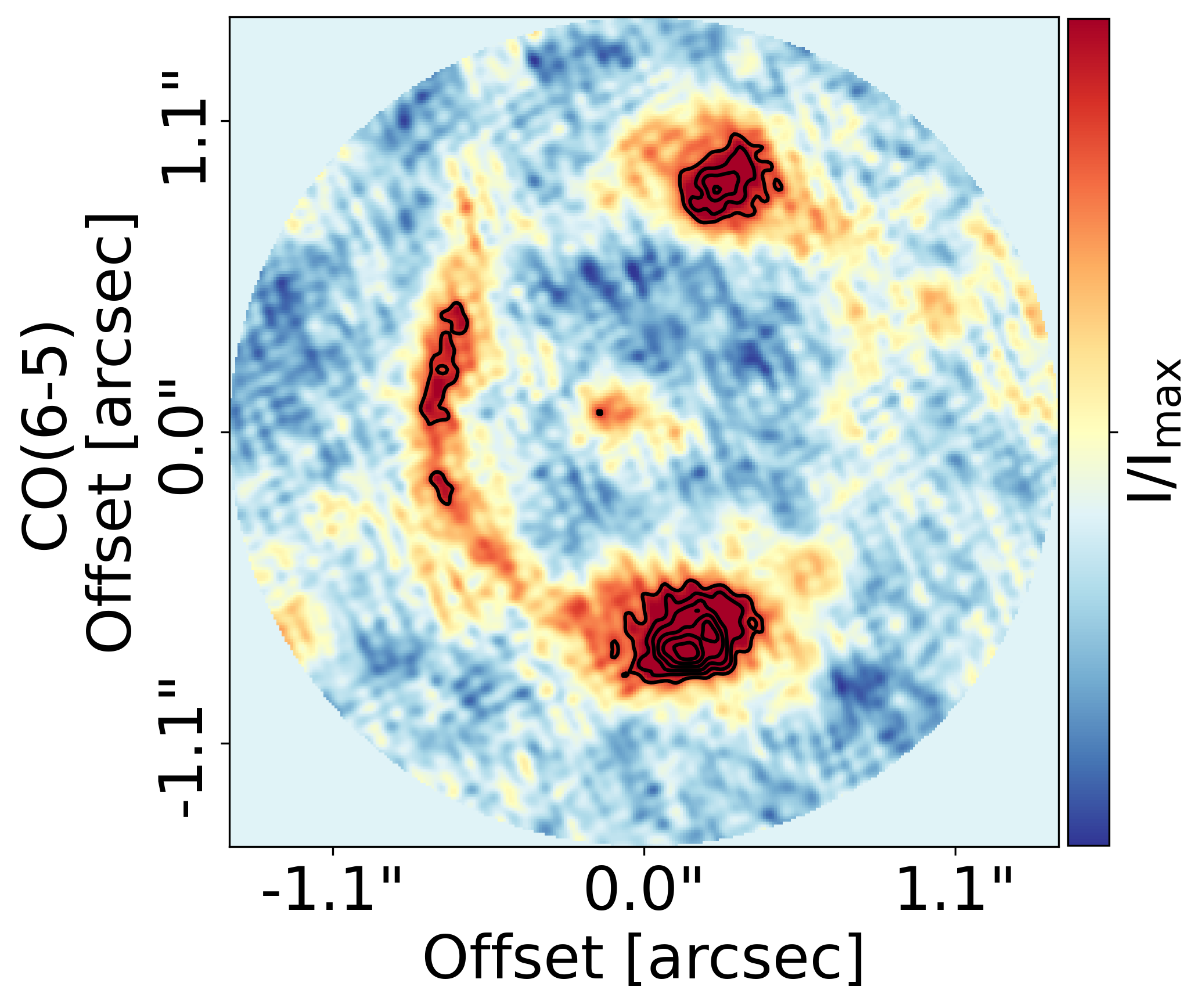}
    \includegraphics[width = 0.19\linewidth]{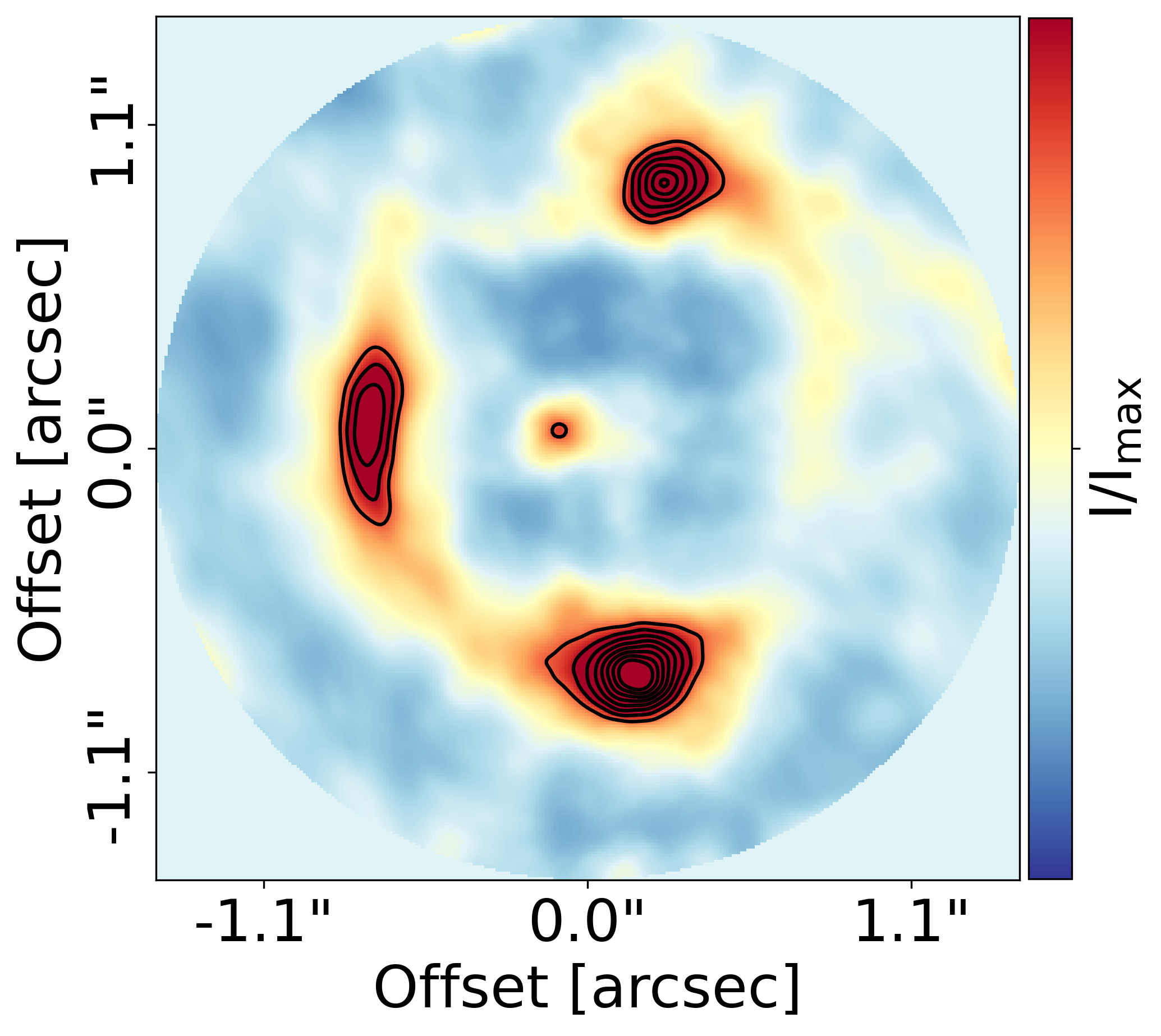}
    \includegraphics[width = 0.19\linewidth]{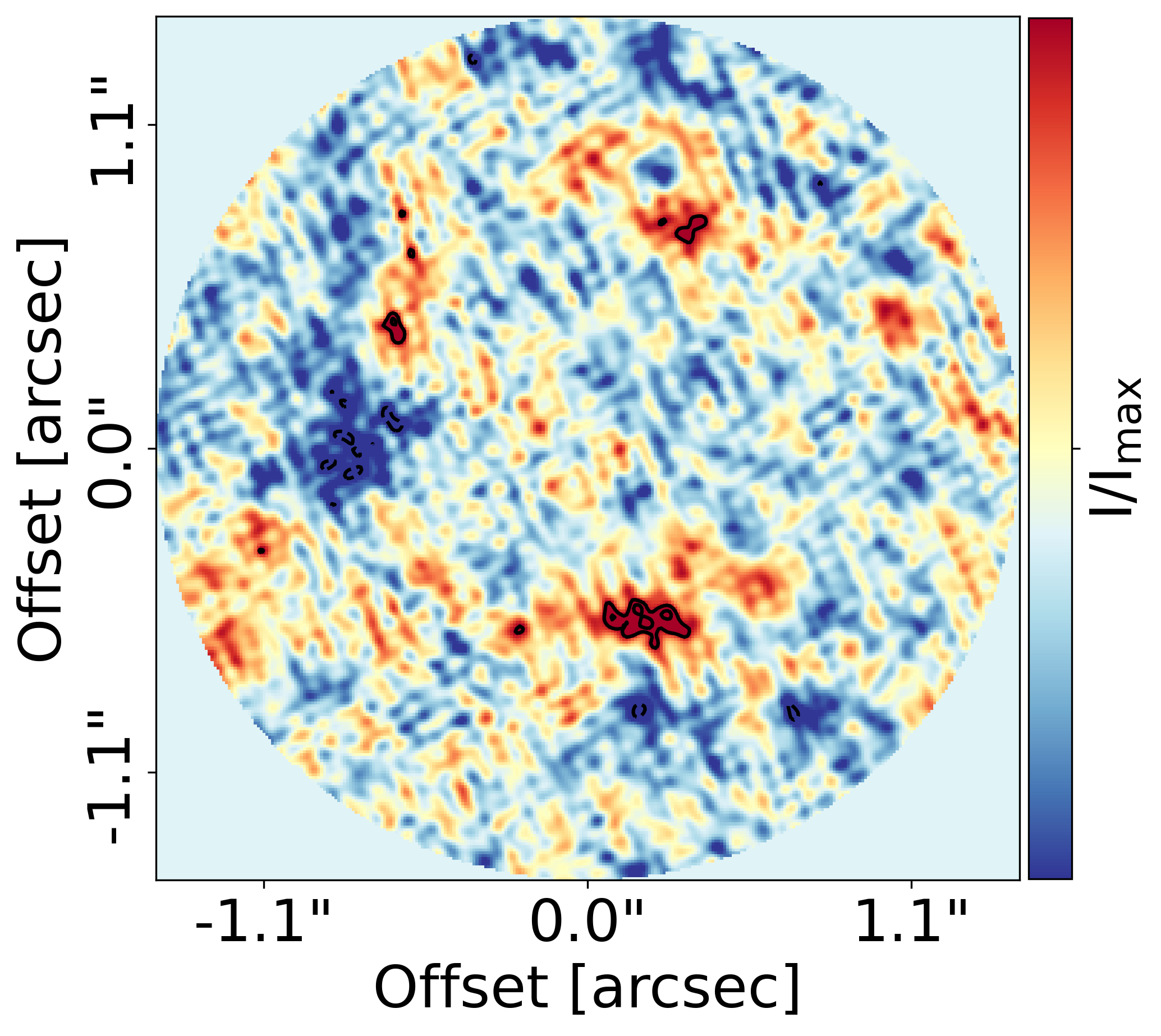}
    \includegraphics[width = 0.19\linewidth]{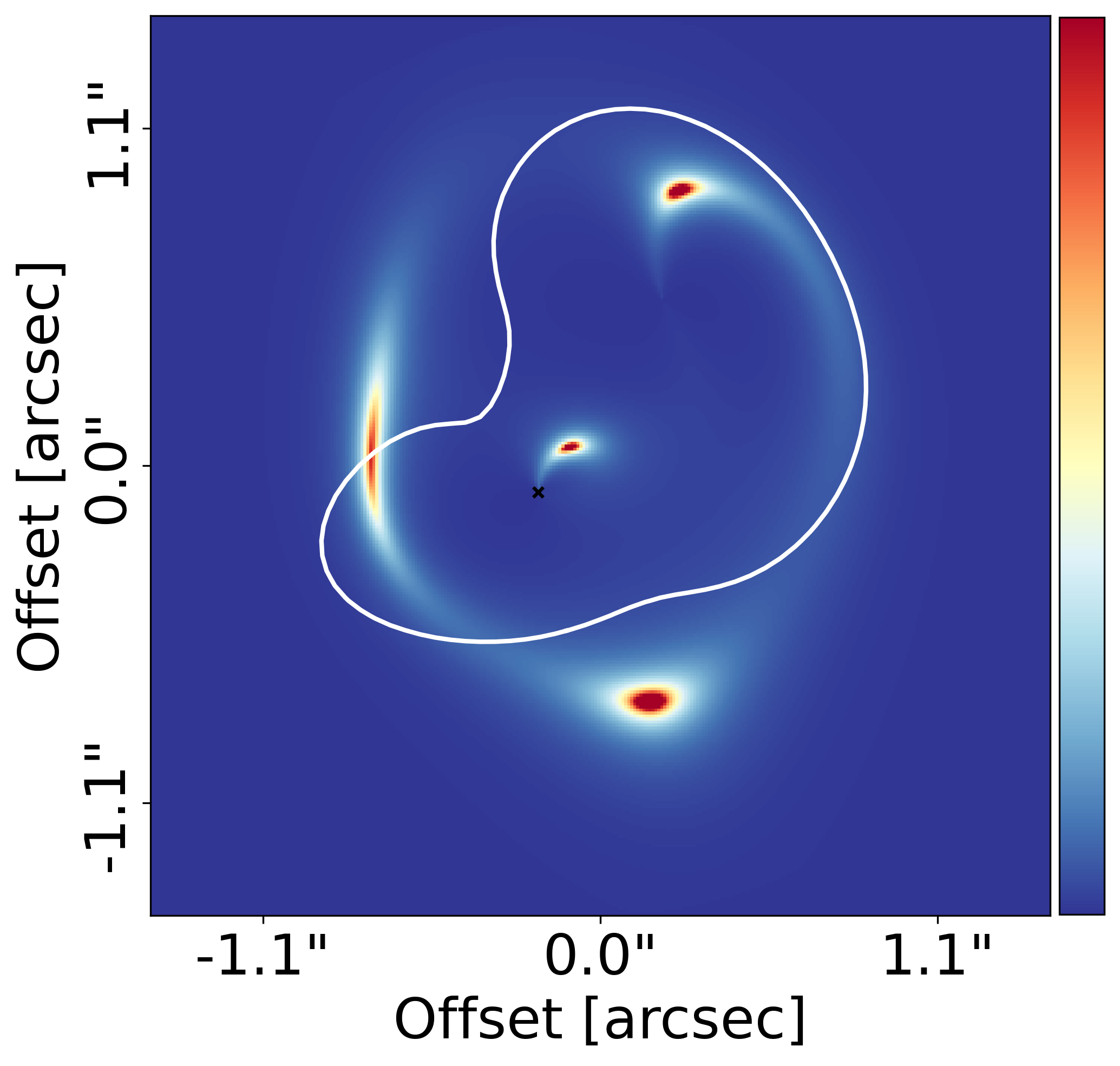}
    \includegraphics[width = 0.19\linewidth]{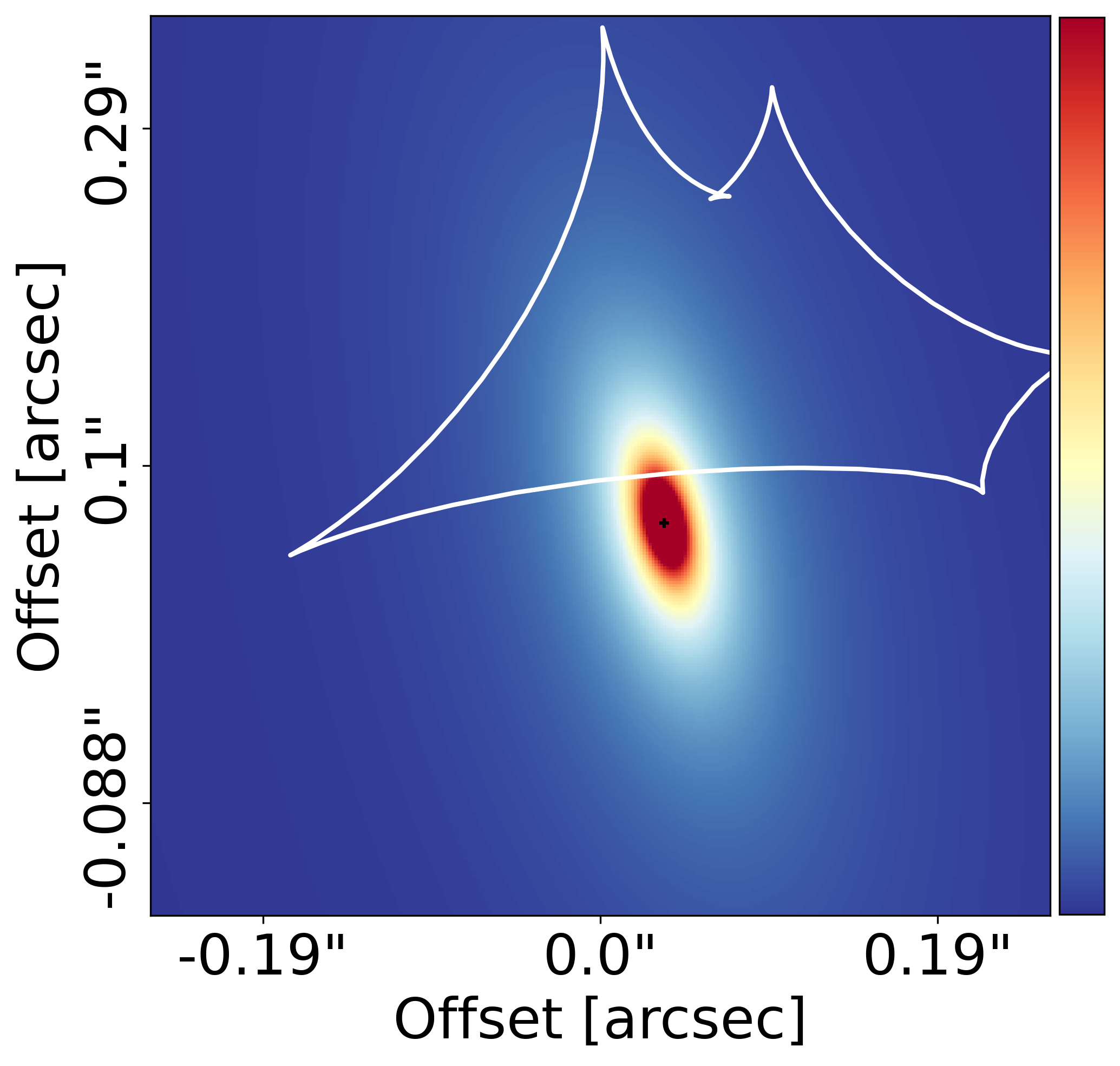}
    
    \includegraphics[width = 0.19\linewidth]{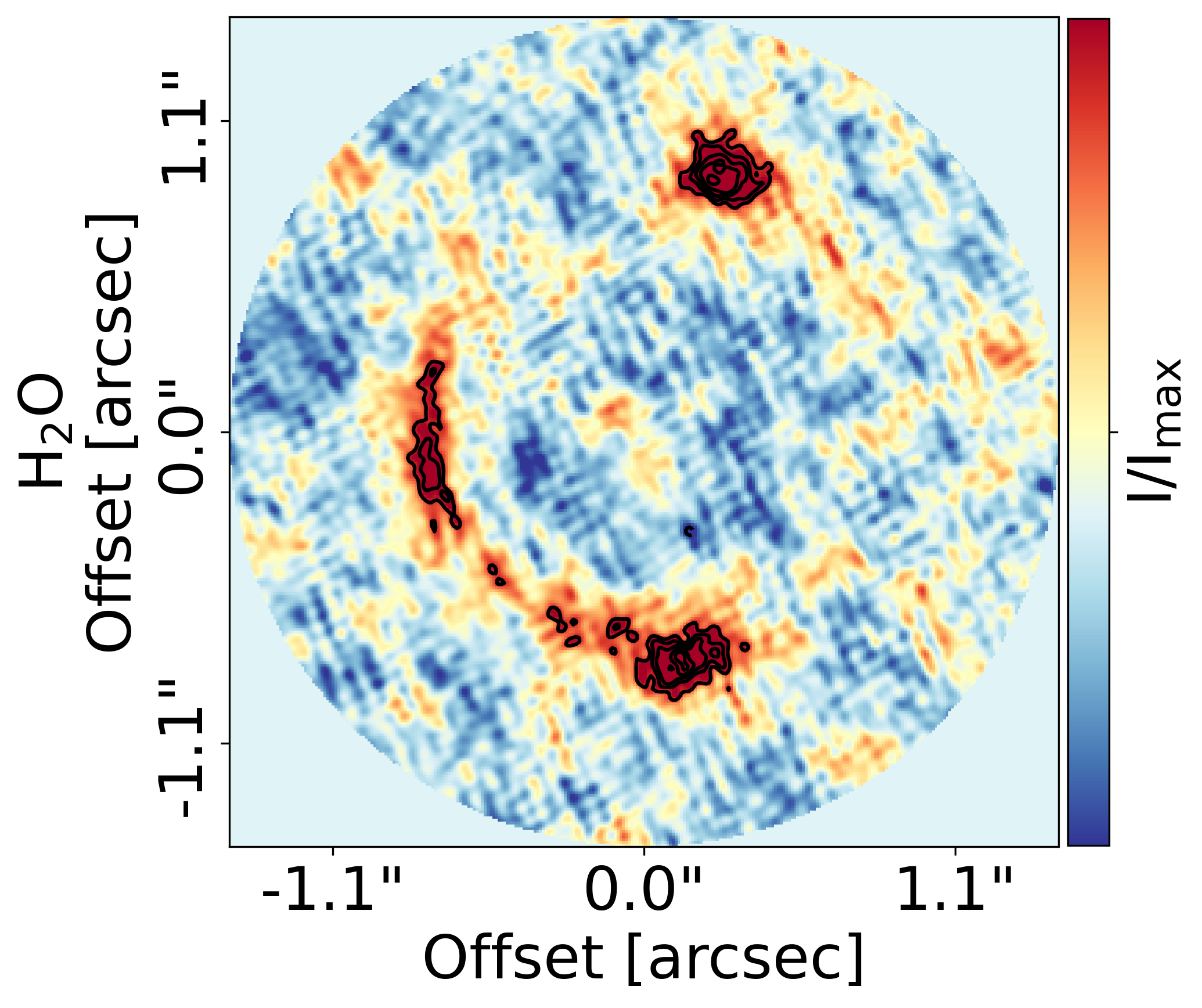}
    \includegraphics[width = 0.19\linewidth]{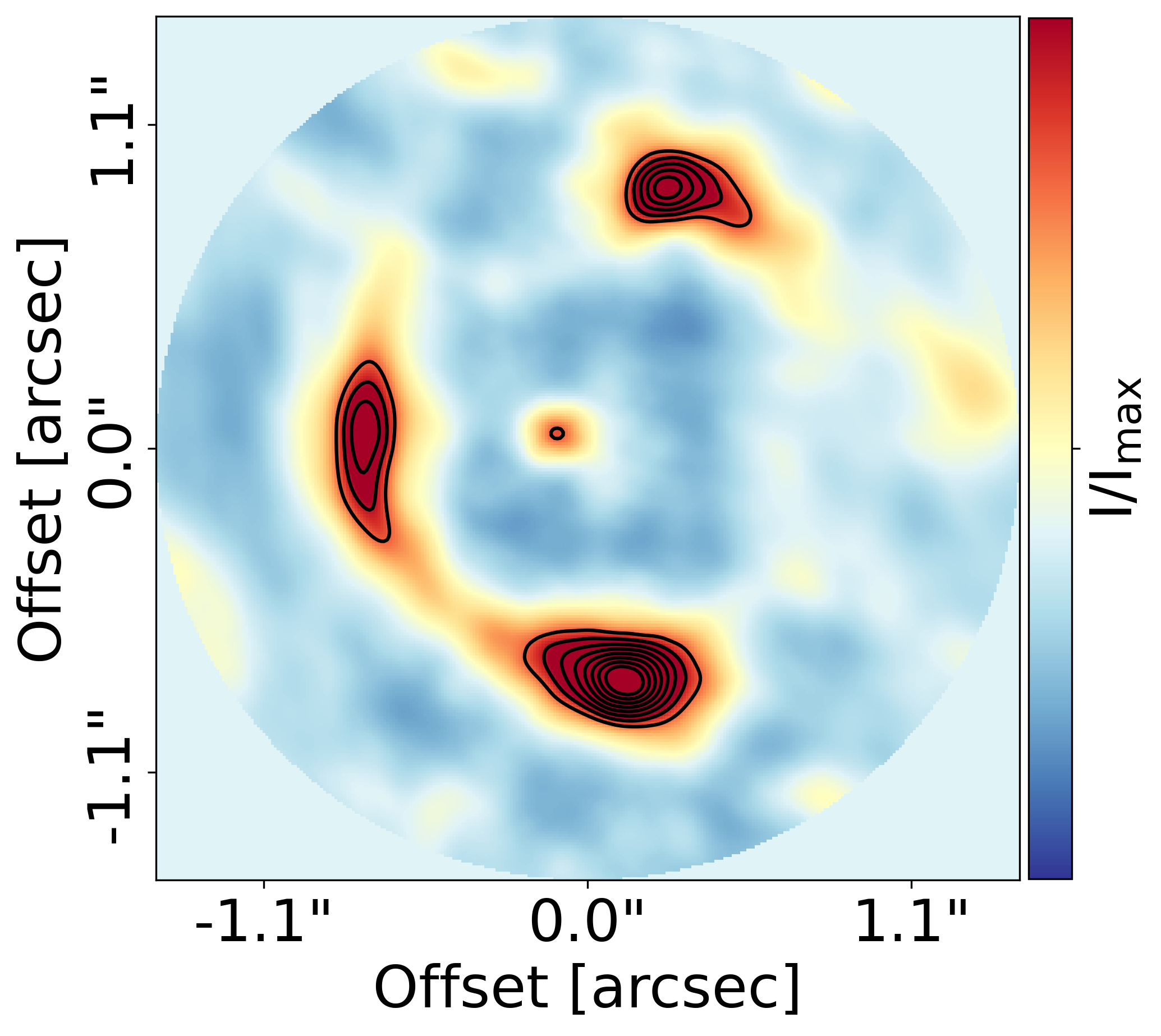}
    \includegraphics[width = 0.19\linewidth]{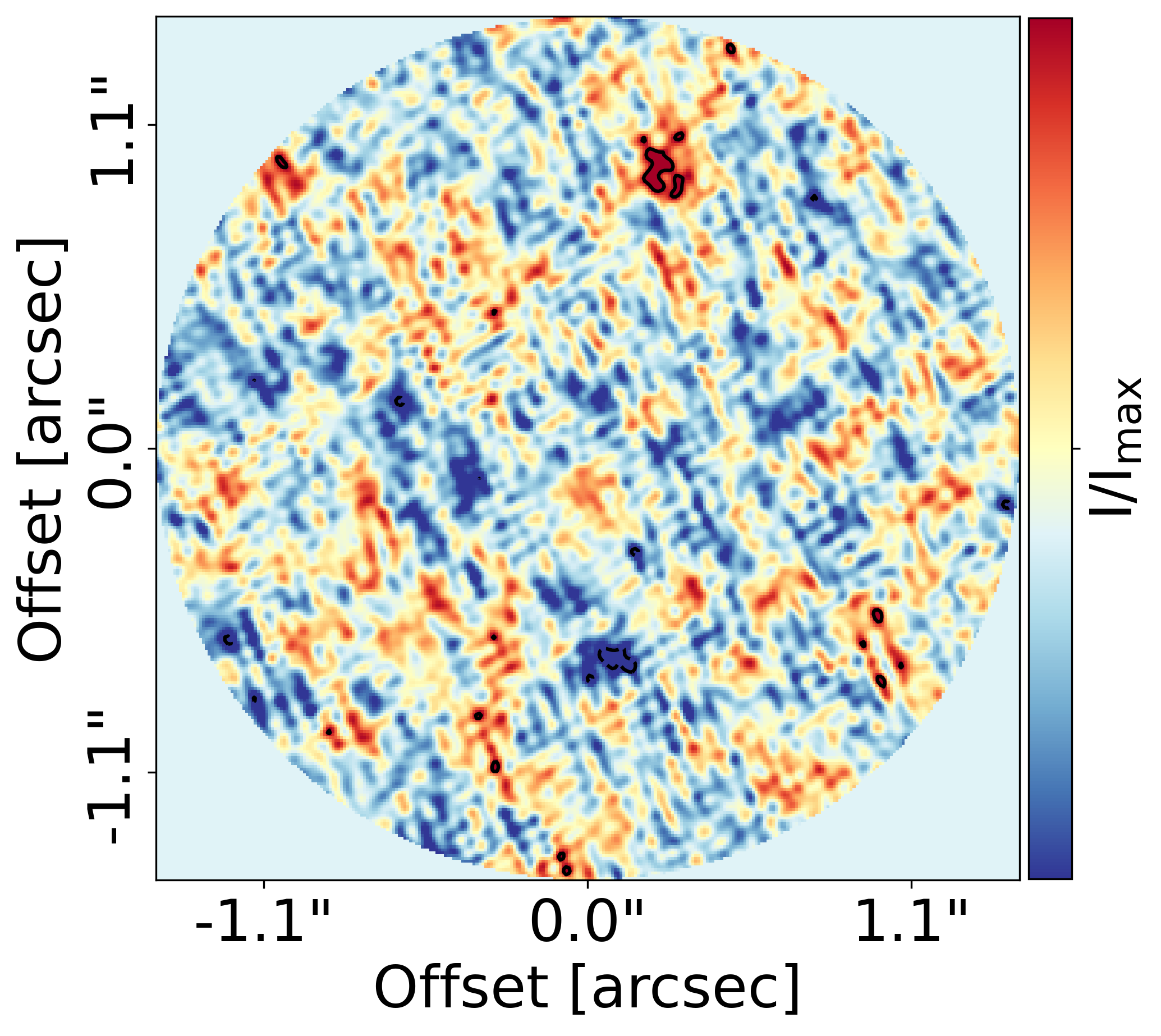}
    \includegraphics[width = 0.19\linewidth]{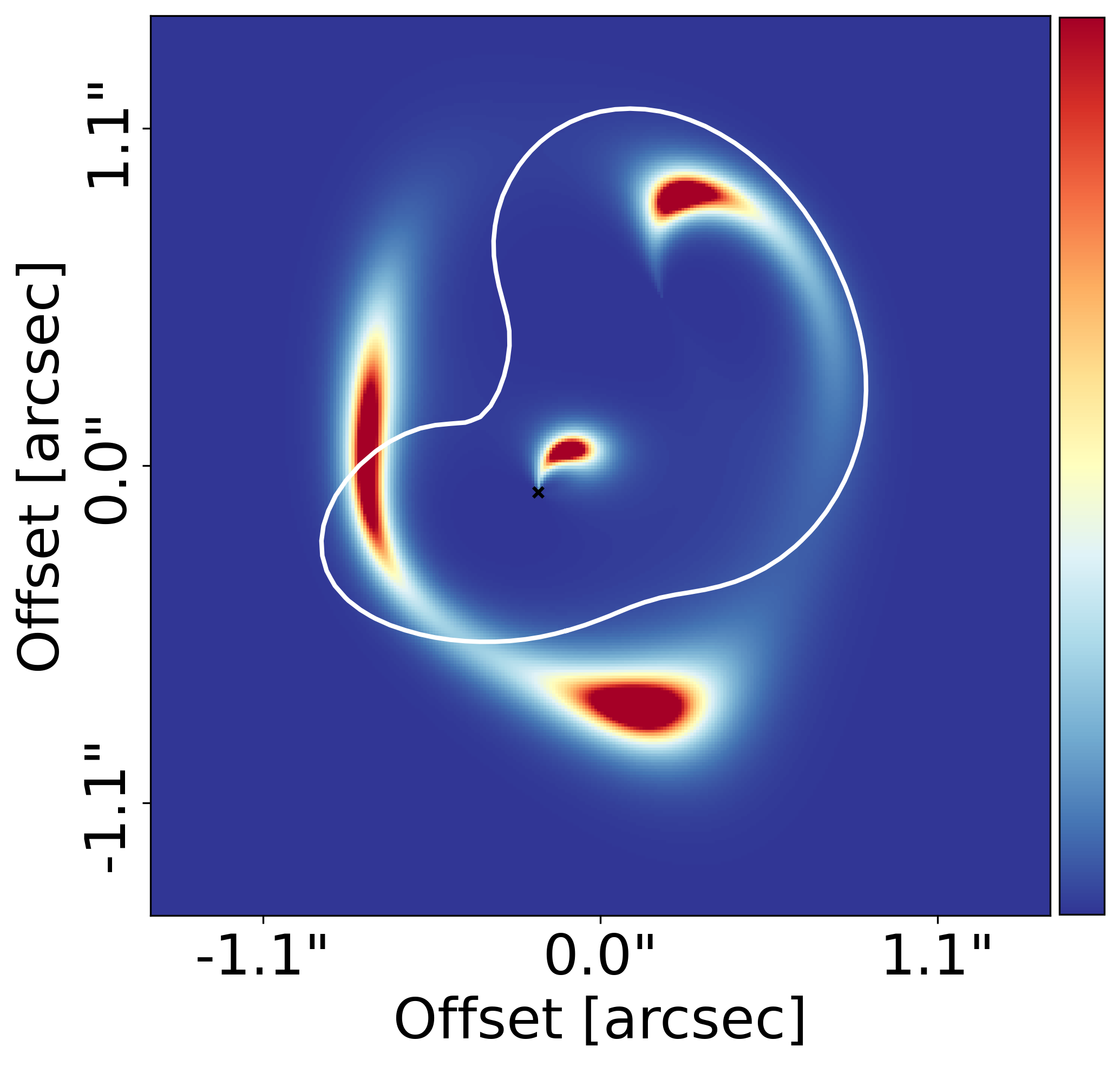}
    \includegraphics[width = 0.19\linewidth]{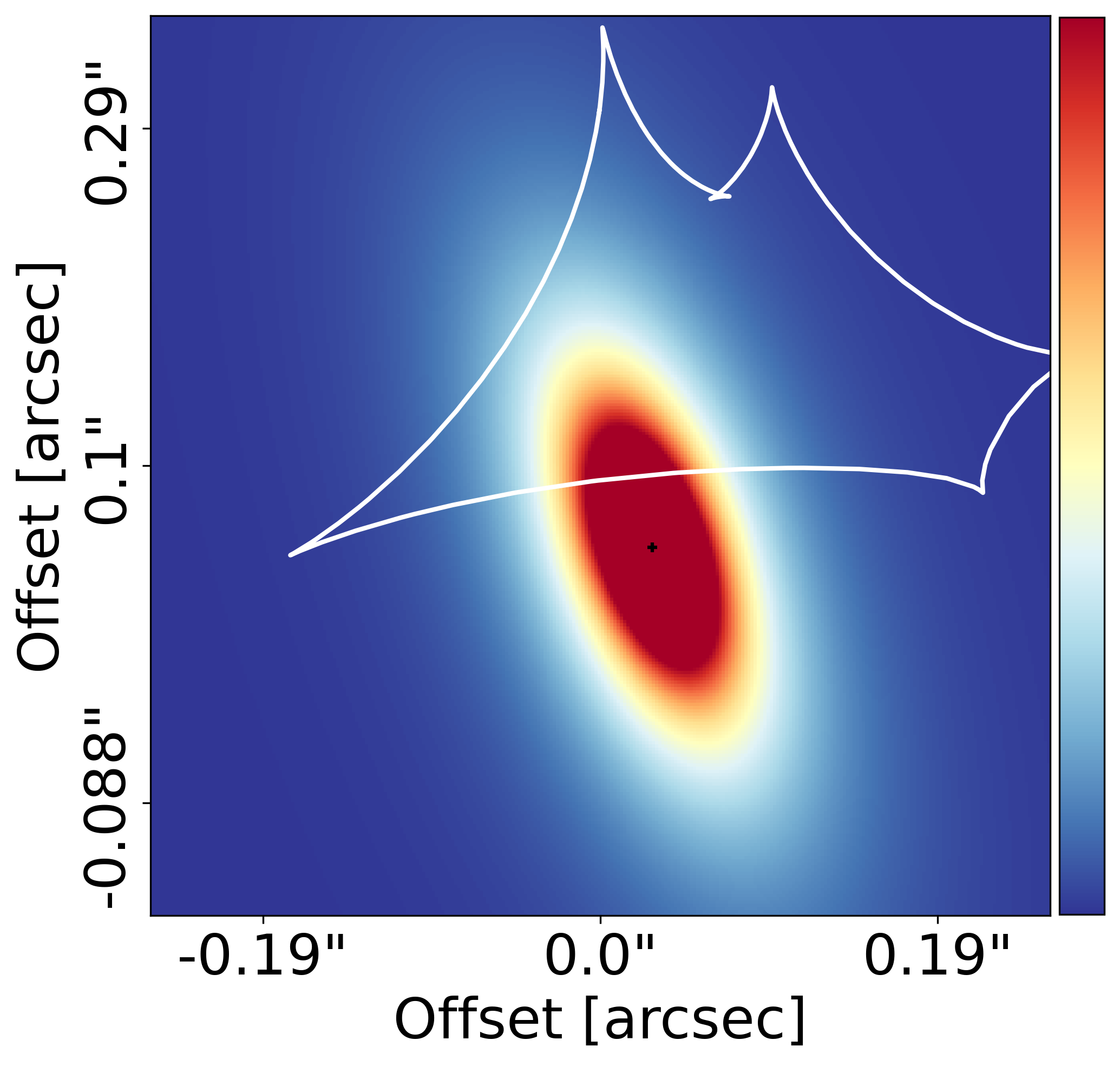}

    \includegraphics[width = 0.19\linewidth]{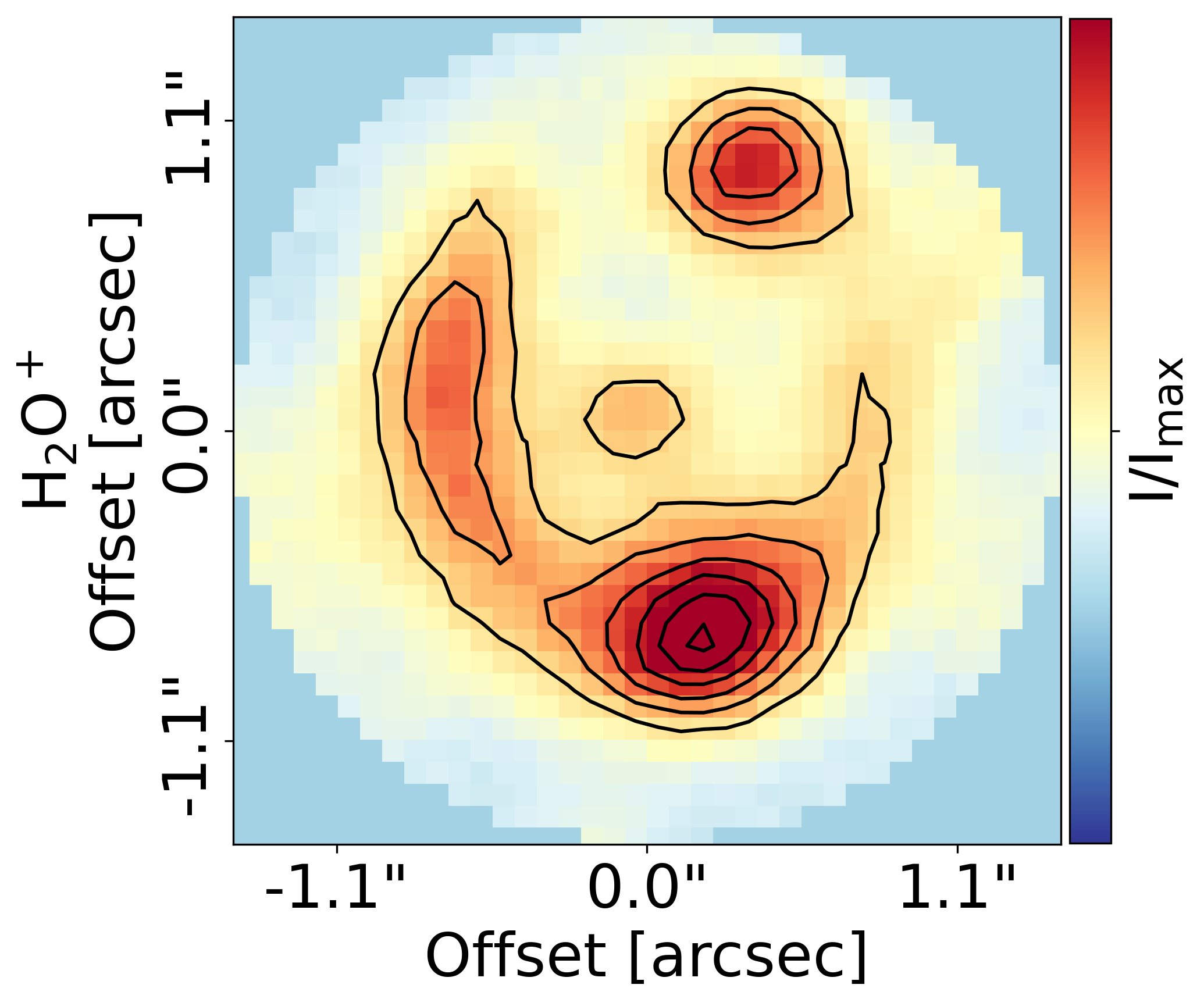}
    \includegraphics[width = 0.19\linewidth]{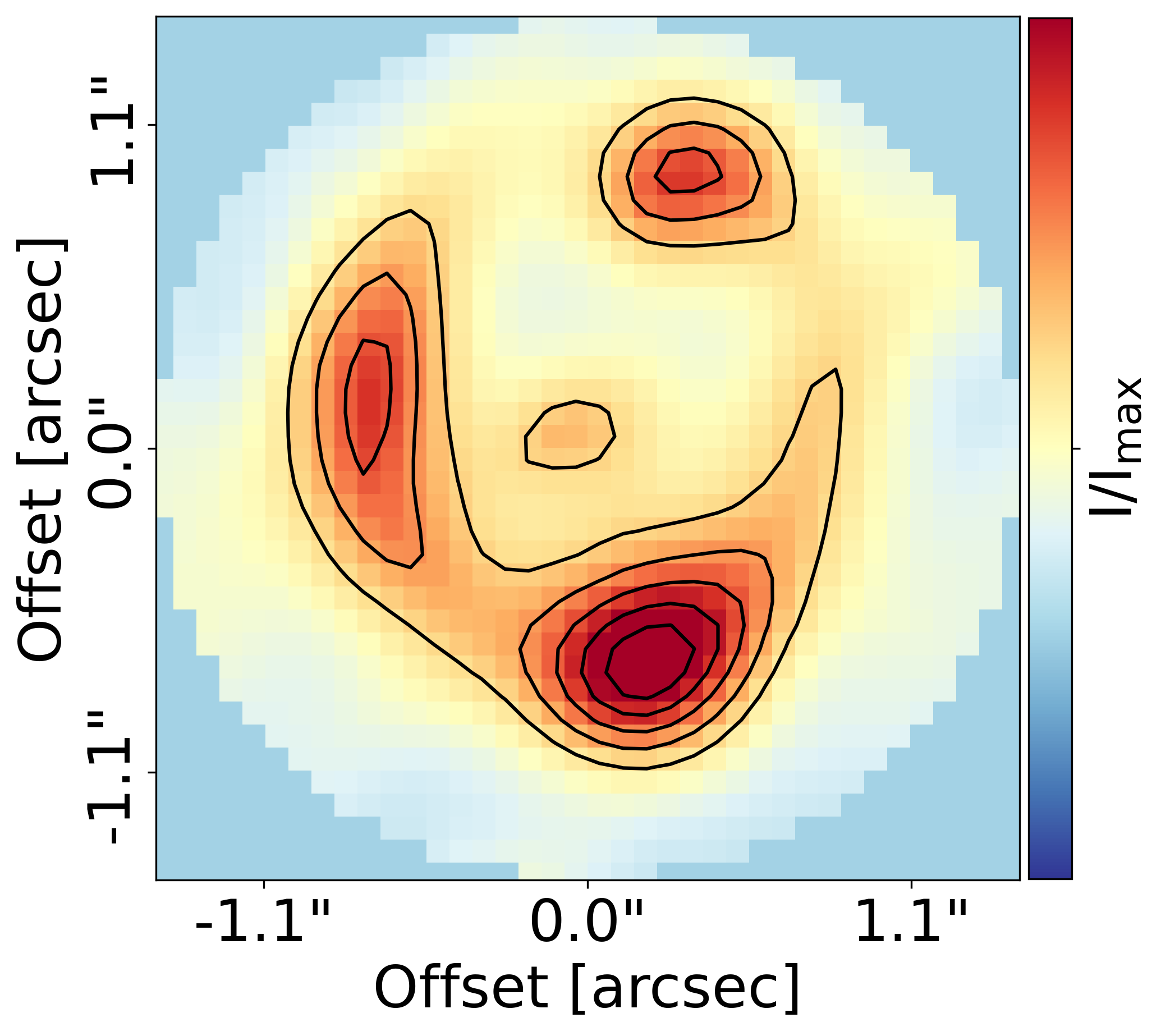}
    \includegraphics[width = 0.19\linewidth]{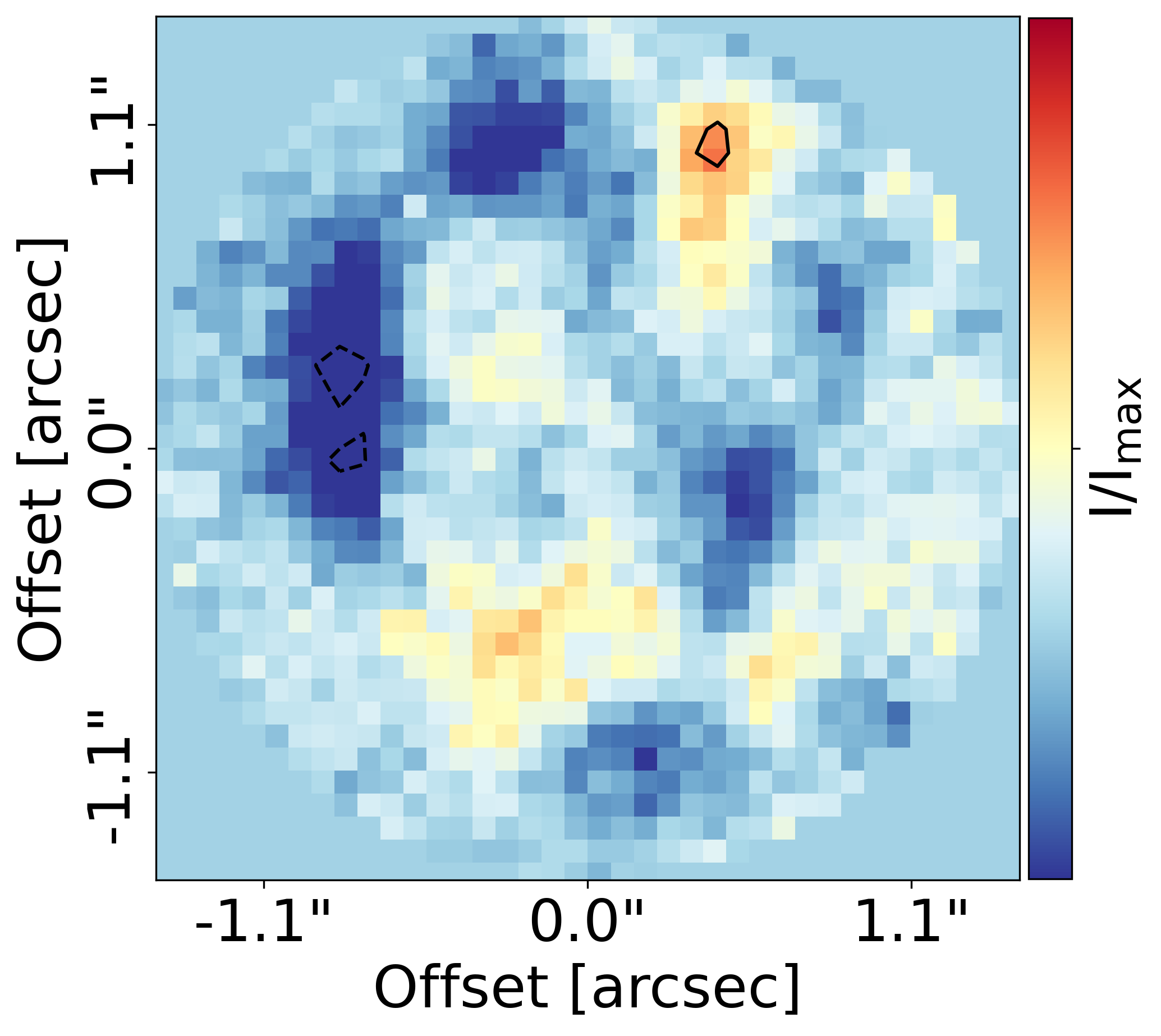}
    \includegraphics[width = 0.19\linewidth]{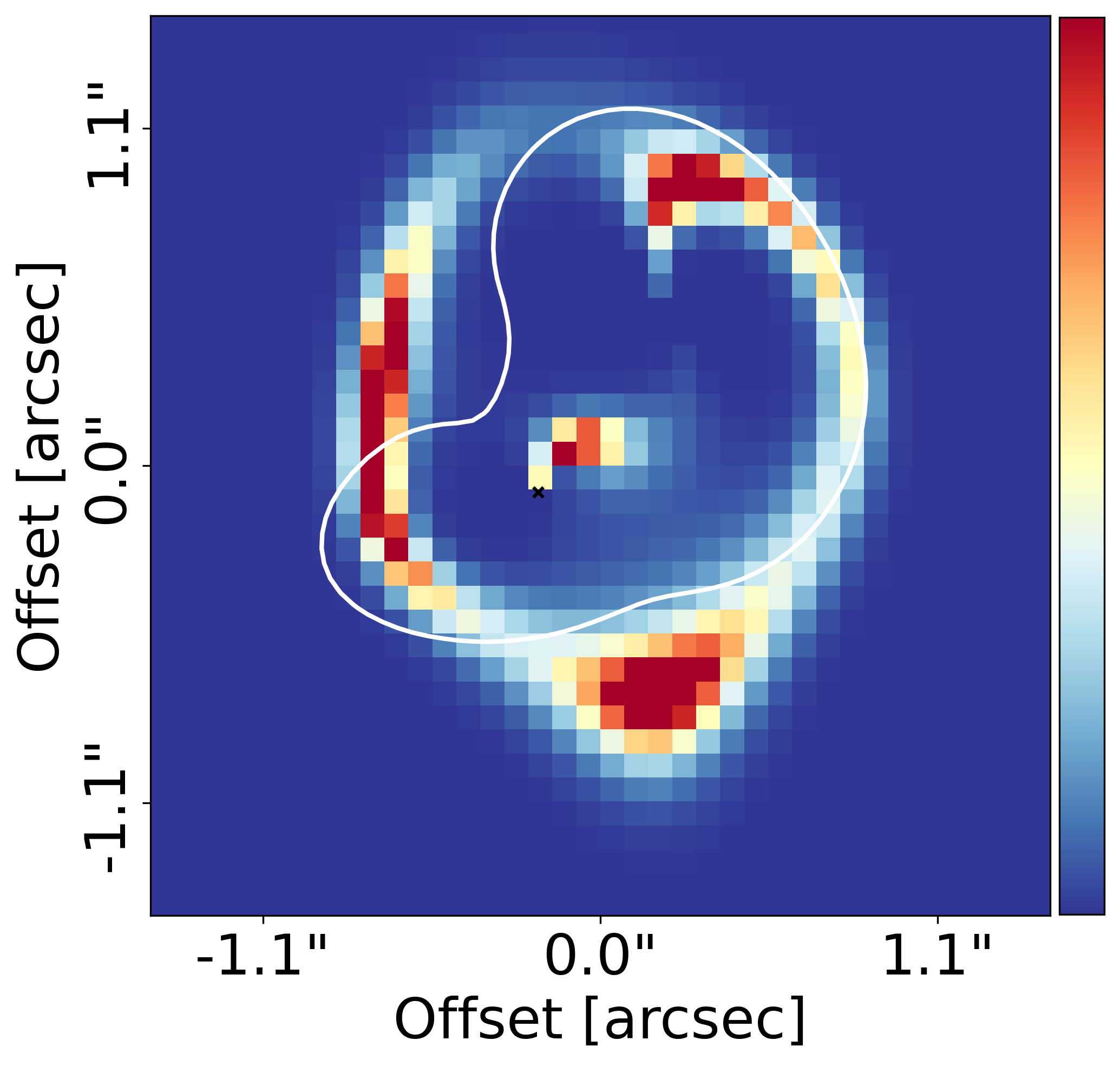}
    \includegraphics[width = 0.19\linewidth]{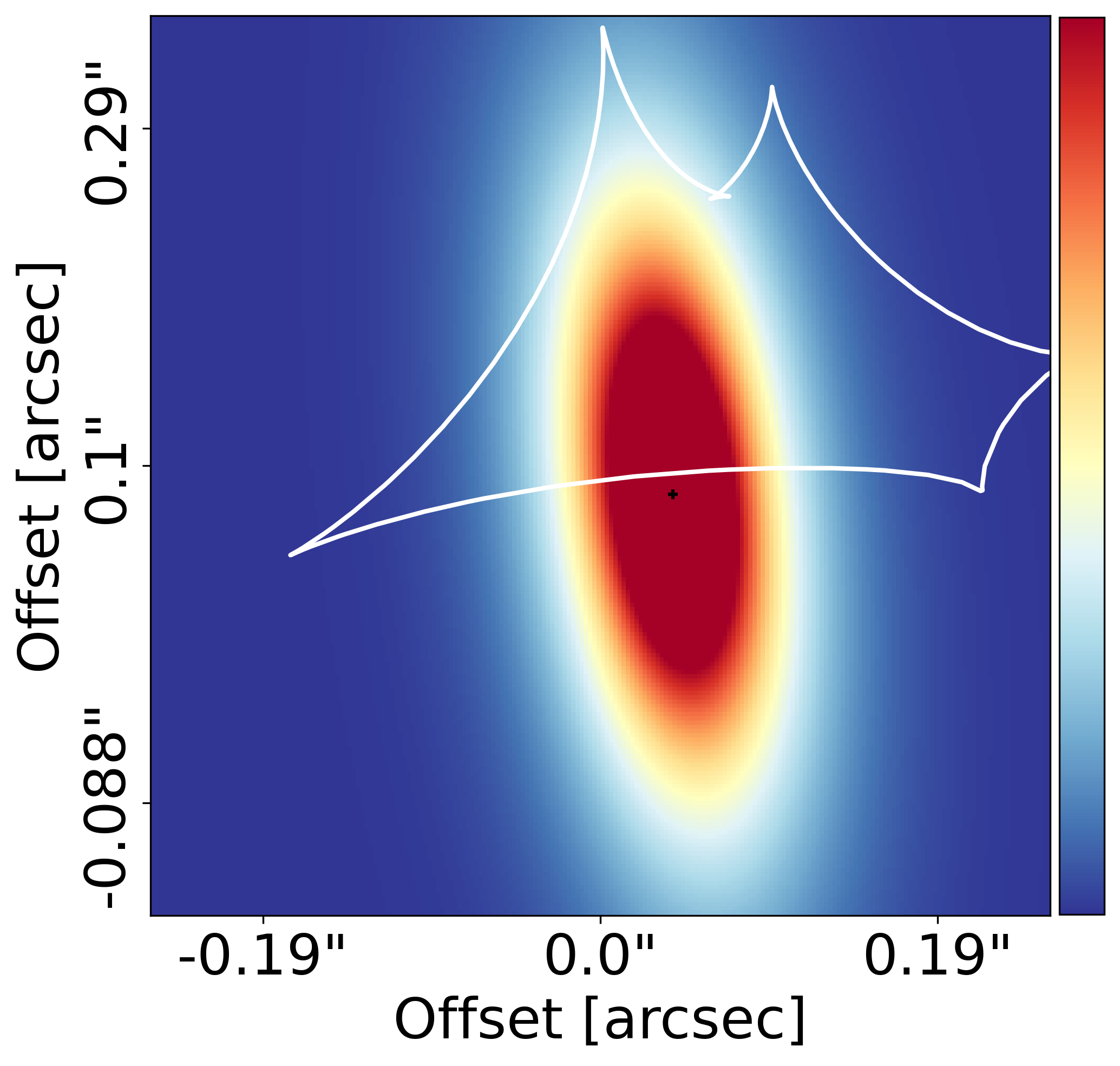}
    
    \caption{Parametric lens and source modeling results for the detected emission towards G09v1.97, first row: dust continuum, second row: CO(6--5), third row: H$_2$O, fourth row: \waterp emission. The first column shows the dirty image as produced by {\sc PyAutoLens} with contours shown at $3, 4, 5, 6, 7, 8, 9, 10\sigma$ levels where $1\sigma$ is the rms of a blank region of the image, note that this is not a cleaned image and structures may look slightly different than those shown in cleaned images. The second column shows the dirty model image as produced by {\sc PyAutoLens} with contours shown at $3, 4, 5, 6, 7, 8, 9, 10\sigma$ levels. The third column shows the dirty residual image produced by {\sc PyAutoLens} with contours shown at $-3, -2, 2, 3, 4, 5\sigma$ levels. The fourth column shows the image plane emission parametric model of the data produced by {\sc PyAutoLens}. The white line represents the critical line. The fifth column shows the source plane emission parametric model of the data produced by {\sc PyAutoLens}. The white line shows the caustic line. All images are centered around the ALMA phase center for each image. Note that the \waterp emission is of significantly lower angular resolution as the lensing model was created using the combined data, as described in Section \ref{sec:observation_details}.}
    \label{fig:parametric_lensing_images}
\end{figure*}

\subsubsection{Parametric source modeling}

The SMG G09v1.97 is lensed by two galaxies with spectroscopic redshifts of $z = 0.636$ and $z = 1.002$, visible in optical wavelengths \citep{Yang19}. We modeled these two galaxies as single isothermal ellipsoid (SIE) mass distributions. These mass profiles are characterized by their ($x, y$) offset from the phase center of the observations in arcseconds, Einstein radius, axis ratio (q), and position angle (PA). During lens modeling, the redshifts of the lens galaxies were fixed while we allowed all other parameters to vary, assuming uniform priors. The ($x, y$) offset and Einstein radius were allowed to vary around previously reported values from \citet{Yang19} and \citet{Maresca22}. Although lens models have been successfully created for G09v1.97 using various lens modeling codes \citep[e.g.,][]{Yang19, Maresca22}, including {\sc PyAutoLens}, discrepancies between different tools and observations necessitated allowing the parameters describing the lens mass models to vary rather than directly adopting values from previous works.

We modeled the source as a single S\'ersic profile using a linear light profile, allowing the intensity parameters to be solved via linear algebra. This minimizes the degeneracy between parameters such as the intensity and the effective radius of the source. This model was parameterized by its ($x, y$) offset from the observation phase center, axis ratio (q), position angle (PA), effective radius, and S\'ersic index. We note that the parametric modeling performed directly on the cleaned images reported in \citet{Yang19} required multiple S\'ersic source profiles to provide a good lens model. In this work, we do not find that we require more than a single source profile to obtain good lens models for each of the emission types, as described below. 

We first model the dust continuum emission to obtain an optimized lens model and proceed to use this to reconstruct the molecular line emission. We allowed all S\'ersic parameters, including centroid coordinates, to vary in the fit to account for possible morphological differences between emission types. However, we note that the values found for different emission types are fairly consistent and are all within errors of each other. Further discussion regarding the extent of the different emission regions is reserved until Section \ref{subsubsec:nonparametric_lensmodeling} where the non-parametric source modeling is free from a priori assumptions.

We calculated the parametric modeling magnification errors through sampling the posterior distribution of the maximum likelihood solution for the lens mass model fit from the continuum data. We performed the sampling 1000 times for each emission type, obtained magnification factors for each of these samples, and calculated the standard deviation of the magnification factors from each sample, thus obtaining the magnification factor error. This approach is further utilized and described in Section \ref{subsubsec:nonparametric_lensmodeling}.

\paragraph{Dust continuum emission} \label{subsubsec:parametric_dust_lens_model} 
We begin by modeling the dust continuum emission using a parametric model for both lenses and the source to optimize the lens mass model. This optimized model is then extended to parametric models of the molecular line emission and non-parametric source modeling. The two lenses and the source profile were fitted as described above, with the best-fit lens mass model results provided in Table \ref{tab:bestfit_lens_model} and the best-fit source parameters in Table \ref{tab:lensing_bestfit_sourceparams}. The best-fit lens model yields a very good fit to the data, as shown in Fig. \ref{fig:parametric_lensing_images}. Through this methodology, we obtain a lensing magnification factor of $\mu_{\rm continuum} = 10.55 \pm 0.26$. This is in very good agreement with the magnification factor found by \citet{Yang19} of $\mu = 10.2 - 10.5$. 

\paragraph{CO(6--5) emission} \label{subsubsec:parametric_co65_lens_model} 
In order to verify that our optimized lens mass model works well for the molecular line emission, we modeled the CO(6--5) emission using the continuum-optimized lens model (i.e., the lens model was fixed with parameters given in Table \ref{tab:bestfit_lens_model}) and a single S\'ersic profile for the source. The lens model derived from the dust continuum emission provided an excellent fit to the CO(6--5) emission, indicating that the lens model is optimal. We provide best-fit source parameters for the CO(6--5) emission in Table \ref{tab:lensing_bestfit_sourceparams}. We show the CO(6--5) image, model, and residual in Fig. \ref{fig:parametric_lensing_images}. Through this methodology, we obtain a lensing magnification factor of $\mu_{\rm CO(6-5)} = 9.72 \pm 0.27$.

We find a slight residual in both the northern and southern images of the source; see Fig.\,\ref{fig:parametric_lensing_images}. These residuals are small, $\leq 3\sigma$. We attempted to improve the model by including a second source, also described by a S\'ersic profile and as motivated by \citet{Yang19}. However, the inclusion of this second source did not improve the residuals, and we discarded this model. We suggest that this residual is mainly due to the `clumpy' nature of the CO(6--5) emission. Indeed, smoother image-plane emission in the dust continuum does not exhibit this residual emission. Therefore, we suggest that this residual is primarily due to limitations in the lens modeling itself, primarily the inability of a single, smooth (by definition) S\'ersic profile to provide a fully realistic model of a galaxy. This is evident in the smoothness of the dirty model images from {\sc PyAutoLens}, see Fig. \ref{fig:parametric_lensing_images}. We suggest that non-parametric modeling would likely provide a better fit to the data and is explored in Section \ref{subsubsec:nonparametric_lensmodeling}.

\begin{figure*}[h]
    \centering
    \includegraphics[width = 0.19\linewidth]{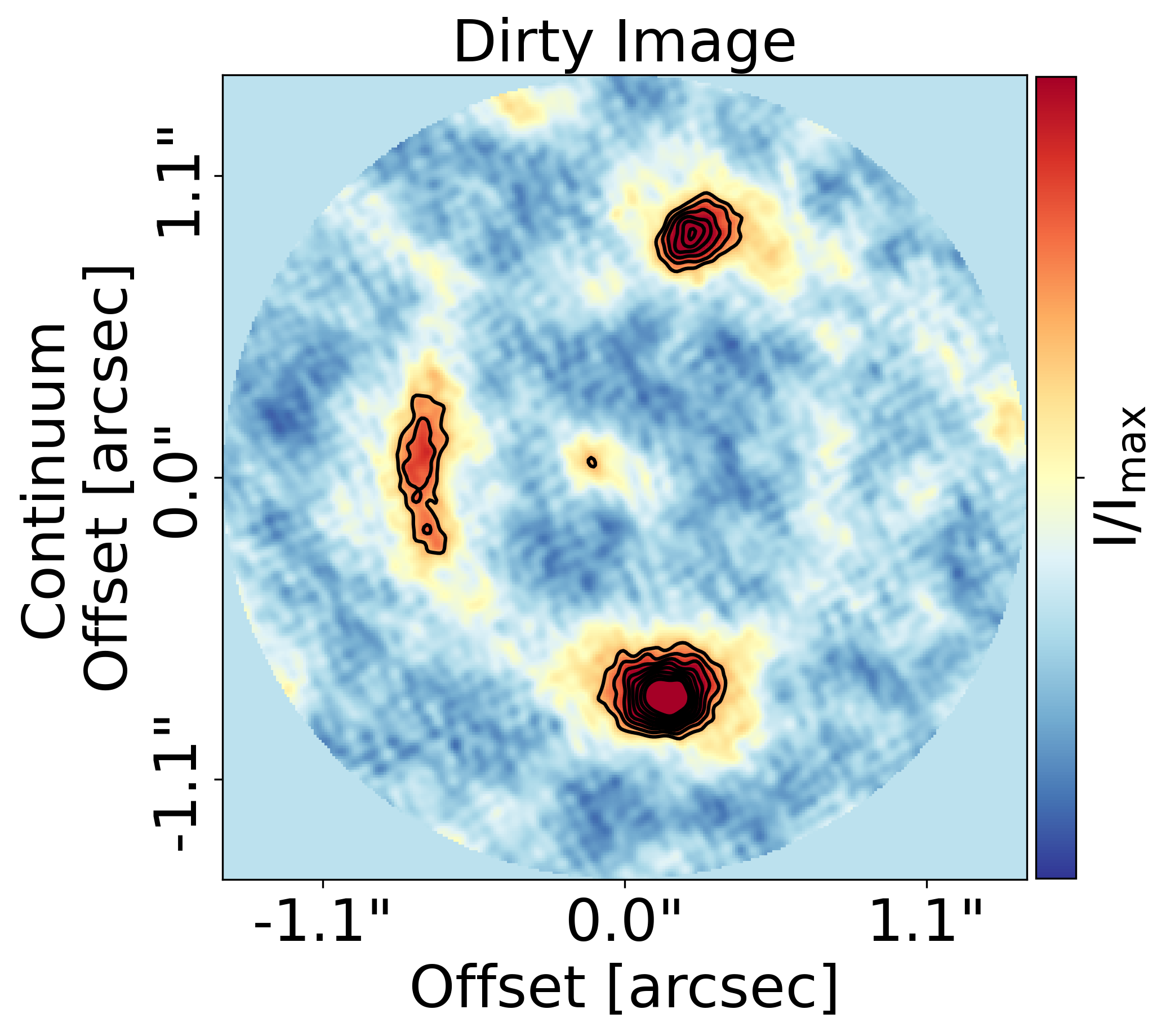}
    \includegraphics[width = 0.19\linewidth]{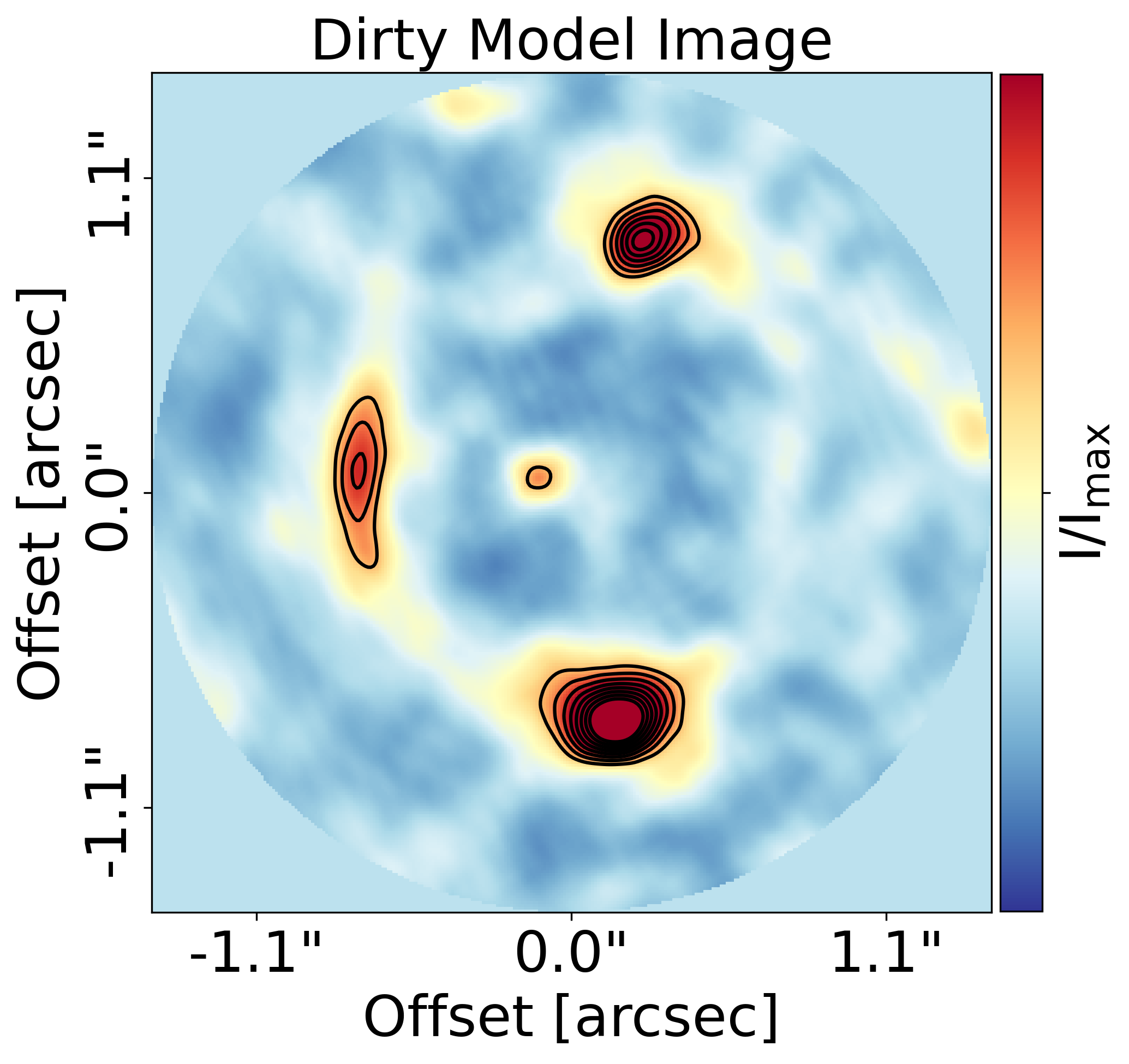}
    \includegraphics[width = 0.19\linewidth]{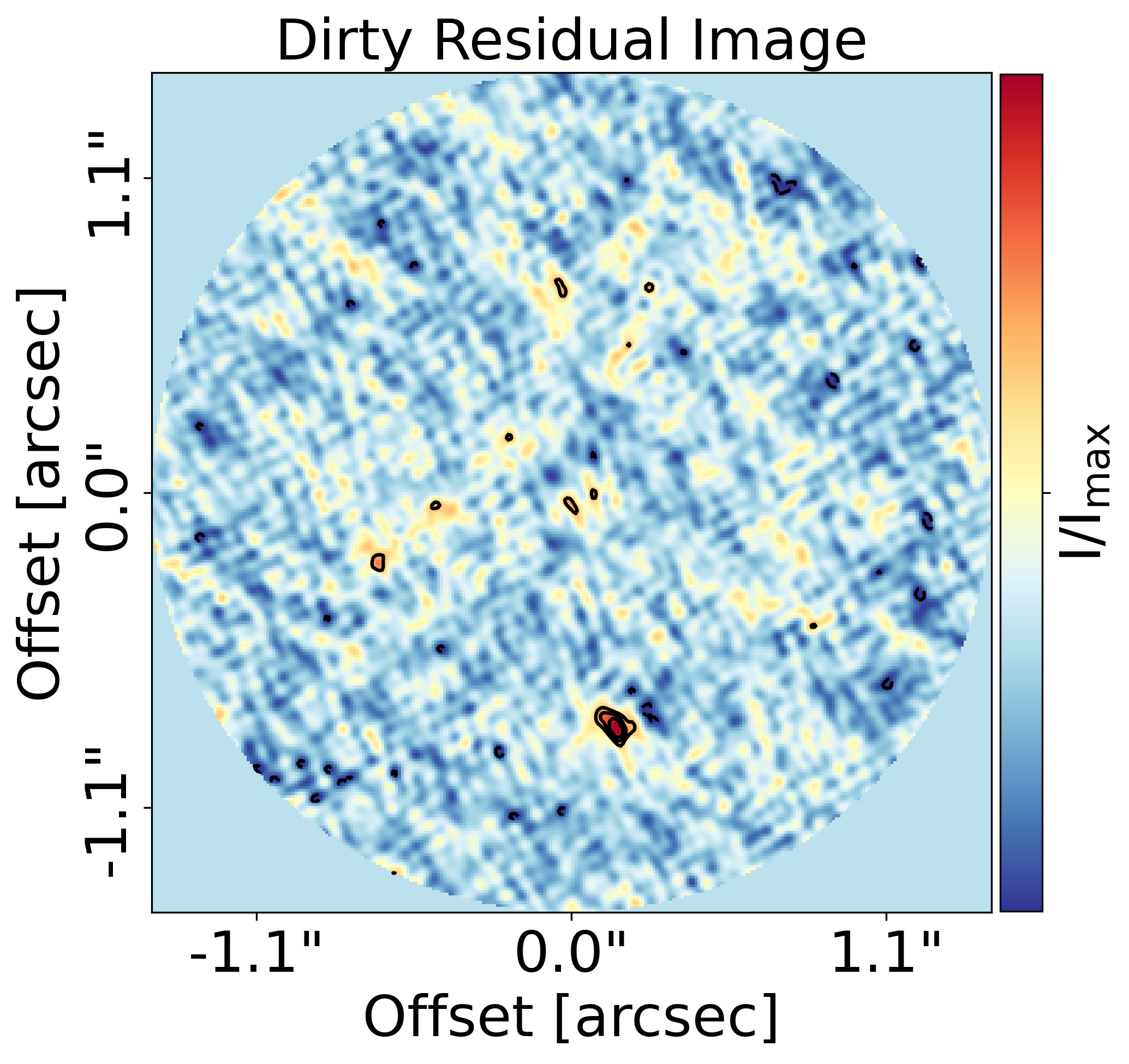}
    \includegraphics[width = 0.19\linewidth]{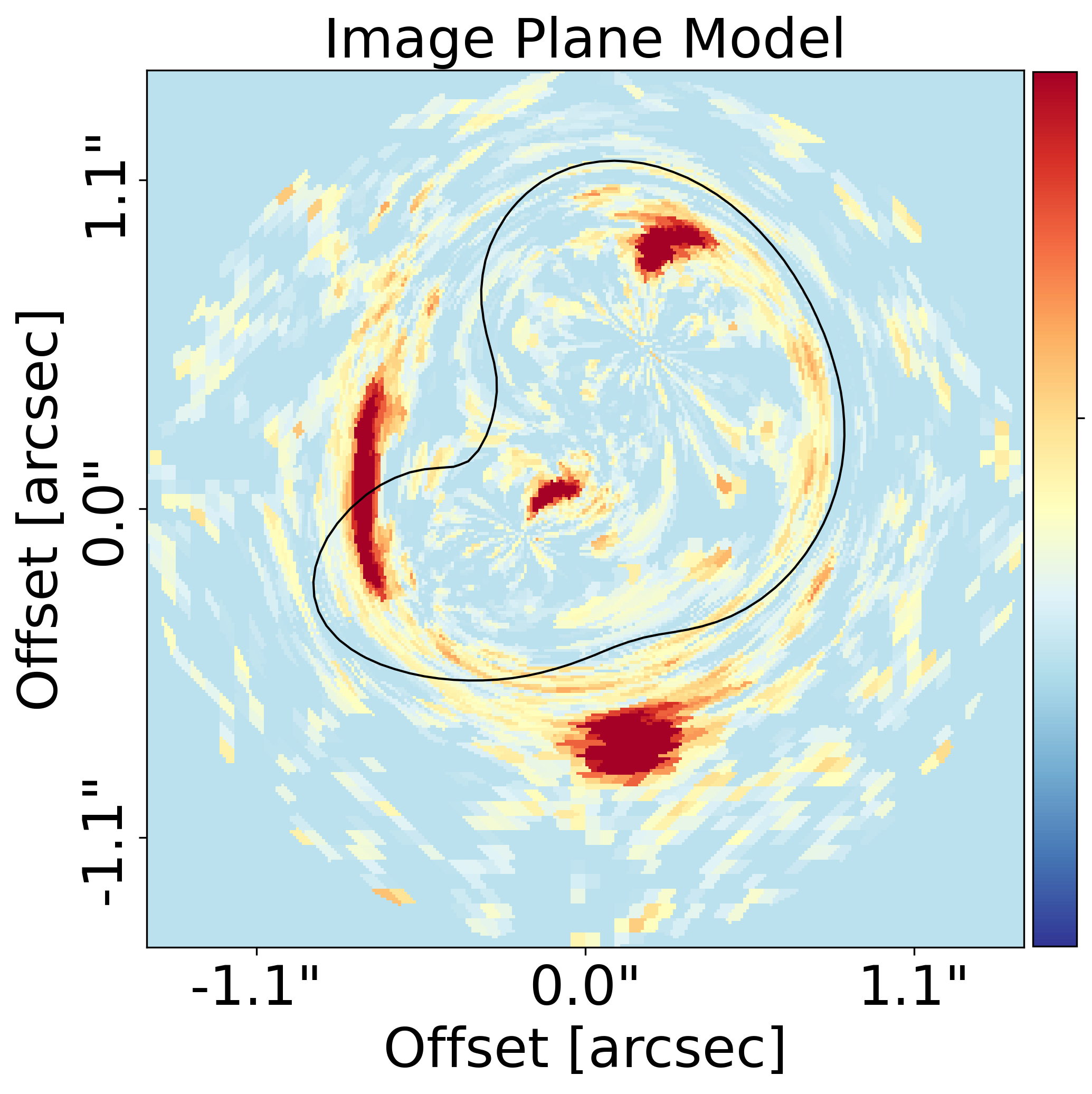}
    \includegraphics[width = 0.19\linewidth]{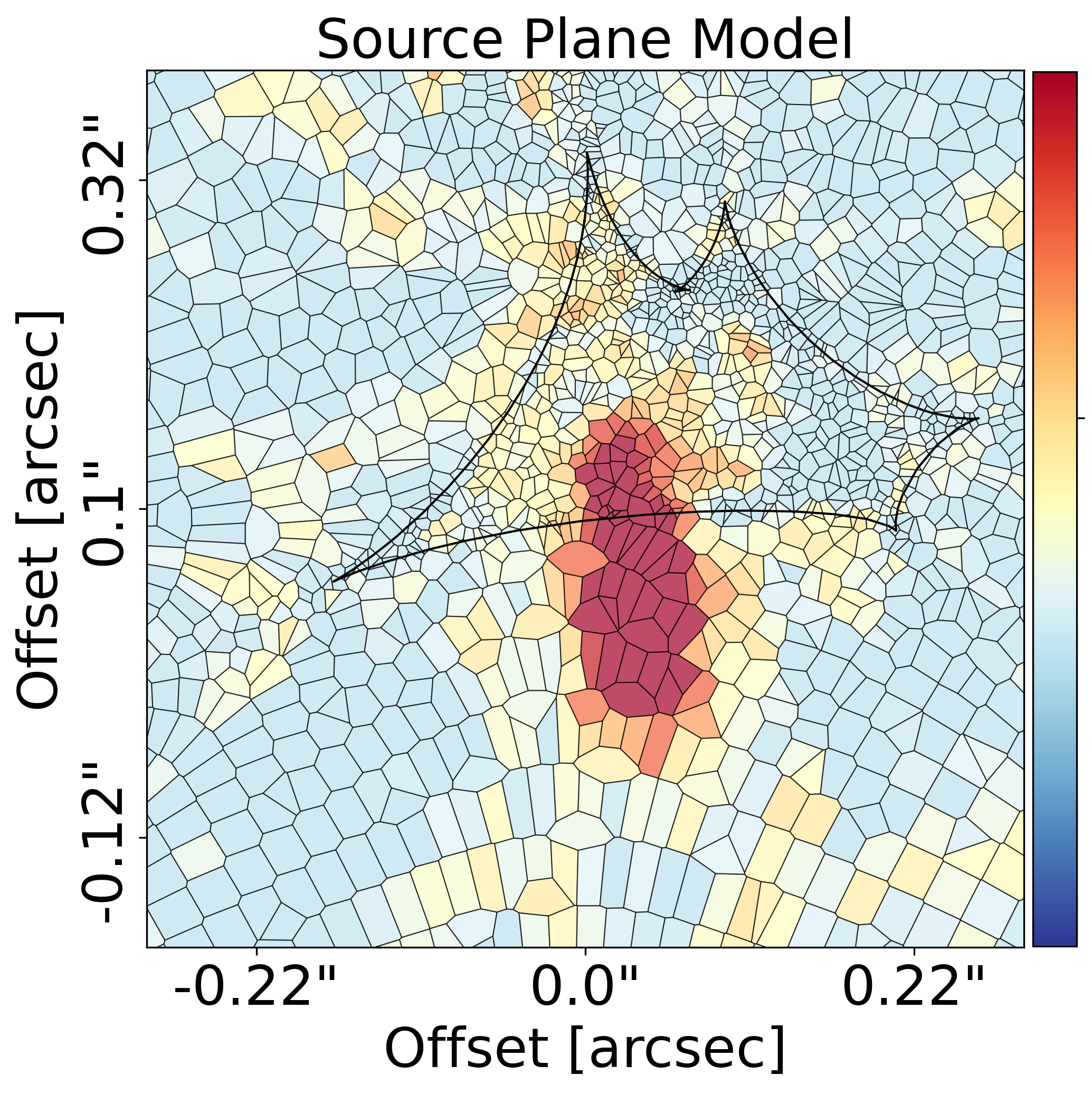}

    \includegraphics[width = 0.19\linewidth]{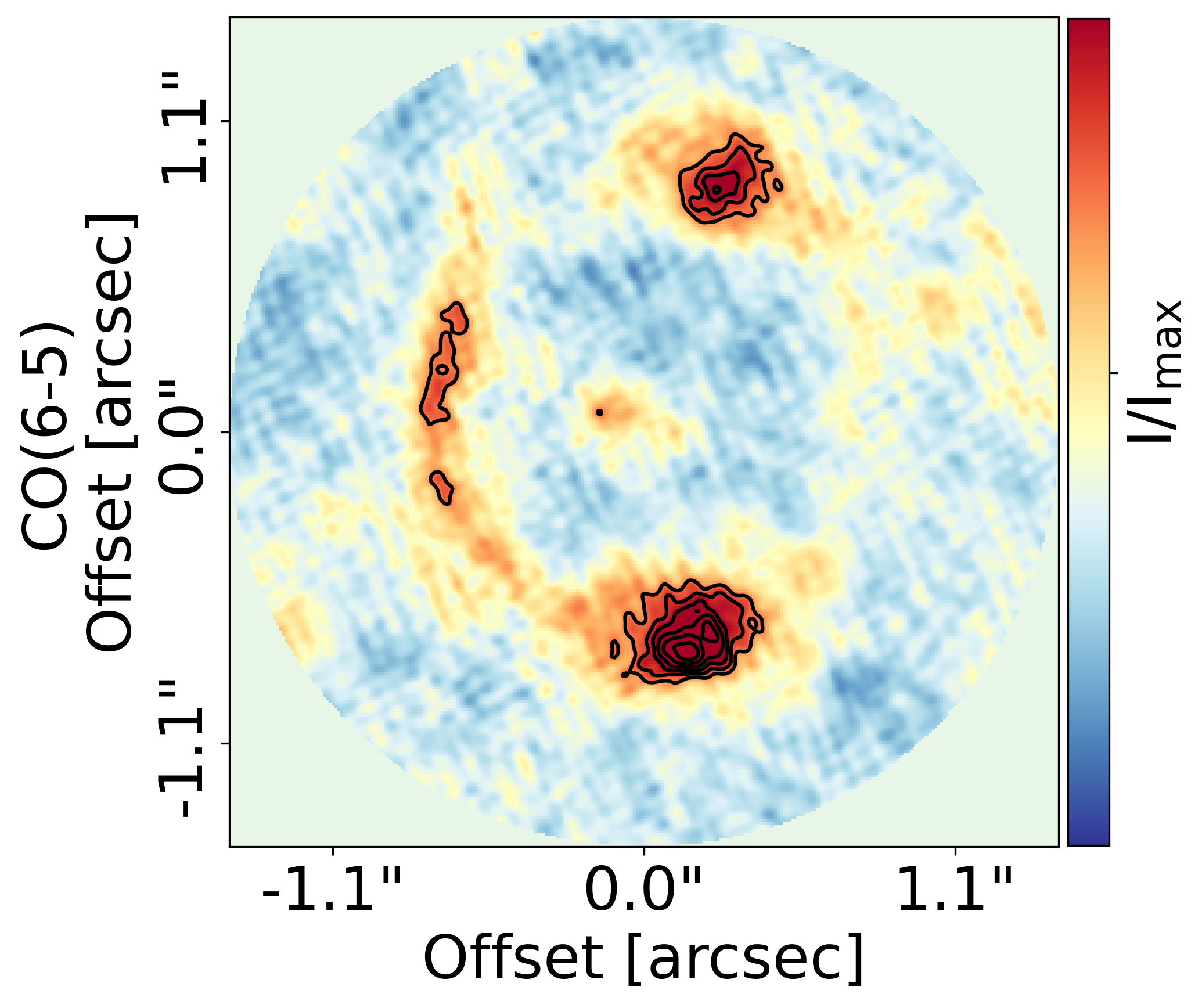}
    \includegraphics[width = 0.19\linewidth]{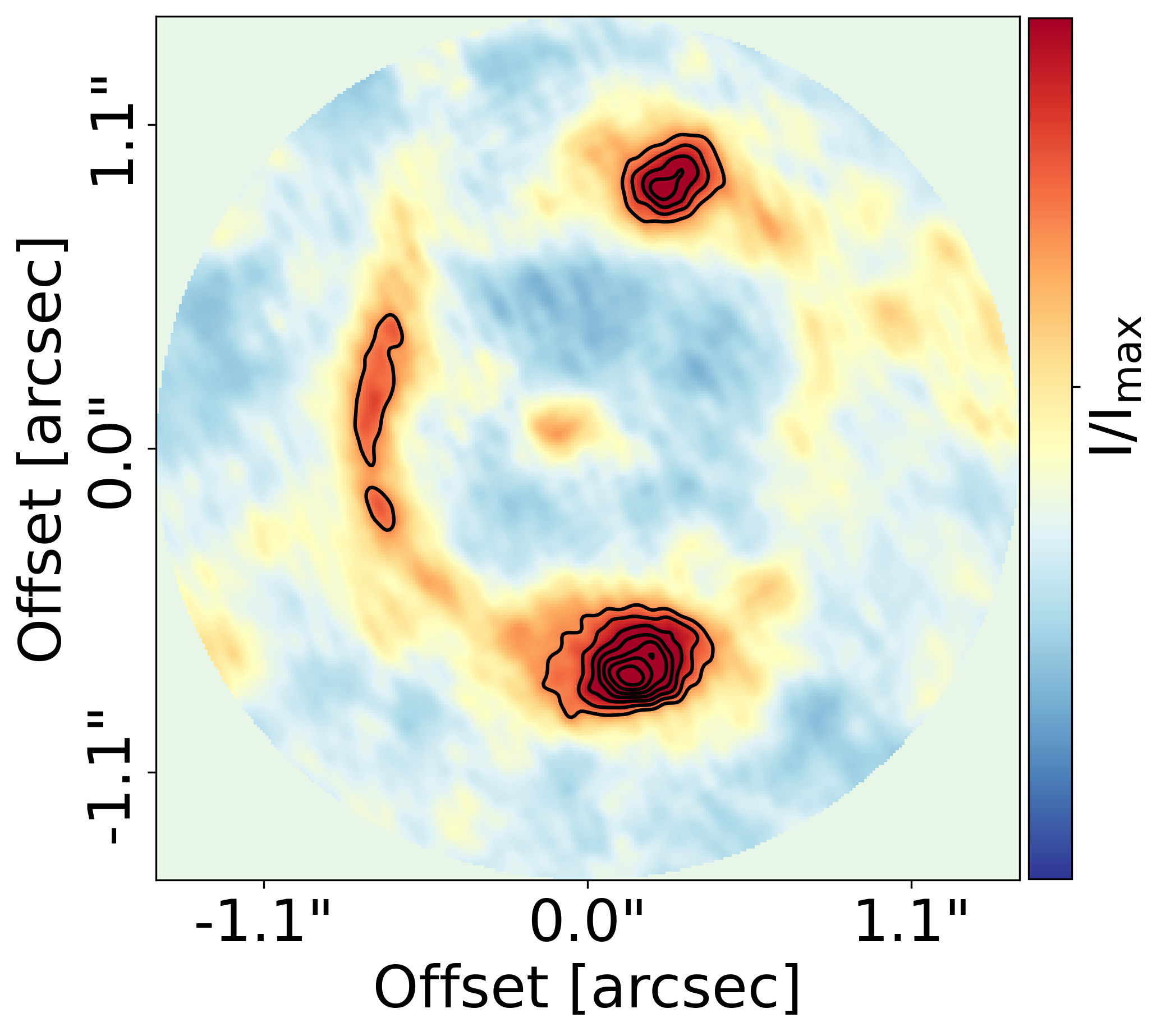}
    \includegraphics[width = 0.19\linewidth]{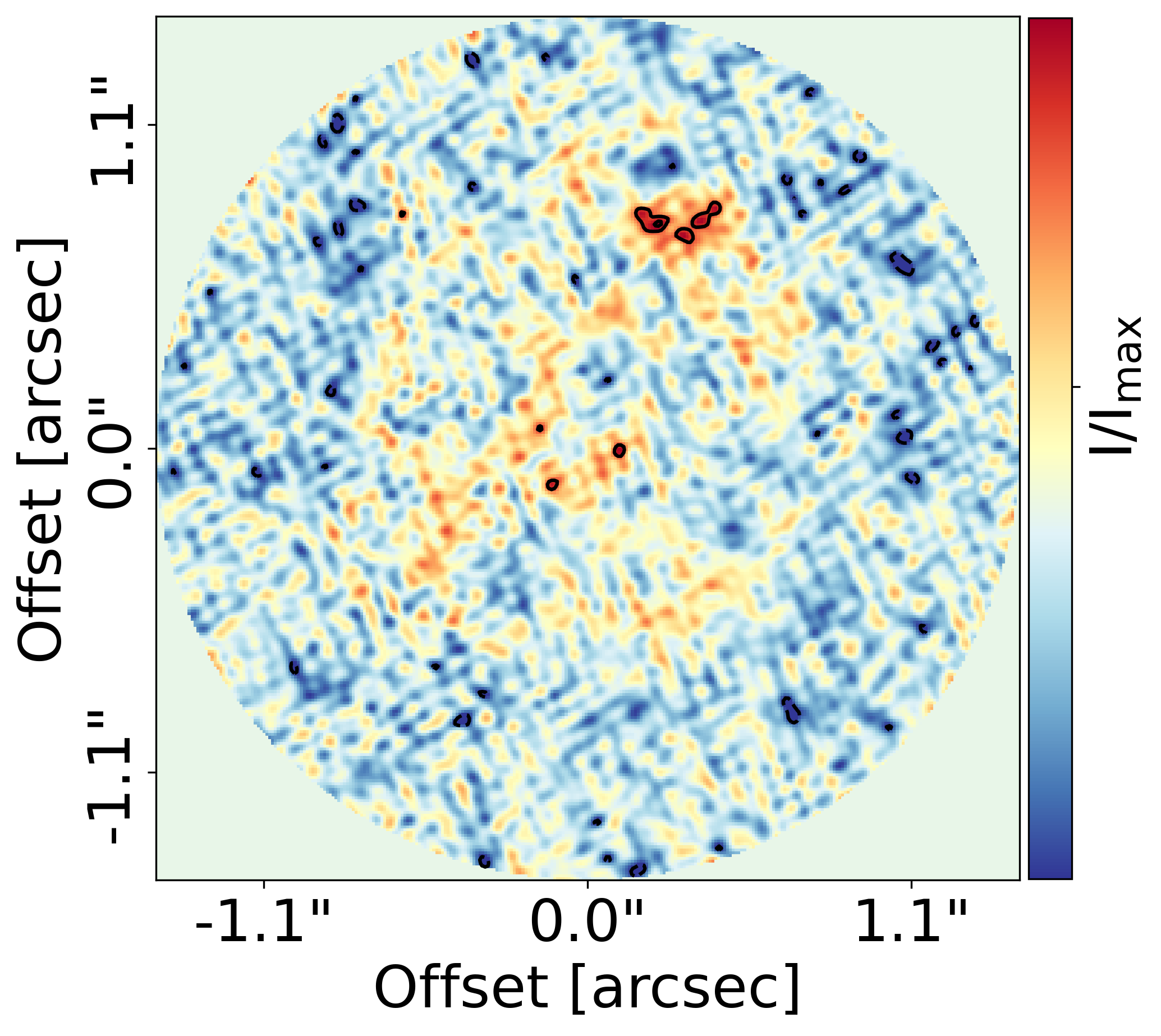}
    \includegraphics[width = 0.19\linewidth]{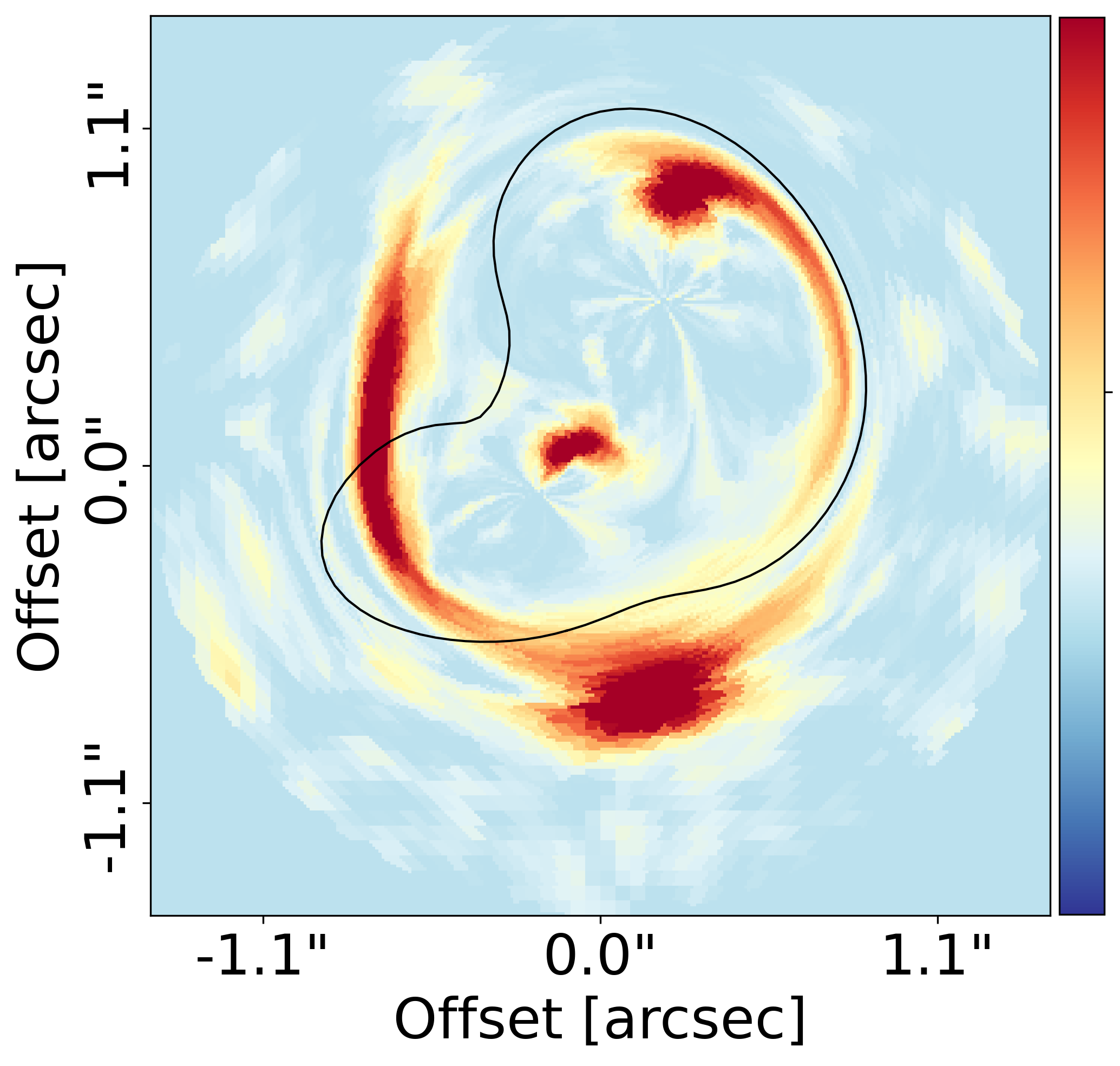}
    \includegraphics[width = 0.19\linewidth]{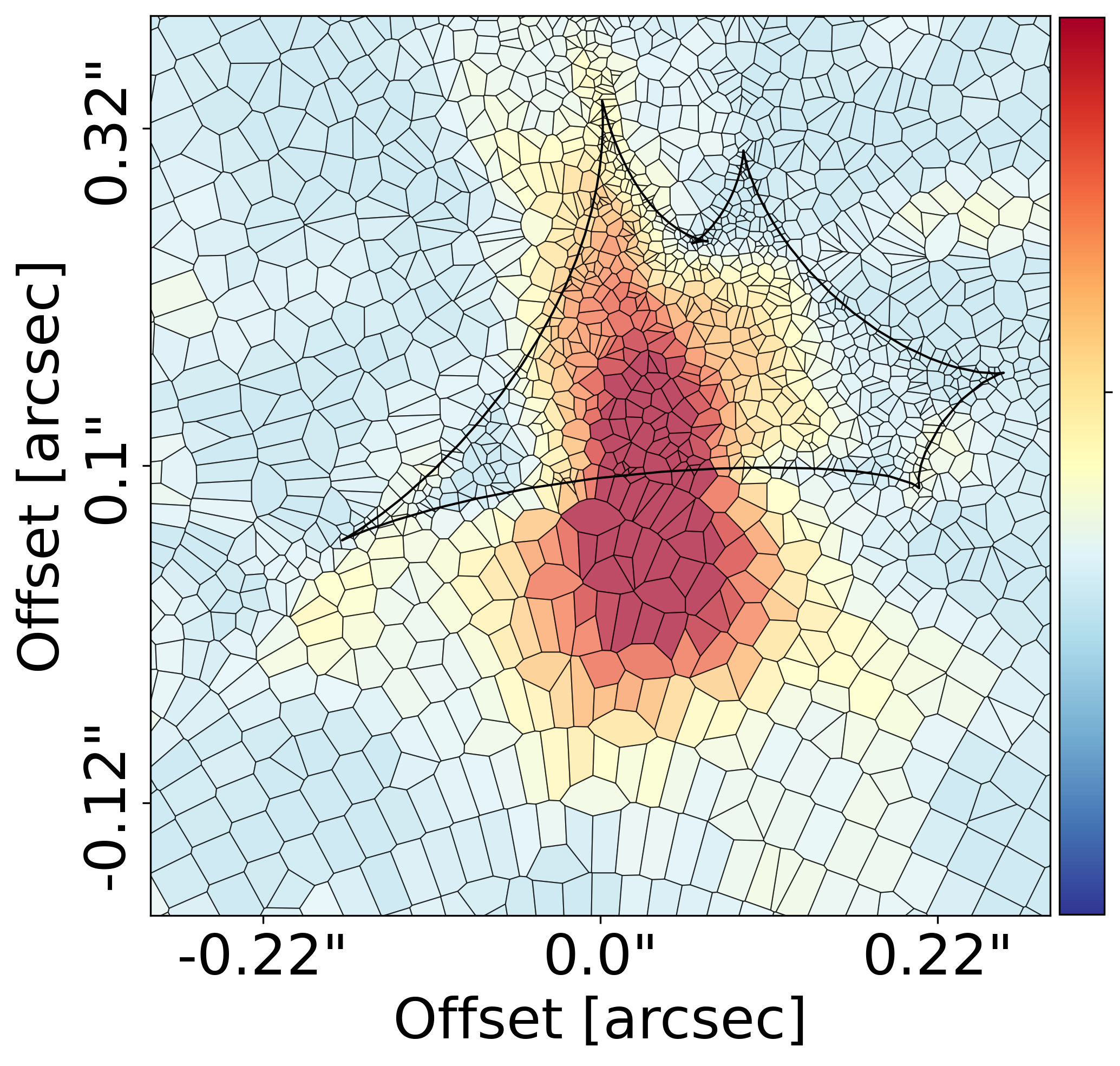}
    
    \includegraphics[width = 0.19\linewidth]{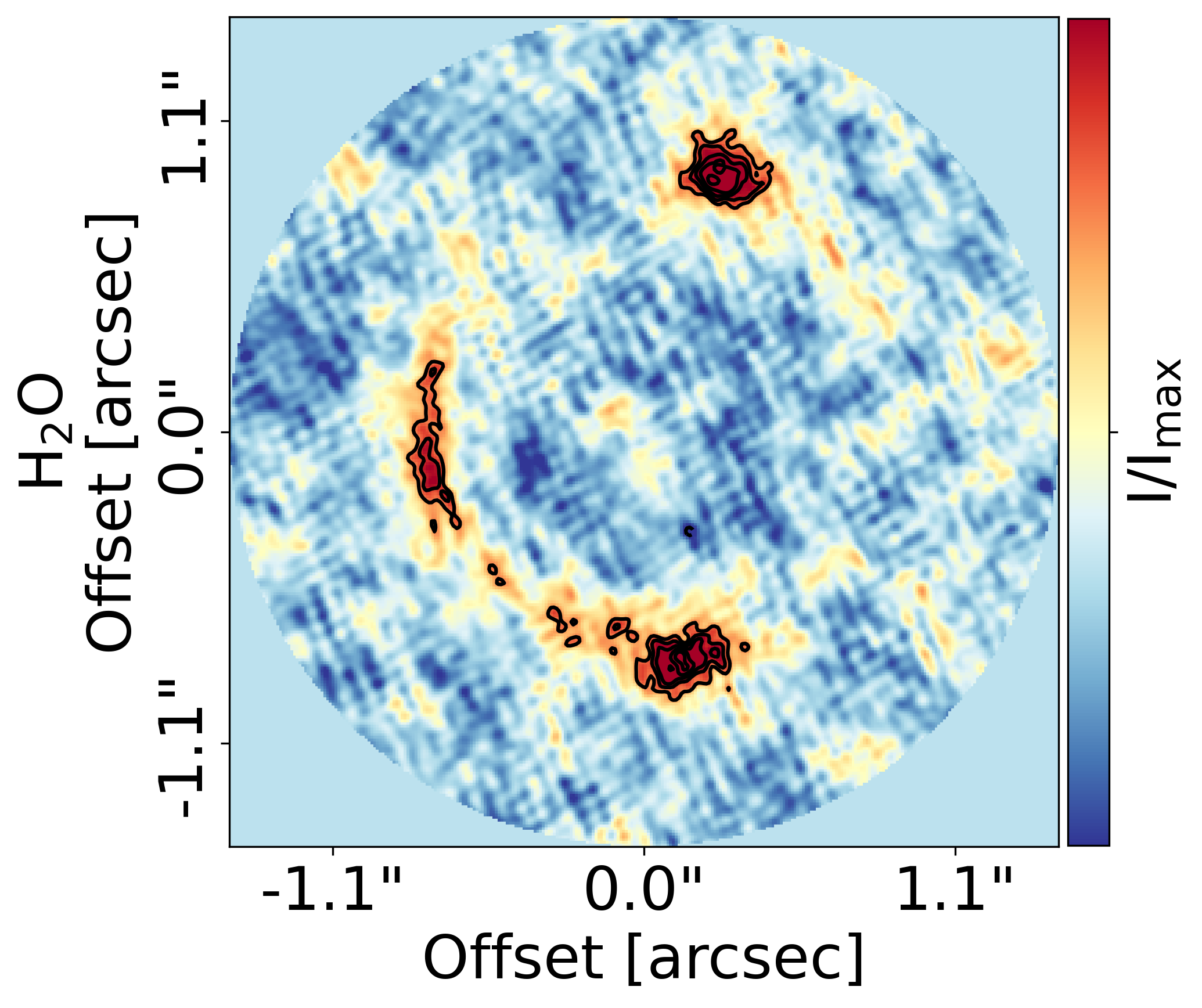}
    \includegraphics[width = 0.19\linewidth]{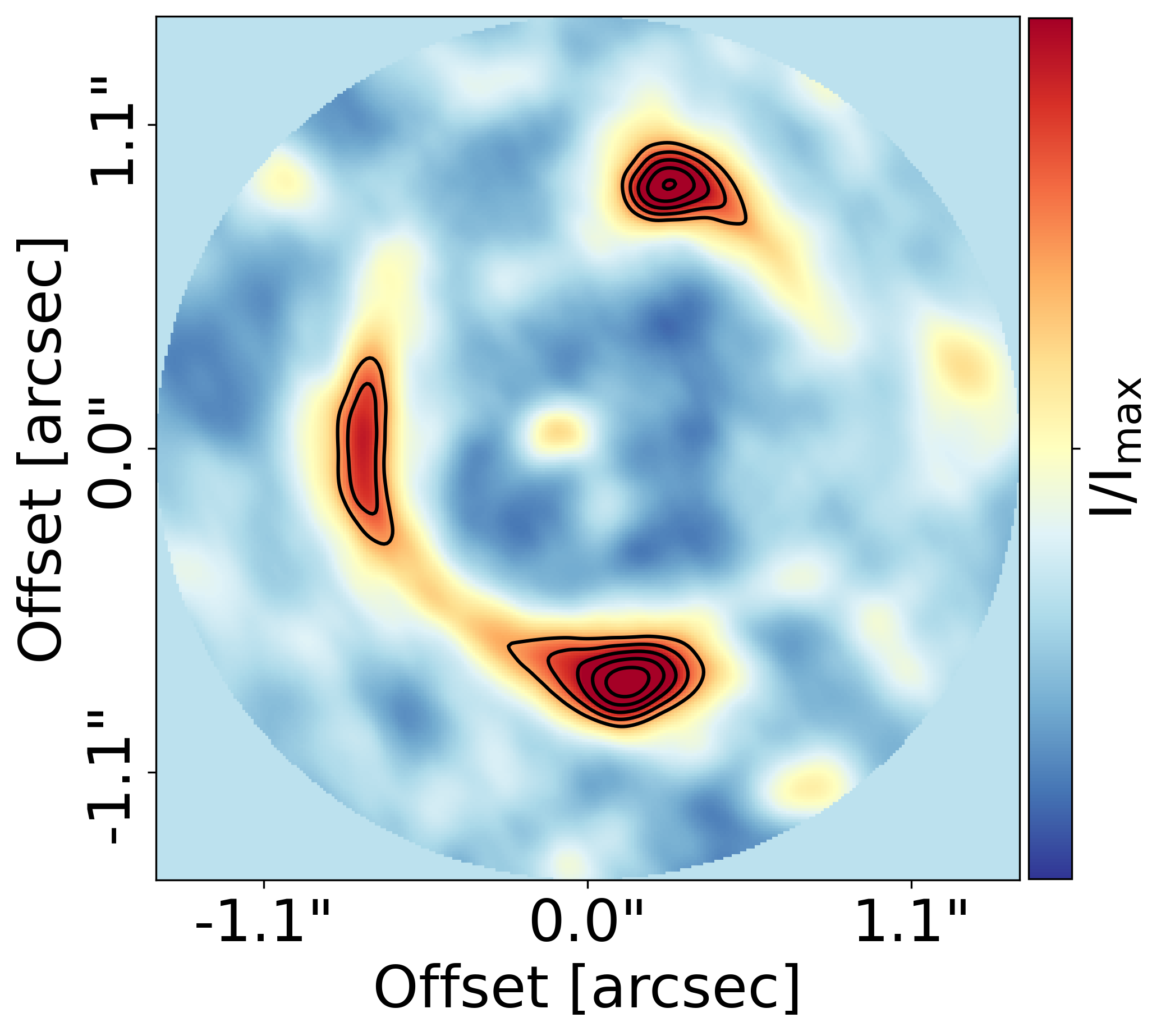}
    \includegraphics[width = 0.19\linewidth]{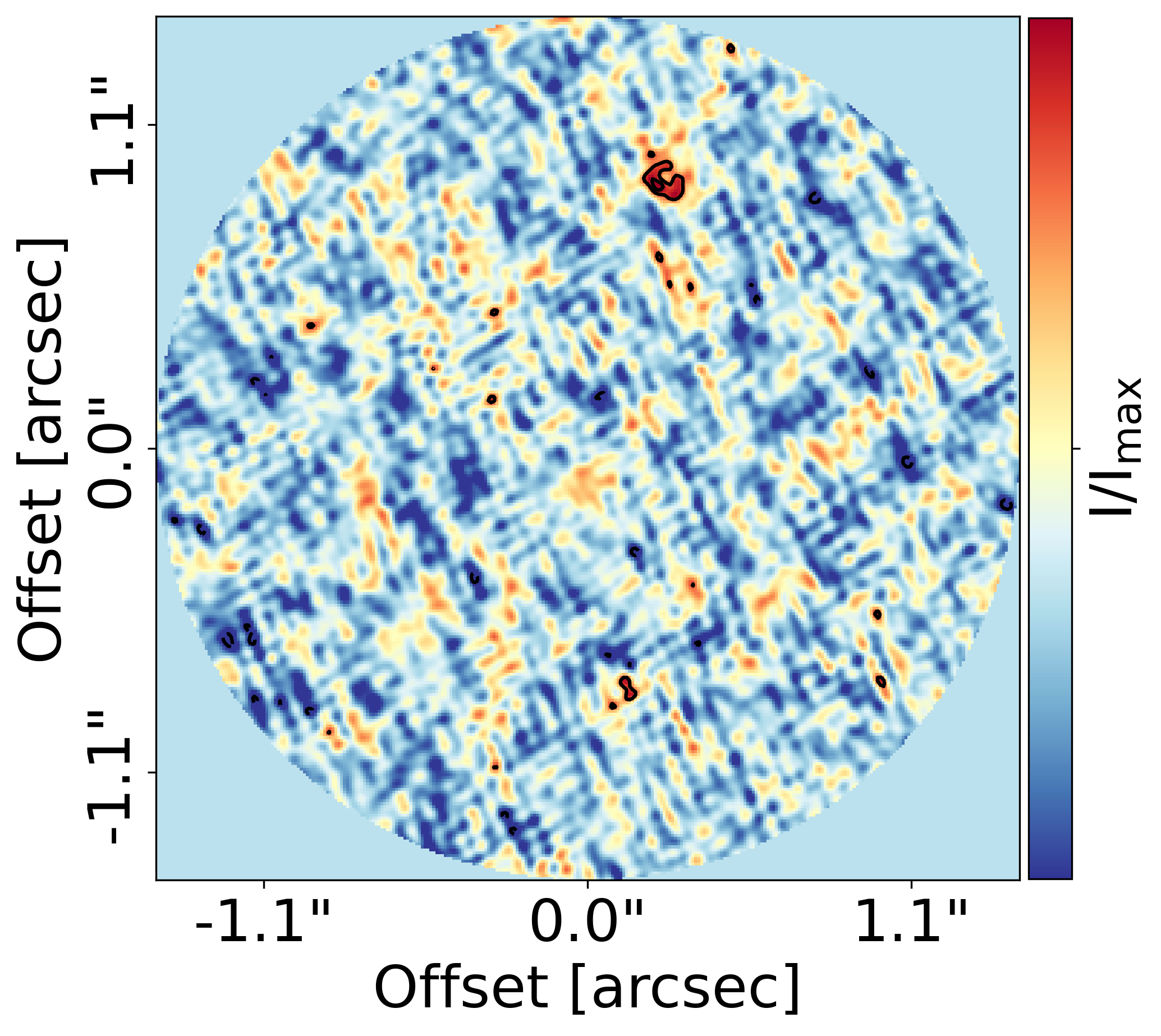}
    \includegraphics[width = 0.19\linewidth]{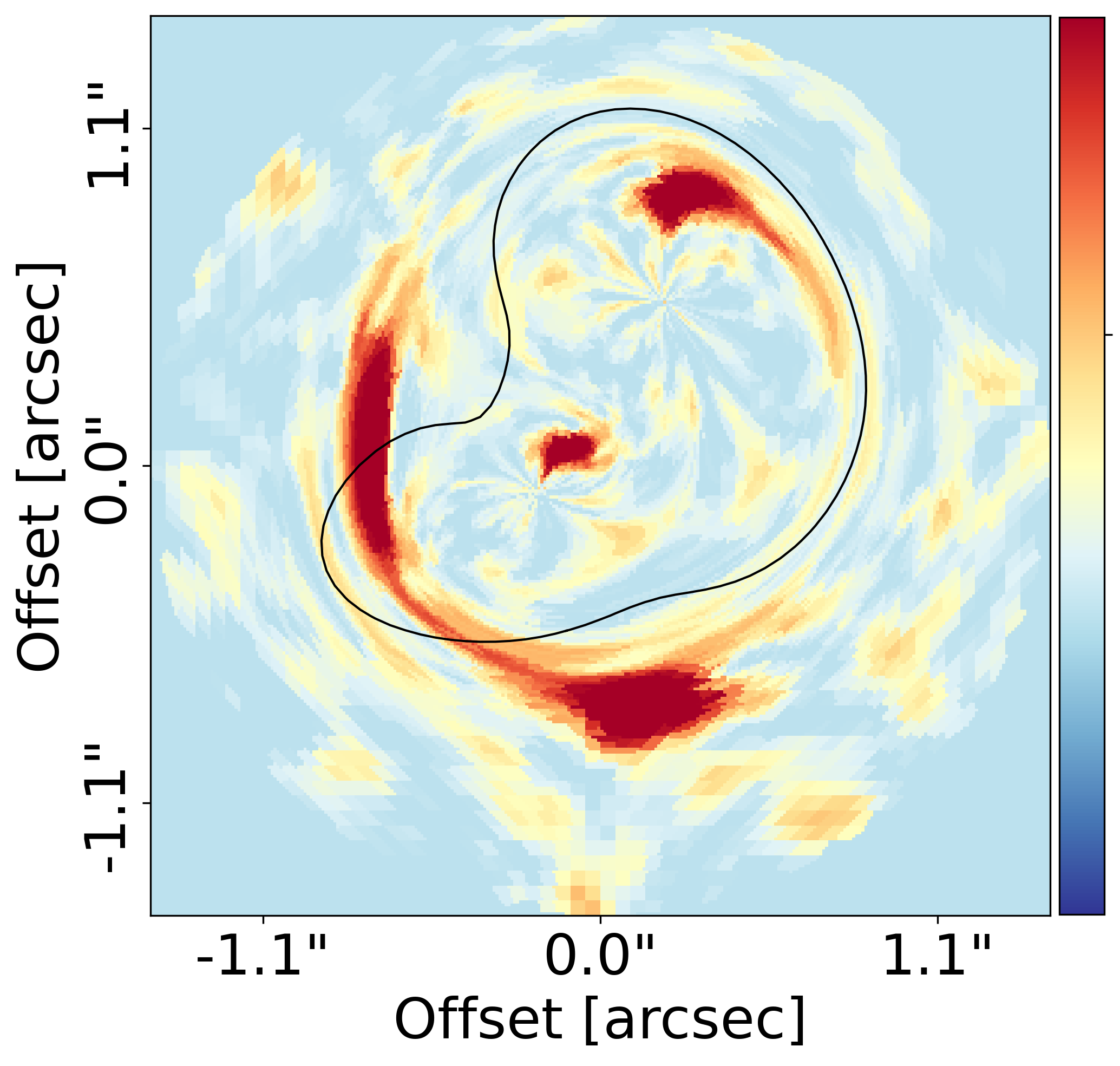}
    \includegraphics[width = 0.19\linewidth]{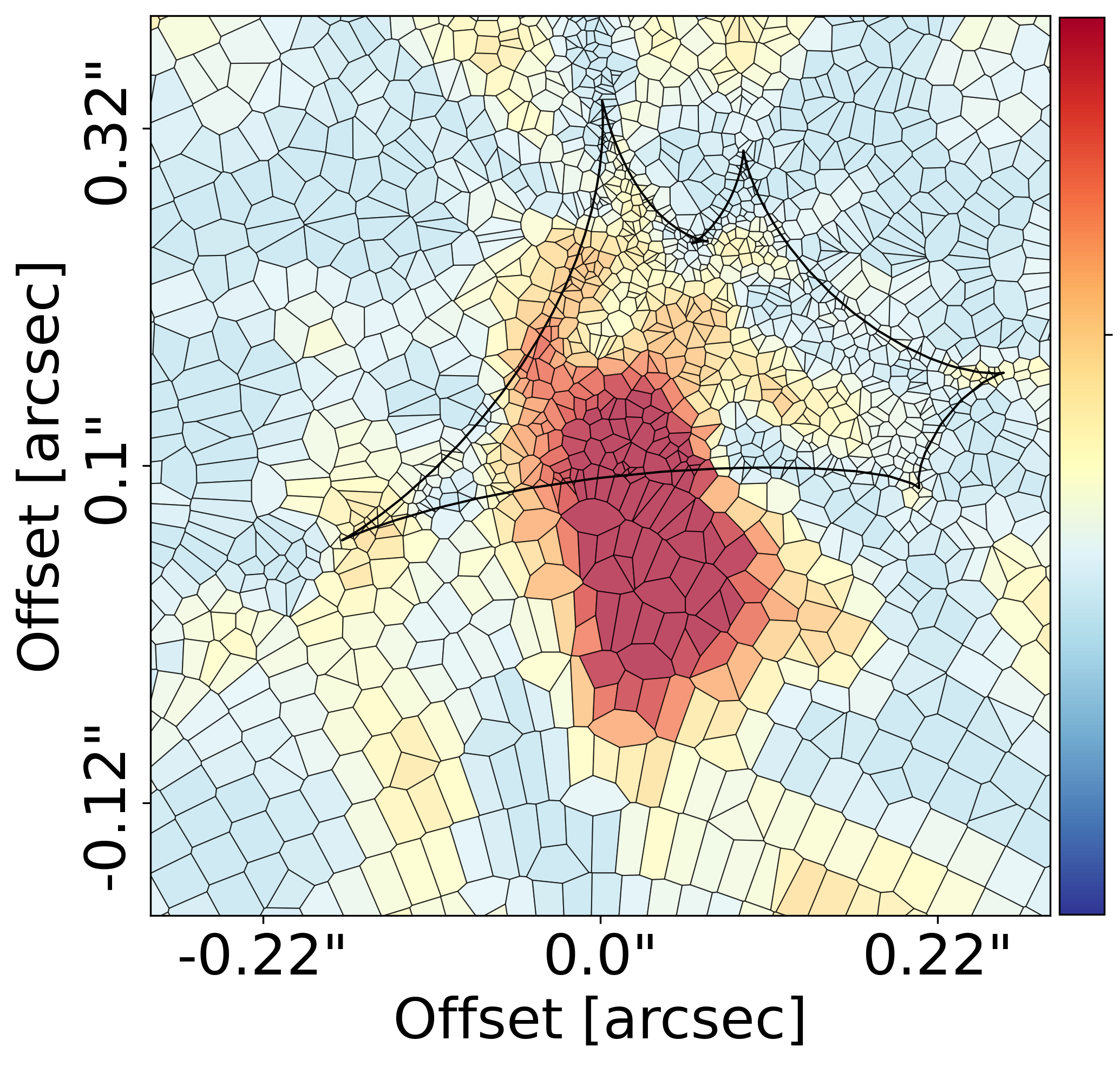}

    \includegraphics[width = 0.19\linewidth]{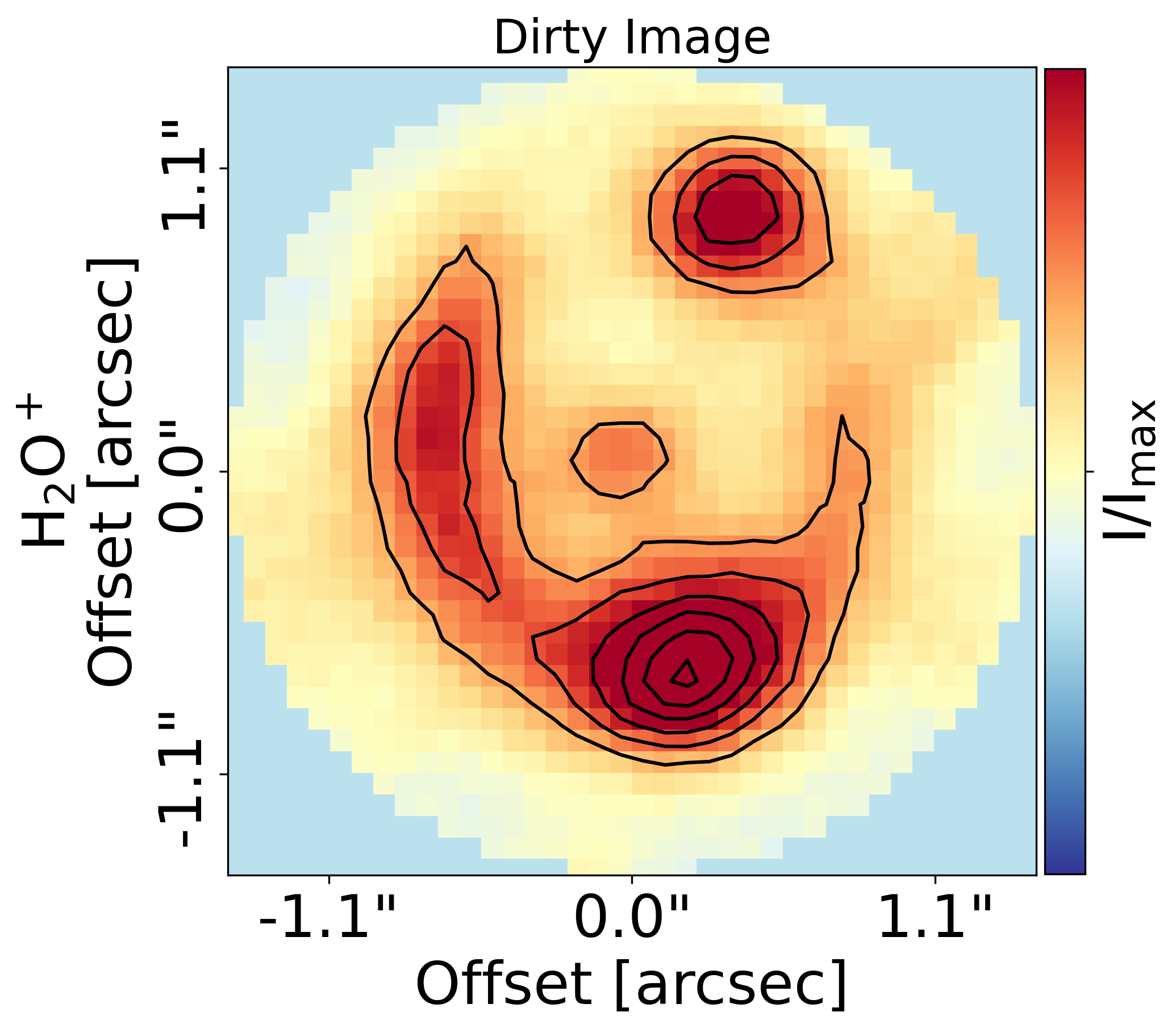}
    \includegraphics[width = 0.19\linewidth]{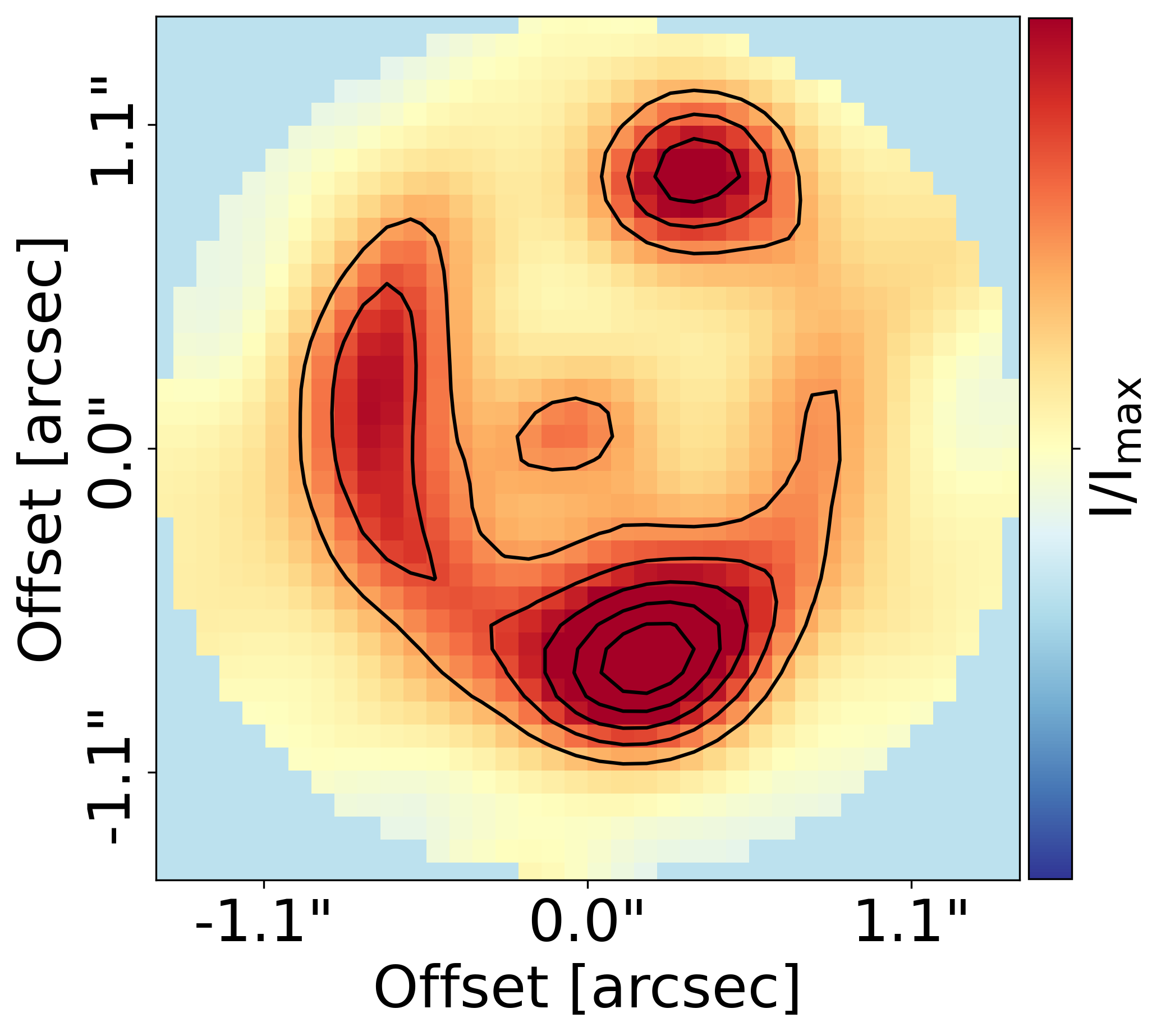}
    \includegraphics[width = 0.19\linewidth]{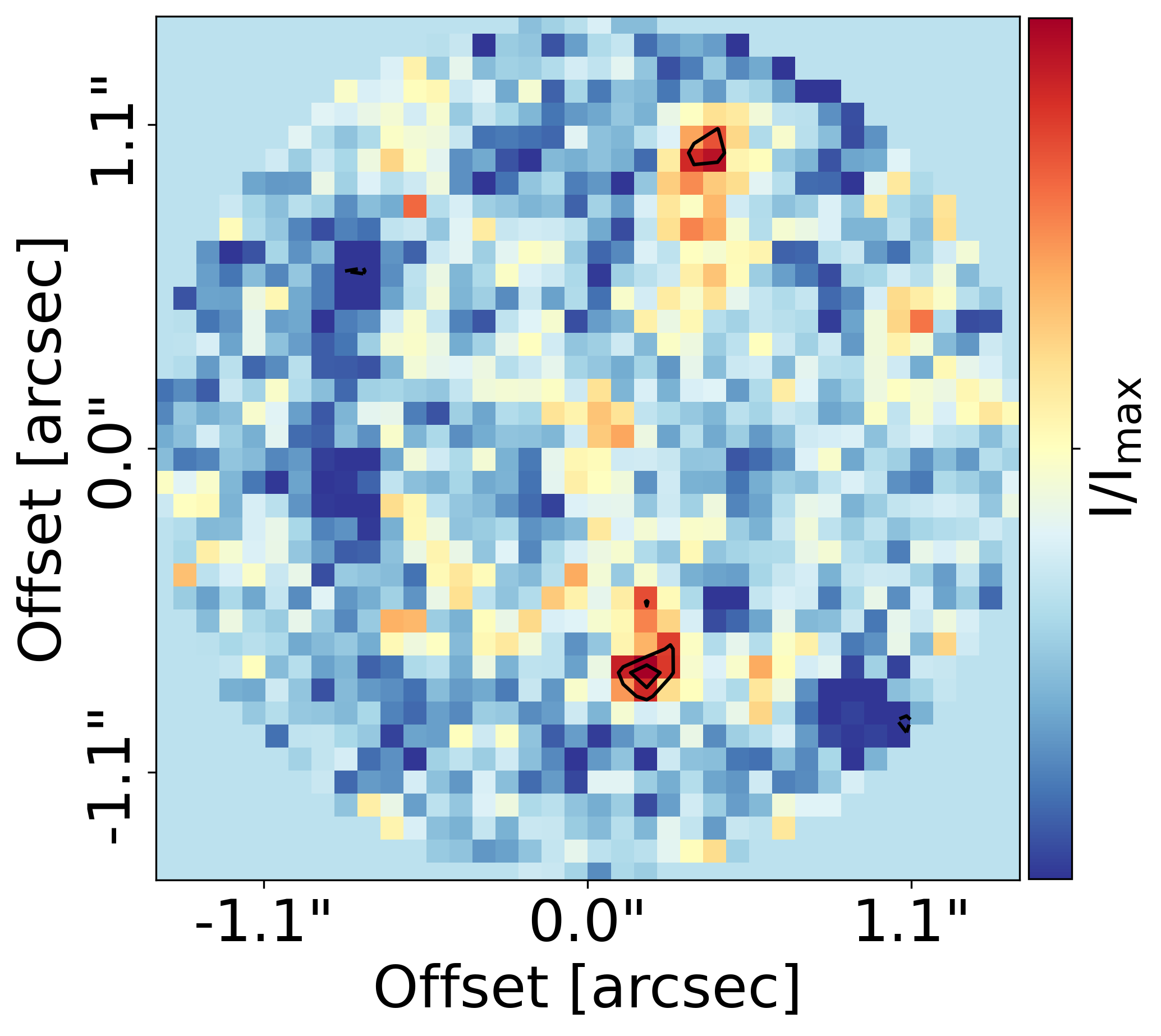}
    \includegraphics[width = 0.19\linewidth]{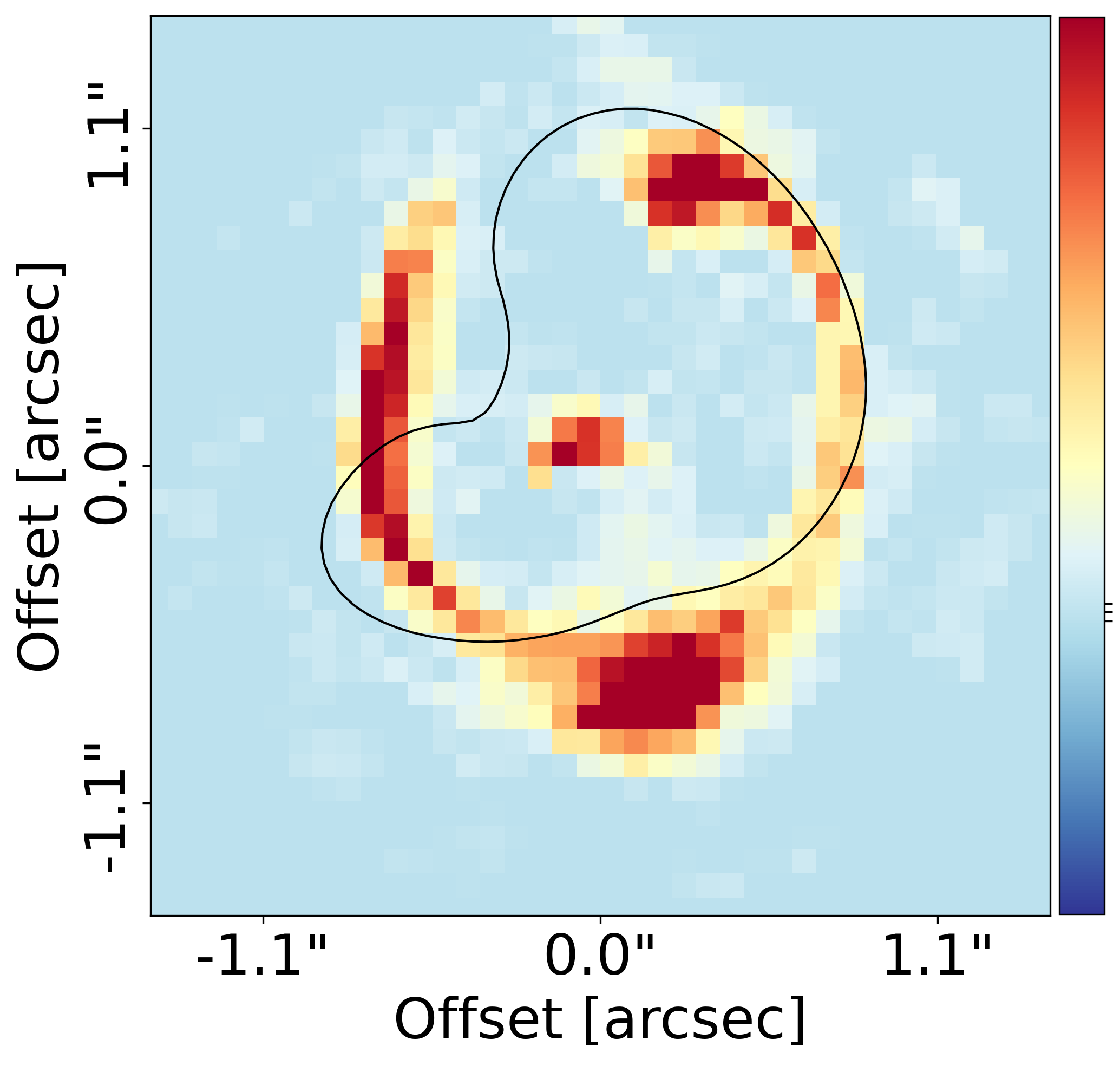}
    \includegraphics[width = 0.19\linewidth]{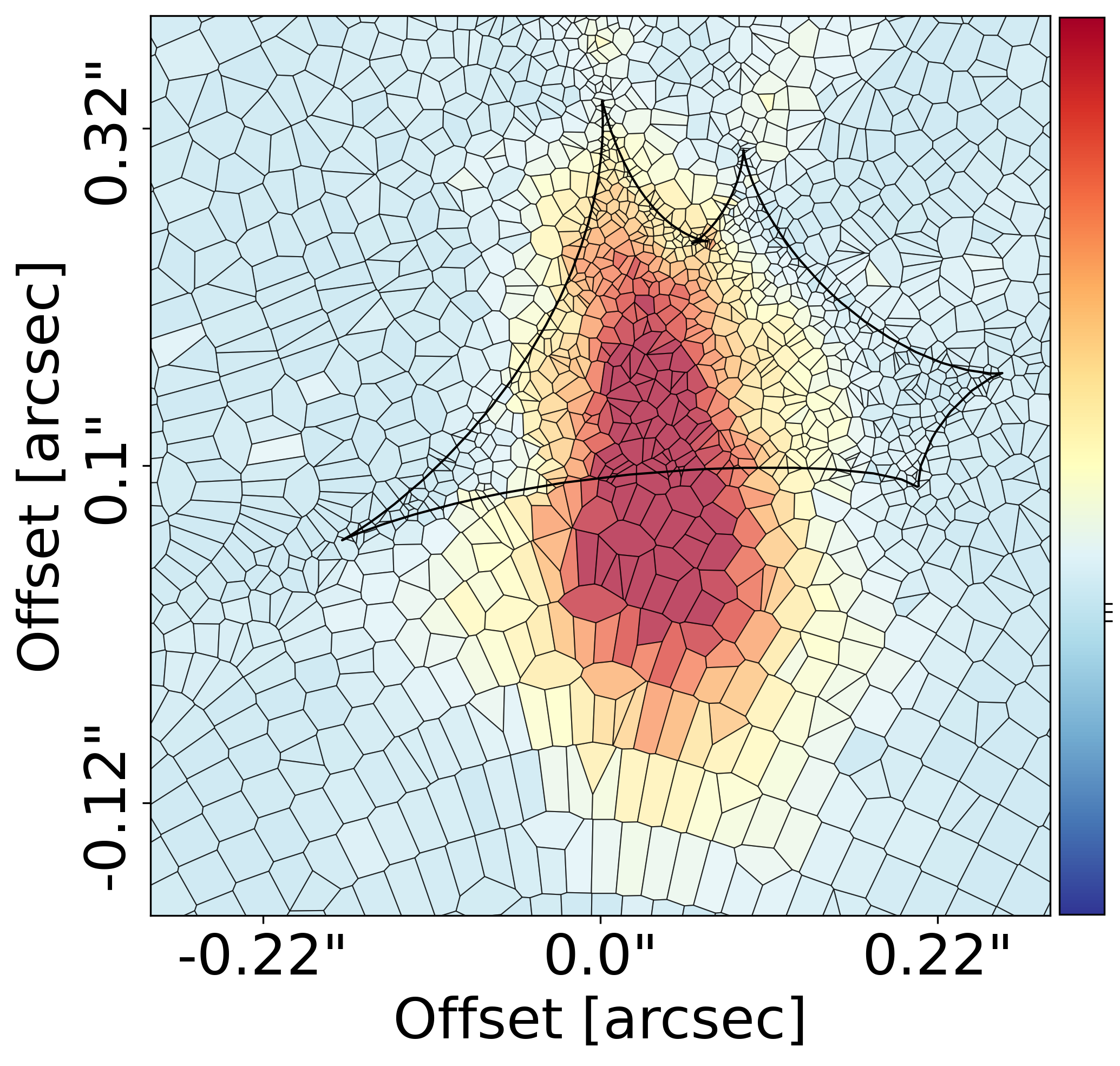}
    
    \caption{Non-parametric source modeling results for the detected emission towards G09v1.97, first row: dust continuum, second row: CO(6--5), third row: H$_2$O, fourth row: \waterp emission. The first column shows the dirty image as produced by {\sc PyAutoLens} with contours shown at $3, 4, 5, 6, 7, 8, 9, 10\sigma$ levels where $1\sigma$ is the rms of a blank region of the image. Note that this is not a cleaned image, and structures may look slightly different than those shown in cleaned images. The second column shows the dirty model image as produced by {\sc PyAutoLens} with contours shown at $3, 4, 5, 6, 7, 8, 9, 10\sigma$ levels. The third column shows the dirty residual image produced by {\sc PyAutoLens} with contours shown at $-3, -2, 2, 3, 4, 5\sigma$ levels. The fourth column shows the image plane emission non-parametric model of the data produced by {\sc PyAutoLens}. The black line shows the critical line. The fifth column shows the source plane emission non-parametric model of the data produced by {\sc PyAutoLens}. The black lines show the caustic line. All images are centered around the ALMA phase center for each image. Note that the \waterp emission is of significantly lower angular resolution as the lensing model was created using the combined data, as described in Section \ref{sec:observation_details}.}
    \label{fig:nonparametric_lens_modeling_images}
\end{figure*}

\paragraph{\water emission} \label{subsubsec:parametric_h2o_lens_model} 
Similar to the CO(6--5) emission, we modeled the \water emission using the continuum-optimized lens model and a single S\'ersic profile for the source. We provide the best-fit source parameters in Table \ref{tab:lensing_bestfit_sourceparams}. We show the \water image, model, and residual in Fig. \ref{fig:parametric_lensing_images}. Through this methodology, we obtain a lensing magnification factor of $\mu_{\rm H_{2}O} = 11.60 \pm 0.41$.

Similar to the CO(6--5) emission, we find a slight residual in the northern image of the source, see Fig. \ref{fig:parametric_lensing_images}, at a $\approx 2\sigma$ level. We suggest the same explanation for this residual as for the CO(6--5) model and explore non-parametric models of the \water emission in Section \ref{subsubsec:nonparametric_lensmodeling}. 

\paragraph{\waterp emission} \label{subsubsec:parametric_h2op_lens_model} 
Similar to both the CO(6--5) and \water emission, we modeled the \waterp emission using the continuum-optimized lens model with a single S\'ersic profile to describe the source. We show the \waterp image, model, and residual in Fig. \ref{fig:parametric_lensing_images}. Through this methodology, we obtain a lensing magnification factor of $\mu_{\rm H_{2}O^{+}} = 14.86 \pm 0.48$. 

We note that the four different emission types have different magnification factors. The magnification factors reported here are the average across the source and are, therefore, dependent on the spatial extent of the source. Given that we did not impose constraints on the different emission types, the differential lensing between different emission types is to be expected\footnote{The large spatial scale of the \waterp emission is also expected due to the significantly lower angular resolution of those observations.}. A similar effect has been found in previous studies in which lens modeling was performed on both different line emission and dust continuum emission at different wavelengths \citep[e.g.,][]{Giulietti23, Perrotta23}. 

\begin{table*}[]
    \centering
    \caption{Best-fit parametric source modeling results for continuum and detected molecular line species.}
    \begin{tabular}{l c c c c c c c c} \hline \hline
        Line & $x_{\mathrm{off}}^{a}$ & ${y_\mathrm{off}}^{b}$ & q$^{c}$ & PA$^{d}$ & Effective Radius & S\'ersic Index & $\mu_{\rm parametric}^{e}$ & $\mu_{\rm pixel}^{f}$\\
         & [''] & [''] & & [degrees] & [''] & & \\ \hline
         
         Continuum & $0.04^{+0.01}_{-0.01}$ & $0.05^{+0.01}_{-0.01}$ & $0.38 ^{+0.04} _{-0.04}$ & $-80.93 ^{+3.06} _{-3.09}$ & $0.08^{+0.02}_{-0.01}$ & $2.12^{+0.67}_{-0.45}$ & $10.55 \pm 0.36$ & $9.83 \pm 0.26$\\
         
         \vspace{1.0mm}
         
         CO(6--5) &  $0.01^{+0.49}_{-0.51}$ &  $0.02^{+0.48}_{-0.51}$ & $0.48 ^{+0.48} _{-0.50}$ & $-0.29 ^{+89.57} _{-89.98}$ & $0.51^{+0.49}_{-0.50}$ & $2.86^{+2.13}_{-2.06}$ & $9.72 \pm 0.27$ & $12.23 \pm 0.38$ \\
         
         \vspace{1.0mm}
         
         \water & $0.03^{+0.01}_{-0.01}$ &  $0.06^{+0.02}_{-0.02}$  & $0.46 ^{+0.19} _{-0.28}$ & $-68.06 ^{+19.23} _{-21.20}$ &  $0.17^{+0.64}_{-0.11}$ & $3.23^{+1.77}_{-2.39}$ & $11.60 \pm 0.41$ & $10.23	\pm 0.33$\\

         \vspace{1.0mm}
         
         \waterp & $0.040^{+0.002}_{-0.002}$ & $0.084^{+0.005}_{-0.005}$ & $0.35 ^{+0.02} _{-0.02}$ & $-83.76 ^{+1.52} _{-1.55}$ & $0.094^{+0.003}_{-0.003}$ & $0.80^{+0.024}_{-0.0029}$ & $14.86 \pm 0.48$ & $13.39 \pm 0.32$\\
         
     \hline 
    \end{tabular}
    \tablefoot{
        \tablefoottext{a}{x-position of the source defined in offset from the observational phase center.}
        \tablefoottext{b}{x-position of the source defined in offset from the observational phase center.}
        \tablefoottext{c}{Minor-to-major axis ratio of the source.}
        \tablefoottext{d}{Position angle of the source, defined by {\sc PyAutoLens} as counter-clockwise from the x-axis.}
        \tablefoottext{e}{Magnification factor of the emission calculated through the ratio of the image plane to source plane emission in the parametric regime.}
        \tablefoottext{f}{Magnification factor of the emission calculated through the ratio of the image plane to source plane emission in the pixelized regime.}
        }
    \label{tab:lensing_bestfit_sourceparams}
\end{table*}

\subsubsection{Non-parametric source modeling} \label{subsubsec:nonparametric_lensmodeling}

Previous studies of G09v1.97 have not included non-parametric source modeling results for molecular or atomic line emission. In this work, we use the optimized lens mass model derived from the dust continuum emission (described in Section \ref{subsubsec:parametric_dust_lens_model}) to create a pixelized reconstruction using a magnification weighted adaptive Voronoi pixel mesh. This approach has two main benefits: it imposes no constraints on the source morphology (apart from a regularization coefficient, see below) and maintains regions with higher magnification and, thereby, angular resolution. The reconstruction employs a regularization coefficient that corresponds to the smoothness of the image and attempts to balance between overfitting and oversmoothing \citep{Nightingale18}. We used a constant regularization scheme parameterized by the regularization coefficient. This method has been employed in numerous recent studies to reconstruct high-redshift galaxies observed with ALMA and the \textit{James Webb Space Telescope} \citep[\textit{JWST};][]{Maresca22, Giulietti23, Lei23, Perrotta23, Amvrosiadis24}. 

We used this methodology to create pixelized reconstructions of the dust continuum emission, CO(6--5), \ce{H_2O}, and \waterp emission using a Voronoi mesh covering a circular region of radius 1.5\,arcseconds (11\,kpc) in the image plane around G09v1.97. We show the images, models, and residuals for all modeled emission in Fig. \ref{fig:nonparametric_lens_modeling_images}. We find that the non-parametric source models provide a similarly good fit to the data as the parametric source models. We find small residuals at the same spatial regions as in the parametric models for the CO(6--5) and \water emission. However, these residuals are found at lower significance ($\approx 2\sigma$ levels) in the non-parametric models, suggesting that the non-parametric models provided an improved fit to the emission. Further, the discrepancies in the morphology between the parametric and non-parametric source plane models suggest that the parametric modeling should be viewed with caution as this is a likely indication that the emission contains structures not well represented by parametric models. We find that the non-parametric source plane morphology of the different emission types is very similar, with no indications of complex morphologies indicative of multiple sources. This provides further verification that there appears to be a single background source, not multiple sources, as found in \citet{Yang19}. We discuss the emission morphologies and extents further in Section \ref{sec:G09_in_source_plane}. 

Finally, we created source plane error maps for each of the emission sources. The source plane errors have two contributions: i) the uncertainties on the lens mass model ($\sigma_{l}$) and ii) the uncertainties on the source-plane pixelized reconstruction ($\sigma_{\rm data}$). The total source plane error map is then given by $\sigma = \sqrt{\sigma_{l}^2 +\sigma_{\rm data}^2}$. Here we describe how each contribution is determined.

We estimated the lens mass model errors ($\sigma_{l}$) by creating source plane pixelized emission maps for the CO(6--5) emission using lens mass models sampled from the posterior distribution of each of the lens parameters. We assume that the error on the lens mass model will not be significantly different for different emission types, and therefore only perform this sampling on the CO(6--5) emission\footnote{Note that this is also due to time limitations from the pixelization process. Using the CO(6--5) was the most logical choice given that the source plane error maps are of most interest in regard to the CO(6--5) source plane kinematics.}. Given that the formal statistical uncertainties in the lens mass model are small (see Table \ref{tab:bestfit_lens_model}), we do not expect the contribution from ($\sigma_{l}$) to be significant. The reported errors on the parameters are comparable to those reported in Table 4 of \citet{Maresca22}, despite differences in the source modeling assumptions used in their analysis. Such small formal uncertainties are common in group-scale strong lenses like G09v1.97, particularly when adopting simple mass models such as the Singular Isothermal Ellipsoid (SIE), which are prone to model misspecification. We sample the posterior of the maximum-likelihood solution for the lens mass model (fit from the continuum data) 1000 times, thereby creating 1000 different samples of the CO(6-5) emission in the source plane. We then created the $\sigma_{l}$ map by calculating the standard deviation of the brightness in each pixel across the 1000 samples. This approach is valid only if the lens model parameters form Gaussian distributions over the 1000 samples. We confirmed that this assumption holds.

We estimated the source-plane pixelized reconstruction error ($\sigma_{\rm data}$) by using {\sc PyAutoLens}'s infrastructure. In order to estimate the uncertainty in our non-parametric source-plane reconstructions, we estimate the root-mean-square (RMS) noise for each pixel in the source plane. Assuming a fixed lens mass model, the RMS error in each pixel is derived from the diagonal elements of the covariance matrix, following Equation 10 of \citet{Warren03}. In our case, this matrix is computed directly from the interferometric visibilities in the uv-plane and corresponding visibility errors extracted using CASA's STATWT task, as described in Equation 8 of \citet{Enia18}. This approach gives us an RMS noise value for each Voronoi pixel in the source plane, which was then interpolated onto a regular square grid using the same method applied in the source reconstruction. 

We show the total error map ($\sigma = \sqrt{\sigma_{l}^2 +\sigma_{\rm data}^2}$) for each emission source in Fig. \ref{fig:src_plane_errors}. We have normalized the error maps by the maximum value in the intensity map, so the error maps can be effectively interpreted as a percentage error in each pixel. As expected, the contribution from the lens mass model errors ($\sigma_{l}$) are small and generally do not provide a meaningful contribution to the total error. The errors for both the continuum and \waterp emission are, as expected due to the larger volume of data used for fitting, smaller than for the other emission types. Fig. \ref{fig:src_plane_errors} also provides signal-to-noise maps (i.e., intensity divided by intensity error) for clarity, in particular for later analyses performed in the paper. Generally, the emission in the source plane is at SNR $\gtrapprox 3$ meaning that the errors are generally modest. However, the outer regions of emission can be at significantly lower SNR. This is further discussed in Section \ref{subsec:G09_merger_or_disk}.

Similar to the parametric models, we calculated the magnification factor from the pixelized regime. The pixelization magnification factor error comes from sampling the maximum-likelihood lens model 100 times, and creating pixelized image and source plane emission maps for each sample where the magnification error is the standard deviation of the magnification factor across these 100 samples. We note that we only performed the posterior sampling 100 times due to time constraints when performing the reconstructions. Generally, the parametric vs. pixelized model magnification factors are similar, within $\approx2\sigma$, with the exception of the CO(6--5) emission where the magnification factor is stronger in the pixelized regime. We do not further investigate the magnification factor discrepancies from the two modeling regimes. For the purposes of this paper, we use the average of the magnification factors where relevant for magnification correction of physical properties.

\subsubsection{Spectral cube reconstructions} \label{subsec:cube_and_spec_reconstructions}

We used the {\sc PyAutoLens}'s non-parametric source modeling capabilities to create a source plane de-magnified emission cube of the CO(6--5) emission. We limited this analysis to the CO(6--5) due to the very high per-channel SNR ratio necessary to create robust de-magnified cubes. Indeed, as the main goal of this analysis was to perform kinematic modeling on the de-magnified cubes, the CO(6--5) emission was of primary interest. 

To create de-magnified source-plane emission cubes, we divided the emission line into bins of equivalent size and non-parametrically modeled the emission in each bin.  We used a spectral resolution of $\approx 50$\,km/s, allowing for a reasonable balance between maintaining SNR in each bin while maintaining a relatively high spectral resolution. The images for each bin were then interpolated onto a square pixel grid, a built-in capability of {\sc PyAutoLens}, and combined to form a spectral cube. The interpolation to a square pixel grid resulted in the loss of non-static pixel size across the image, necessitating the adoption of a beam size for future kinematic modeling. We assumed a standard beam size of five pixels per source-plane beam, where the source-plane beam was defined as $\theta_{\rm maj}/\sqrt{\mu} \times \theta_{\rm min}/\sqrt{\mu}$, with $\mu$ given by the parametrically modeled CO(6--5) emission (see Table \ref{tab:lensing_bestfit_sourceparams}). Thus, the static angular resolution achieved in the source plane was approximately $0''.035 \times 0''.027$ ($\approx 260 \times 200$\,pc) for the reconstructed CO(6--5) spectral cube.

The CO(6--5) reconstructed source plane cube was used to create moment-0 and moment-1 maps for the CO(6--5) emission, shown in Fig. \ref{fig:src_plane_mom0_mom1}. These can be compared with moment-0 and moment-1 maps made on the non-static images, which utilize a Voronoi mesh. We computed moment-0 and moment-1 maps for the Voronoi pixels and visually verified that the interpolation to square pixels did not significantly affect the moment-0 or moment-1 maps. Therefore, for the remainder of the analysis, we use the reconstructed source plane cube with a static beam unless otherwise stated. 

Similar to the parametric modeling and the non-parametric models of the dust continuum and molecular line emission, we find no evidence of two sources in the moment-0 maps (Fig. \ref{fig:src_plane_mom0_mom1}), contrary to the findings of \citet{Yang19}. Combined with the smooth velocity gradient seen in the moment-1 map (Fig. \ref{fig:src_plane_mom0_mom1}), this indicates a single, rotating disk. This interpretation is further investigated in Section \ref{subsec:kin_modeling} and Section \ref{subsec:G09_merger_or_disk}. 

We show the error maps for the CO(6--5) emission line using the same method as described in Section \ref{subsubsec:nonparametric_lensmodeling}. In this case, the $\sigma_{\rm data}$ is a moment-0 map created from an error cube for the CO(6--5) using the visibility errors in each channel. We show the CO(6--5) source plane moment-0, total error moment-0, and SNR maps in Fig. \ref{fig:src_plane_errors}. We find that inner regions of the image are at $\geq 5$ whereas the outer regions of the emission are at $\approx 2$. This highlights the difficulty associated with creating source plane emission cubes of lensed sources as it is very important to have very high quality data with high signal to noise in each channel. Although the data used in this paper is very good, the results drawn from the source plane cube need to be interpreted carefully.

\begin{figure*}[h]
    \centering
    \includegraphics[width = 1.0\linewidth]{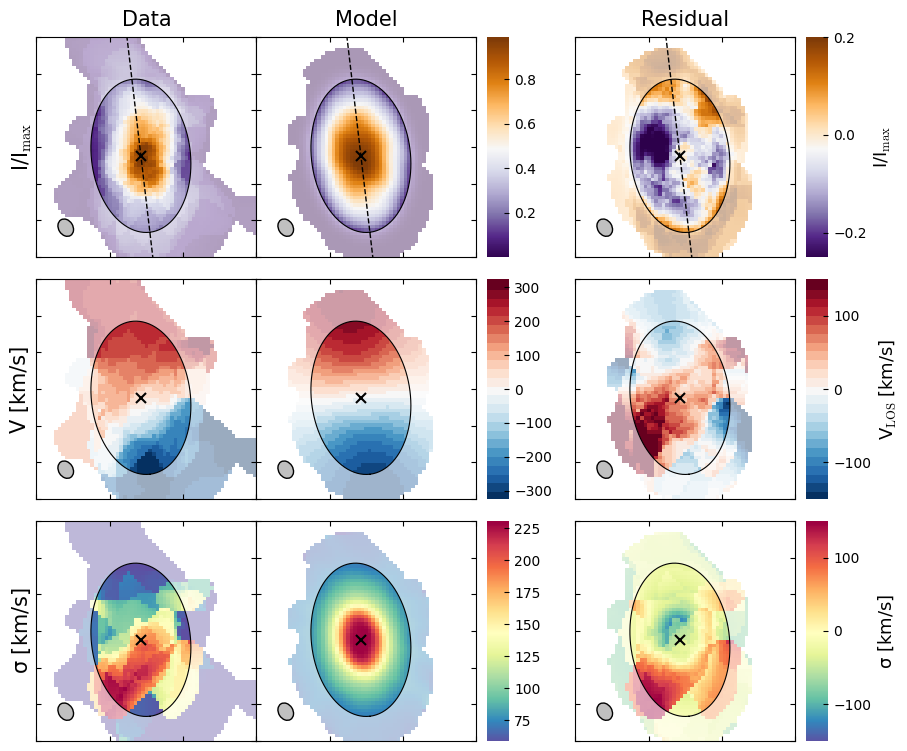}
    \caption{Fitting results from {\sc $^{\rm 3D}$Barolo} with parameters set as described in Section \ref{subsec:kin_modeling}. For all rows, the first column shows the data, the second the model, and the third the residual (data minus model). The first row shows the moment-0 emission (intensity), the second the velocity field (rotational velocity, $\rm V_{rot}$), and the third the velocity dispersion ($\sigma$). The final {\sc $^{\rm 3D}$Barolo} ring is shown as the black ellipse and regions outside this ellipse are made semi-transparent to emphasize the region used for fitting. The major axis is shown in the first two rows by the dashed grey line. The kinematic center is shown by the green line in the first two columns of the second row. The approximate beam, assuming a static magnification factor across the image, is shown in the bottom left of each image.}
    \label{fig:barolo_modeling}
\end{figure*}

\subsection{Kinematic modeling} \label{subsec:kin_modeling}

To investigate the velocity gradient seen in the CO(6--5) data in both image plane and source plane moment-1 maps (see Fig. \ref{fig:img_plane_mom0_mom1} and Fig. \ref{fig:src_plane_mom0_mom1}), we use the kinematic modeling tool {\sc $^{\rm 3D}$Barolo} \citep{Barolo}. We perform all kinematic modeling on the CO(6--5) source plane de-magnified cube (see Section \ref{subsec:cube_and_spec_reconstructions}), assuming a static beam in the source plane. As mentioned in Section \ref{subsec:lens_modeling}, the brightness value of the source plane cube is arbitrary. This does not affect the kinematic modeling as the primary point of interest is the velocity associated with each channel, which is largely not affected by the lens modeling. In this section, we describe the methodology used to model the CO(6--5) emission and report our results from this modeling. 

\subsubsection{Modeling with {\sc $^{\rm 3D}$Barolo}}

We use {\sc $^{\rm 3D}$Barolo} to model the kinematics of the CO(6--5) emission in the source plane. {\sc $^{\rm 3D}$Barolo} is a commonly used kinematic modeling tool that has been used across a wide redshift range \citep[e.g.,][]{Barolo, Mancera19, Di21, Mancera22, Fernanda23, Gray23, Amvrosiadis24, Rowland24}. Rather than employing a parametric approach, this tool fits a user-specific number of tilted rings to the input emission line cube and returns the best-fit model for the rings through a Monte Carlo evaluation methodology. {\sc $^{\rm 3D}$Barolo} attempts to then correct for any beam-smearing effects by convolving the modeled rings to the resolution of the data. 

\begin{figure}[h]
    \centering
    \includegraphics[width = 1.0\linewidth]{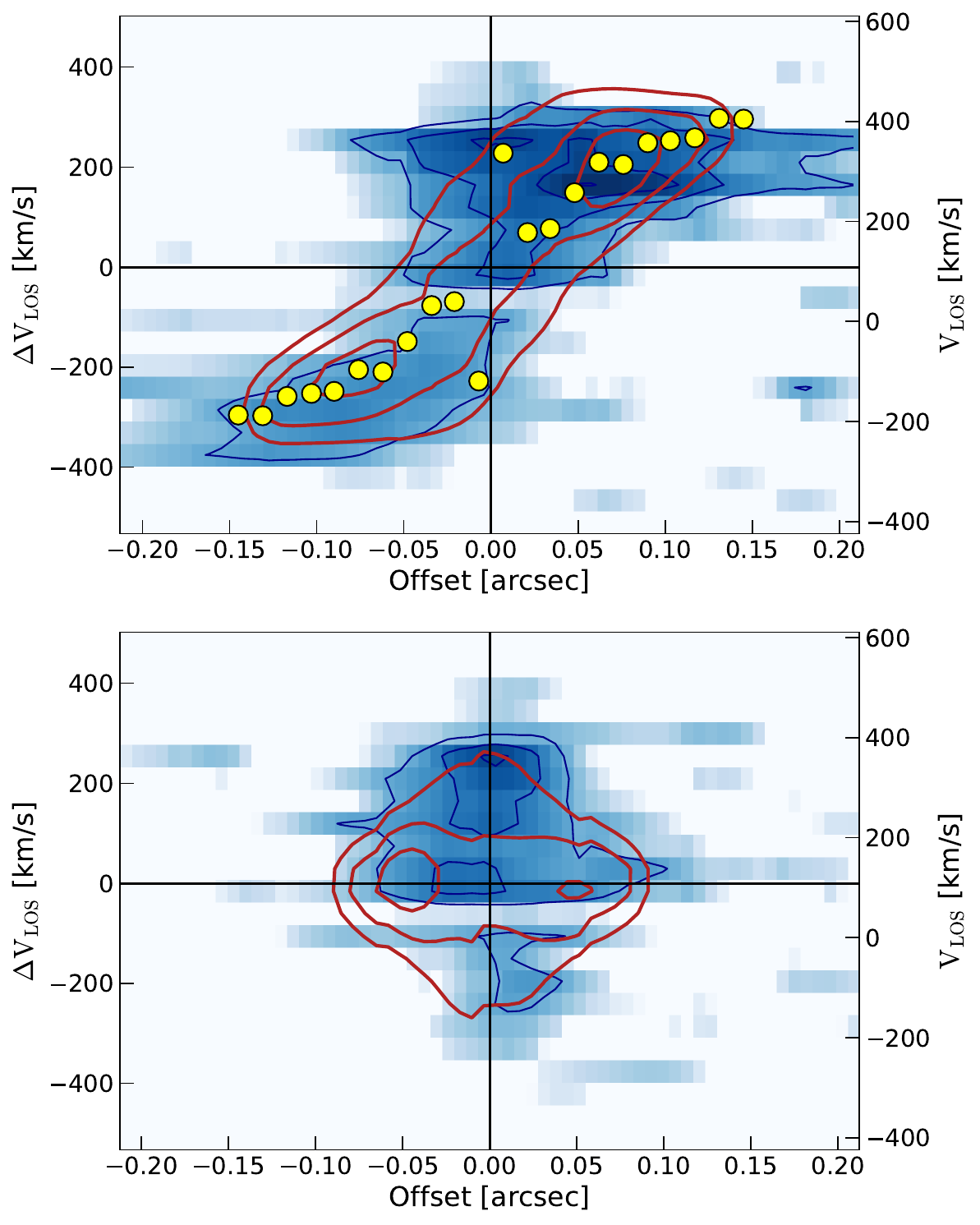}
    \caption{PV diagrams from {\sc $^{\rm 3D}$Barolo}, upper: major axis, lower: minor axis. The blue contours show the contours of the data at $2, 3, 4, 5\sigma$ levels, and the red contours show the best-fit model to the data. In the major axis pv diagram, the yellow dots show the best fit rotational velocity in each of the rings. The right axis ($\rm V_{LOS}$) shows the velocities centered around the systemic velocity of the galaxy at the redshift of G09v1.97 ($z = 3.63$). The left axis ($\Delta \rm V_{LOS}$ ) shows the velocities centered around the user-set systemic velocity, here $120$\,km/s (Table \ref{tab:barolo_params}).}
    \label{fig:barolo_pvs}
\end{figure}

We use \barolo's {\sc Smooth \& Search} algorithm to mask the data. This algorithm first smooths the data, performs a $3\sigma$ cut to the data to mask values at $<3\sigma$, and then searches for sources in the smoothed and masked data. We use a radial separation of $0.0138''$ for the individual rings, calculated by $\theta_{\rm maj}/2.5$ where $\theta_{\rm maj}$ is the major axis of the source-plane static beam, resulting in 11 rings. We used 2D-Gaussian fitting to find the ($x,y$) center of the moment-0 emission, and employed this value as the center while fitting the tilted rings. Hence, the free parameters in our model were the rotation velocity ($\rm V_{rot}$), the dispersion ($\sigma$), the systemic velocity ($\rm V_{sys}$), the inclination ($i$) of each ring, and the position angle (PA) of each ring. 

We first ran the fitting procedure with both position angle and inclination free (together with the rotation velocity and dispersion). We then took the averages of the position angle and the inclination, respectively, and fixed those values to refit the disk. The ability of {\sc $^{\mathrm{ 3D}}$Barolo} to adjust each ring individually allows for potential non-physical adjustments to non-axisymmetric motions/structures. By fixing the position angle and inclination, we ensure that the final fit most closely resembles a disk, and the residuals retain the ability to reveal the presence of non-axisymmetric structures such as outflows, tidal tails, etc. For the fitting of this particular galaxy, the variation of the position angle and inclination was small. The inclination ($i$), at 50 degrees, did not vary at all, and the position angle varied from $\approx 1$ to $\approx 19$ degrees, resulting in an average of $(6.7 \pm 1.7)^{\circ}$. 

\begin{table}[]
    \centering
    \caption{Best-fit parametric from{\sc $^{\rm 3D}$Barolo} fitting.}
    \begin{tabular}{l c} \hline \hline
      Parameter & Value \\ \hline
      PA$_{\rm kin}^{a}$ & 6.7 \\
      i$^{\circ, b}$ & 50 \\ 
      $\rm V_{sys}^{c}$ & 109 \\
      $\rm V_{rot, max}^{d}$\,[km/s] & $388^{55}_{-36}$ \\
      $\bar{\sigma}^{e}$\,[km/s] & $138^{+7}_{-6}$\\
      $\rm V_{rot, max} / \bar{\sigma}^{f}$ & $2.8 \pm 0.4$   \\
         
     \hline 
    \end{tabular}
    \tablefoot{
        \tablefoottext{a}{The position angle of the disk, measured counterclockwise from the y-axis.}
        \tablefoottext{b}{The inclination angle of the disk.}
        \tablefoottext{c}{Systemic velocity of the CO(6--5) line.}
        \tablefoottext{d}{Maximum rotation velocity across all rings.}
        \tablefoottext{e}{Average $\sigma$ value across all rings.}
        \tablefoottext{f}{Ratio of the maximum rotational velocity over the average $\sigma$ with errors propagated.}
        }
    \label{tab:barolo_params}
\end{table}

\subsubsection{Modeling results}

We show the velocity maps from the final model from {\sc $^{\rm 3D}$Barolo} Fig. \ref{fig:barolo_modeling} and the generated PV-diagrams in Fig. \ref{fig:barolo_pvs}. The PV diagram and rotational velocity along the major axis show a clear S-shape, which is a characteristic of rotation. Additionally, the minor axis PV-diagram exhibits a diamond shape, which is also characteristic of rotation. However, we note that in both diagrams, emission in the negative velocities (i.e., the blue region of the CO(6--5) spectrum) is proportionally weaker. This is likely due in part to the offset kinematic center of the data with respect to the model and could be associated with other non-circular motions, discussed below and in Section \ref{subsec:G09_merger_or_disk}. We note that in Fig. \ref{fig:barolo_modeling}, we show regions outside the elliptical region used for the {\sc $^{\rm 3D}$Barolo} fitting as semi-transparent. While the mask used for the {\sc $^{\rm 3D}$Barolo} fitting masks regions below $3\sigma$, the regions outside the elliptical fitting region are at lower SNR ($<2$) regions of significance when considering the source plane SNR map in Fig. \ref{fig:src_plane_errors}. There is a delicate balance to consider when taking into account the different types of errors in the combined analyses, but we find that this method conveys the most information. We report the best-fit values from {\sc $^{\rm 3D}$Barolo} in Table \ref{tab:barolo_params}. We calculate the maximum rotational velocity ($\rm V_{max}$), average velocity dispersion ($\bar{\sigma}$), and $\rm V_{max} / \bar{\sigma}$ for G09v1.97 and report these values in Table \ref{tab:barolo_params}. We can interpret $\bar{\sigma}$ as a proxy for the overall turbulence in the galaxy, and hence $\rm V_{max} / \bar{\sigma}$ provides a measure of the ordered motion to the turbulent motions in a galaxy. We note that we use $\rm V_{max}$, which peaks at the largest probed radii. We have ensured that $\rm V_{max}$ probes regions of the emission with significant emission, those with SNR $\geq 2$ when considering the source plane error and SNR maps, in the galaxy. We note that CO(6--5) might not trace large distances into the disk of the galaxy, given the relatively high temperatures and densities necessary to excite mid-$J$ transitions of the CO molecule, in comparison to those necessary for CO(1--0), HII, or \cii, and is likely to not trace the spatial extent of the bulk colder molecular gas in G09v1.97 \citep[e.g.,][]{Cailli13, Madden20}. However, the rotation curve of the CO(6--5) emission does appear to flatten towards the edges of the CO(6--5) emission, and therefore we assume that using $\rm V_{rot}$ in the outermost ring is the best approximation of $\rm V_{max}$ possible with the current data.

\begin{figure}[h]
    \centering
    \includegraphics[width = 1.0\linewidth]{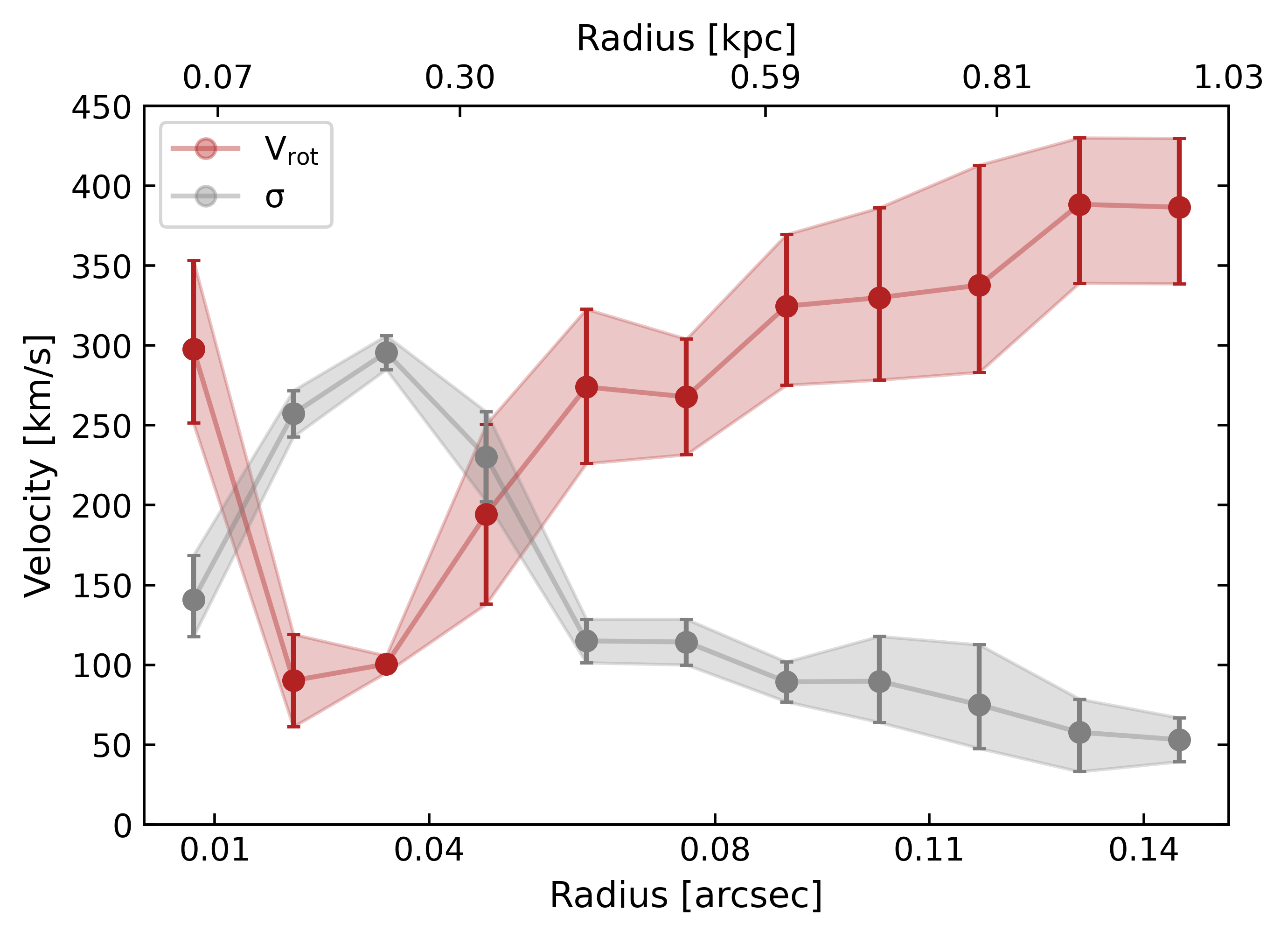}
    \caption{Rotational velocity and dispersion as a function of radius for G09v1.97 for each ring in our {\sc $^{\rm 3D}$Barolo} fit. The red shows the rotational velocity, and the gray shows the dispersion. The errors are represented by the shaded regions for both.}
    \label{fig:v_sig_function_of_radius}
\end{figure}

We plot $\rm V_{rot}$ and $\sigma$ in each ring with increasing radius in Fig. \ref{fig:v_sig_function_of_radius}. Excluding the innermost ring, $\rm V_{rot}$ rises steadily, and $\sigma$ falls with increasing distance from the center of the galaxy, as expected for a rotating disk. Apart from this innermost ring, the decrease in $\sigma$ with increasing radius is expected as the most turbulent motions mostly occur in the innermost regions of the galaxy \citep[e.g.,][]{Fraternali02, Boomsma08, Tamburro09, Sharda19, Rizzo23}. The innermost ring exhibits what appears to be a `bump' in $\rm V_{rot}$ and has a correspondingly lower $\sigma$ value. This bump was also present in the fit where all parameters were left free and is significant (i.e., cannot be explained as an error). This elevated $\rm V_{rot}$ and lower $\sigma$ could be indicative of an additional galaxy component such as a bar or a bulge \citep[e.g.,][]{Amvrosiadis24, Rowland24}. However, the elevated $\rm V_{rot}$ at the innermost ring does not appear as strong in the blue velocity portion major axis pv-diagram (Fig. \ref{fig:barolo_pvs}), so it is possible that this is an asymmetric component in the galaxy, further discussed in Section \ref{subsec:G09_merger_or_disk}. The corresponding increase in $\sigma$ in the innermost ring is likely due to the strong emission in the eastern part of the dispersion map and dispersion residual (Fig. \ref{fig:barolo_modeling}) and is likely dominating \barolo's fitting procedure for the innermost ring. Given the limitations of the data and the inherent errors associated with both the lensing analysis and the kinematic modeling, we can not make conclusive statements about the cause of this bump. 

The best-fit systemic velocity of the CO(6--5) emission from the kinematic fitting, $\rm V_{sys} = 109$\,km/s, corresponds to a redshift of $z = 3.6300017$, slightly higher than previously reported. This offset lies closer to the peak in the red portion of the spectrum, which occurs at $210$\,km/s in the image plane; see Fig. \ref{fig:img_plane_spectra}. This velocity offset from 0 is further discussed in Section \ref{subsubsec:spec_line_shapes} and \ref{subsec:G09_merger_or_disk}. 

Fig. \ref{fig:barolo_modeling} shows an apparent offset in the kinematic center between the observed data and the model. We find a slight residual in the moment-0 map and in approximately the direction of the minor axis in the moment-1 map, see Fig. \ref{fig:barolo_modeling}. Additionally, there is a residual in the velocity dispersion in the region corresponding to the residual in the moment-1 map. We further discuss these residuals in Section \ref{subsec:G09_merger_or_disk}.

We note that although we assumed a static beam across the image (see Section \ref{subsec:cube_and_spec_reconstructions}), as has been done in previous kinematic studies of lensed galaxies \citep[e.g.,][]{Amvrosiadis24}, the magnification across the image is not, in fact, static; this is easily seen in the differing sizes of the Voronoi cells in Figure \ref{fig:nonparametric_lens_modeling_images}. In effect, the beam size should vary across the image, becoming larger in less magnified regions and vice versa in more magnified regions. The effect of this variation has been shown in previous studies \citep[e.g.,][]{Dye22}, but remains difficult to quantify in the case of G09v1.97 \footnote{Note that the Voronoi cells are not a one-to-one mapping of the beam and therefore, for example, an average of the size of the cells cannot be directly employed as the beam size.}. This assumption primarily affects the number of rings used in the {\sc $^{\rm 3D}$Barolo} modeling, which is largely a user-defined preference. 

We performed a simple investigation into the impact of the assumed static beam on our kinematic modeling results by varying the ring separation used in the {\sc $^{\rm 3D}$Barolo} modeling. We vary the separation from $0''.007$ ($\theta_{\rm maj}/5$, corresponding to 23 rings) to $0''.042$ ($\theta_{\rm maj}*1.2$, corresponding to 4 rings) in steps of a factor of two. We followed the same fitting procedure as described in Section \ref{fig:barolo_modeling}. Through this investigation we found that the ring separation has a minimal impact on the results obtained for the fit. Figure \ref{fig:v_sig_radius_perterbbeam} shows the rotation curve and dispersion in each ring for the different ring separations and clearly shows that the variation of the number of used rings largely does not affect the shape of the curves and, thereby, the results of the kinematic modeling. We note that when using either 4 or 5 rings, we do not find the increase in rotational velocity and a corresponding decrease in dispersion found towards the center of the galaxy. We suggest that the feature in the center of the galaxy producing the increase seen when using a higher number of rings is smoothed out when using too few rings; however, we acknowledge that with the current data, it is difficult to determine the nature of this feature. We found that the $\rm V_{max} / \bar{\sigma}$ varied between $2.6 - 3.5$ with an average value of $3.2 \pm 0.15$. Given that the average value found in this investigation was is within errors of our initial fit using 11 rings, we conclude that the size of the beam, and thereby the assumed ring separation, does not largely affect the results of this work.

\section{Discussion}\label{sec:discussion}

\subsection{G09v1.97 in the source plane} \label{sec:G09_in_source_plane}

\begin{figure*}[h]
    \centering
    \includegraphics[width = 0.9\linewidth]{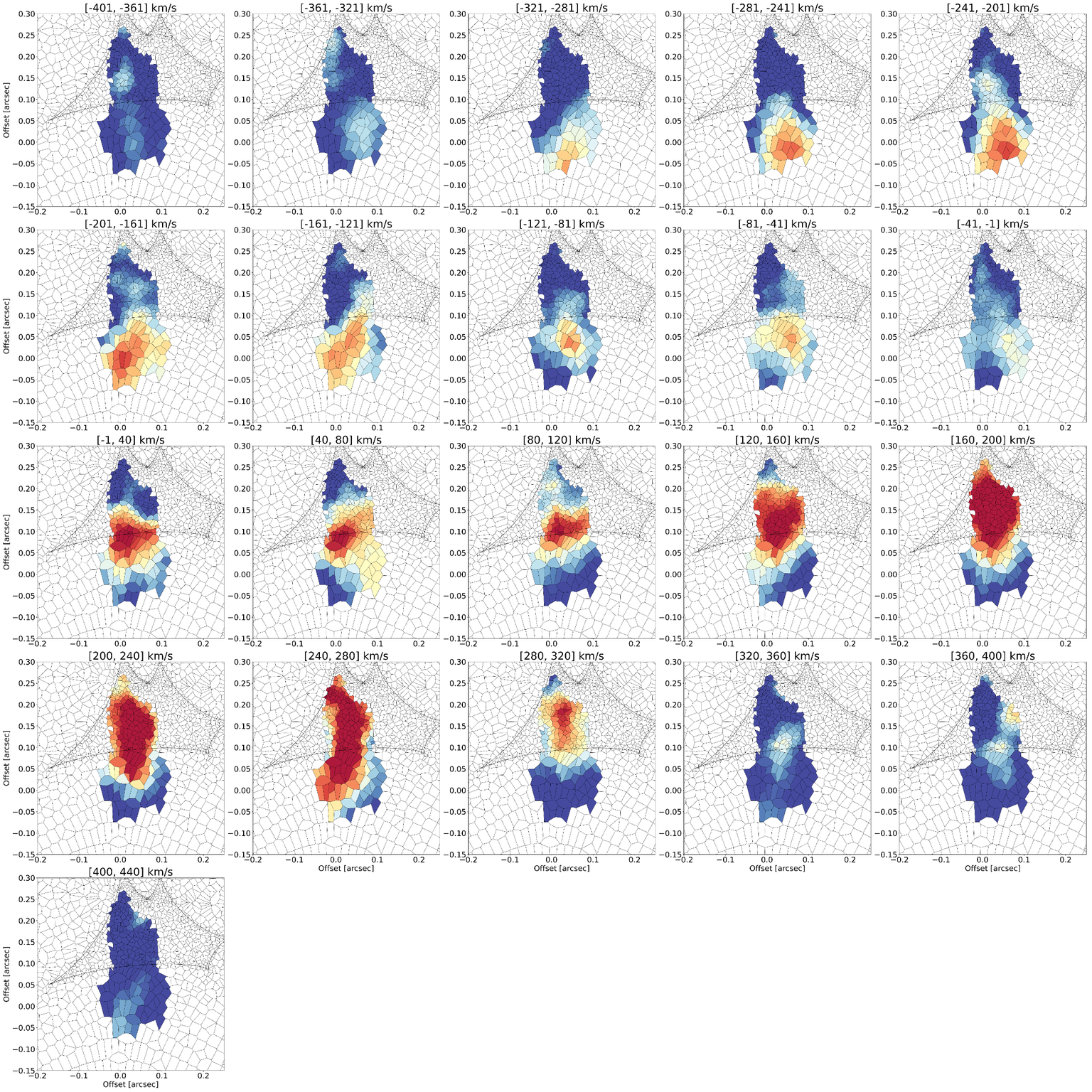}
    \caption{Source plane channel maps for the CO(6--5) emission from the de-magnified source plane emission cube described in Section \ref{subsec:cube_and_spec_reconstructions}. The velocities covered by each bin are shown at the top of the image. The background mesh shows the Voronoi mesh pixels prior to interpolation to a square grid. The black line shows the caustic line. A clear velocity gradient is evident across the emission line.}
    \label{fig:src_plane_channel_maps}
\end{figure*}

\subsubsection{Spectral line shapes} \label{subsubsec:spec_line_shapes}
Motivated by the clear velocity gradient seen in the source-plane reconstruction moment-1 map (Fig. \ref{fig:src_plane_mom0_mom1}), we created channel maps across the reconstructed cube in 50\,km/s bins. We masked these images to only show emission at locations above $5\sigma$ in the moment-0 map (i.e., emission in the region shown in the masked Voronoi moment-0 map shown in Fig. \ref{fig:src_plane_mom0_mom1}), see Fig. \ref{fig:src_plane_channel_maps}. These channel maps clearly indicate a south-north velocity gradient across the line profile, possibly providing another indication of rotation. However, this could also be an indication of merger activity \citep[e.g.,][]{Yang19}.

We investigate the emission line profiles, which are of particular interest given their asymmetric profiles. \citet{Yang19} clearly demonstrated that there is differential magnification occurring across the emission line. This is also clear from the velocity maps (Fig. \ref{fig:src_plane_channel_maps}) as the emission in redder velocities is shifted north towards the caustic line, meaning that the magnification will be higher in that region of the spectrum. We investigate this through two different, respective, methodologies for the \water and the CO(6--5) emission. In the case of the CO(6--5) emission we used the source plane cube as described in Section \ref{subsec:cube_and_spec_reconstructions}. Since we do not create a source plane \water spectral cube we investigate the emission line profile through parametrically modeling the red and blue portions of the emission line, as described below. 

\begin{figure*}[h]
    \centering
    \includegraphics[width = 1.0\linewidth]{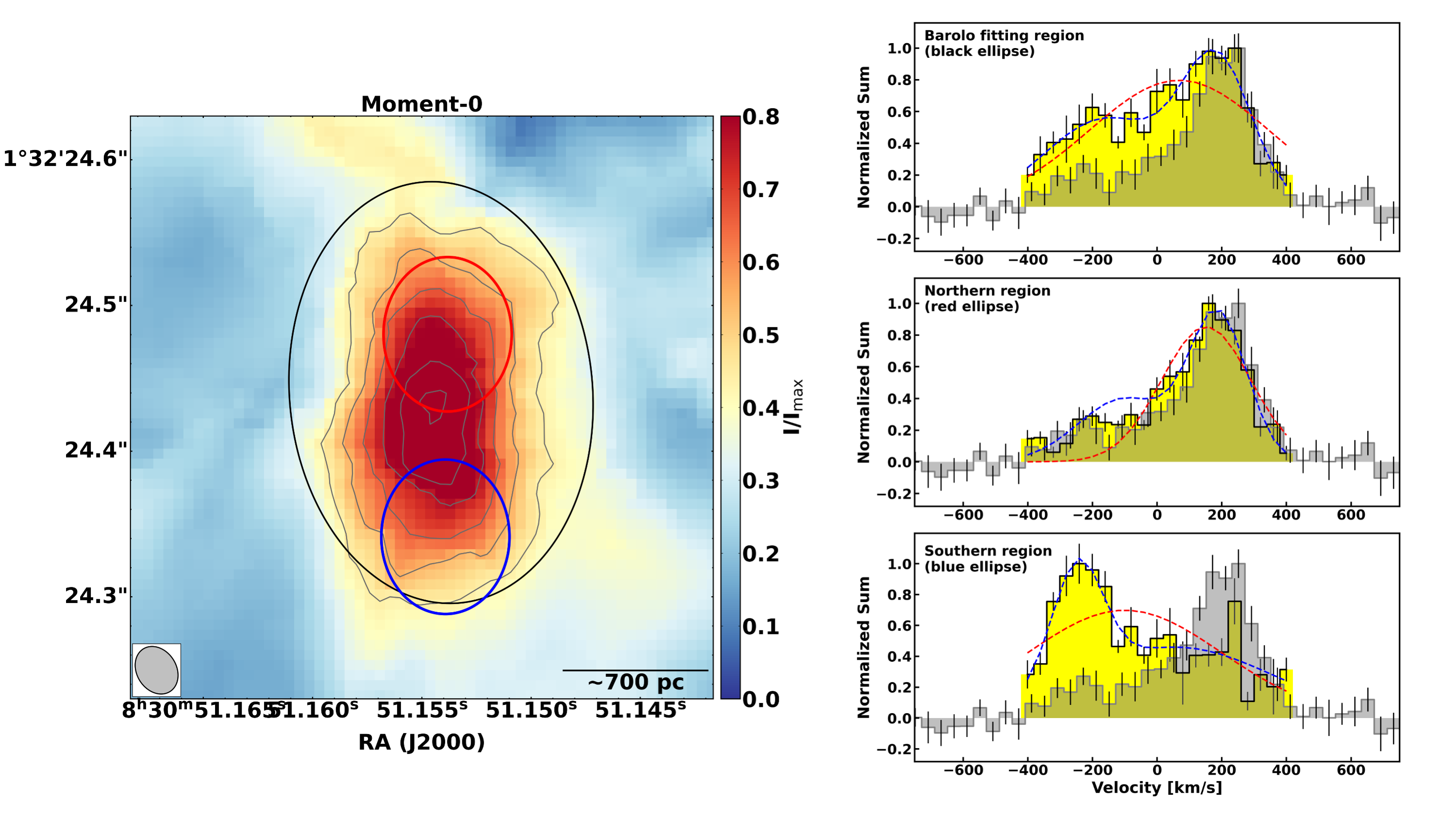}
    \caption{Comparison between the approximate image plane spectrum and source plane spectrum of the CO(6--5) emission in G09v1.97. \textit{Left:} the source plane moment-0 map of the CO(6--5) emission with the different regions used for the source plane spectral extraction. \textit{Right:} the normalized source plane spectrum (in yellow) compared to the normalized image plane spectrum (in grey) extracted from the region used in {\sc $^{\rm 3D}$Barolo} modeling (top), a northern region of the source located inside the caustic line (middle), and a southern region of the source located outside the caustic line (bottom). Note that the per-channel values are the normalized sum in a specific region as described in Section \ref{subsubsec:spec_line_shapes} and not a standard flux. The error bars are shown for each channel. Single Gaussian profiles are shown in red, and double Gaussian profiles are shown in blue.}
    \label{fig:co65_srcplane_spec}
\end{figure*}

We show the approximate source plane spectrum for the CO(6--5) emission in Fig. \ref{fig:co65_srcplane_spec}. Although units in the source plane reconstructions from {\sc PyAutoLens} are arbitrary, qualitative investigations between source plane and image plane spectra can be performed by normalizing both spectra before comparison. We only reconstructed the channels with the brightest emission and, therefore, do not have a full spectrum across all velocities in the source plane. Further, note that both the image and source plane spectrum are normalized per-channel sums of the flux inside a region, and therefore, figures depicting the image-plane CO(6--5) spectrum in units of mJy may appear slightly different. The region used for the spectral extraction of the normalized image plane spectrum is the same as used previously to extract the source plane spectra for \water and CO(6--5) as well as for extracting the underlying dust continuum (see Sections \ref{sec:results}) and is shown in Fig. \ref{fig:continuum}. The region used to extract the source plane spectrum was the same region used in the {\sc $^{\rm 3D}$Barolo} modeling, shown by the black ellipse in Fig. \ref{fig:barolo_modeling} and \ref{fig:co65_srcplane_spec}. In addition, we extracted the source plane spectrum from a northern and southern region G09v1.97, corresponding to above/inside the caustic line and below the caustic line. We calculated the RMS in each channel using the same simple sampling method as described in Section \ref{subsec:lineemission_results}. 

We found that the source plane CO(6--5) spectrum from the {\sc $^{\rm 3D}$Barolo} region was significantly flatter than the image-plane spectrum and does not appear to have two clear red and blue components, as seen in the image plane spectrum. Instead, we find a profile that is more similar to a double-horned profile, indicative of rotation, and in good agreement with our results from {\sc $^{\rm 3D}$Barolo}. We investigated the systemic velocity offset of $109$\,km/s from the {\sc $^{\rm 3D}$Barolo} modeling by fitting the source plane spectrum with a single Gaussian profile and using two Gaussian profiles. We find that the systemic velocity in the source plane spectrum is $68 \pm 29$\,km/s using a single Gaussian profile and that the red component peaks at $186 \pm 18$ when using two Gaussian profiles. This does not fully account for the offset in the systemic velocity in the {\sc $^{\rm 3D}$Barolo} modeling but provides an indication that the cause of this offset may be the asymmetric line profile. 

We further investigated the two components seen in the image plane spectrum by considering the spectrum extracted from the northern and southern regions. It is clear that the red peak in the spectrum corresponds to emission in the north of the image and the blue peak in the spectrum to emission in the south of the image, see Fig. \ref{fig:co65_srcplane_spec}. The red region is located within the caustic lines and therefore subject to higher magnification than the blue region, which accounts for the discrepancy in the peak of the components in the image plane spectrum, see the bottom right panels in Fig. \ref{fig:co65_srcplane_spec}. However, the source plane spectrum from the entire source (i.e., encompassed in the region used for {\sc $^{\rm 3D}$Barolo} region) still appears to be brighter in redder regions of the spectrum. This could be a residual effect of the lens modeling or an observational effect caused by, for example, inclination effects. \citet{Kohandel19} shows that a change in inclination can have unexpected and asymmetric effects on the line profile when the source is not an ideal disk. Future observations and source plane analysis of additional bright, atomic, or molecular emission lines such as \cii could assist in determining the cause of the asymmetric source plane line profile. 

We now consider the line profile of the \water emission. We divided the \water lines into two bins, with each bin covering one of the two respective peaks in the spectrum of both molecules; these bins covered the velocity ranges of $-300$ to 175\,km/s, corresponding to the peak in the blue region of the spectrum, and 175 to 500\,km/s, corresponding to the peak in the red region of the spectrum. We found that the magnification factor is different in the red and blue regions of the spectrum with $\mu_{\rm blue} = 8.87$ and $\mu_{\rm red} = 13.97$, making the ratio of the emission, as described in Section \ref{subsubsec:h2o_line_results}, $\rm peak_{red} / peak_{blue} \approx 1.5$. Both the observed and corrected \water ratios are lower than those found for the CO(6--5) emission, while the magnification factor in the blue part of the spectrum remains approximately constant between the two lines. This suggests that the majority of the differential lensing is occurring in redder portions of both spectra. 

An alternative explanation for the differences in the peak strengths of the emission lines is an inclination effect. \barolo fitting results suggest that G09v1.97 lies at an inclination of $i \approx 50^\circ$. At this inclination, an asymmetric double-horned profile would be expected (e.g., Yttergren et al. in prep). However, if this profile is consistent with rotation or interactions, it is unclear from the spectra alone \citep[e.g.,][]{Rizzo22}. The spectrum of the CO(6--5) in the image plane is qualitatively similar to that of one component of the merger SMG HXMM01 while the source plane spectrum qualitatively resembles another component \citep{Hai18}. Without additional observations, it is difficult to come to robust conclusions about the source of the asymmetric line profile. 

\subsubsection{Emission morphology \& extent}
\begin{figure}[h]
    \centering
    \includegraphics[width = 1.0\linewidth]{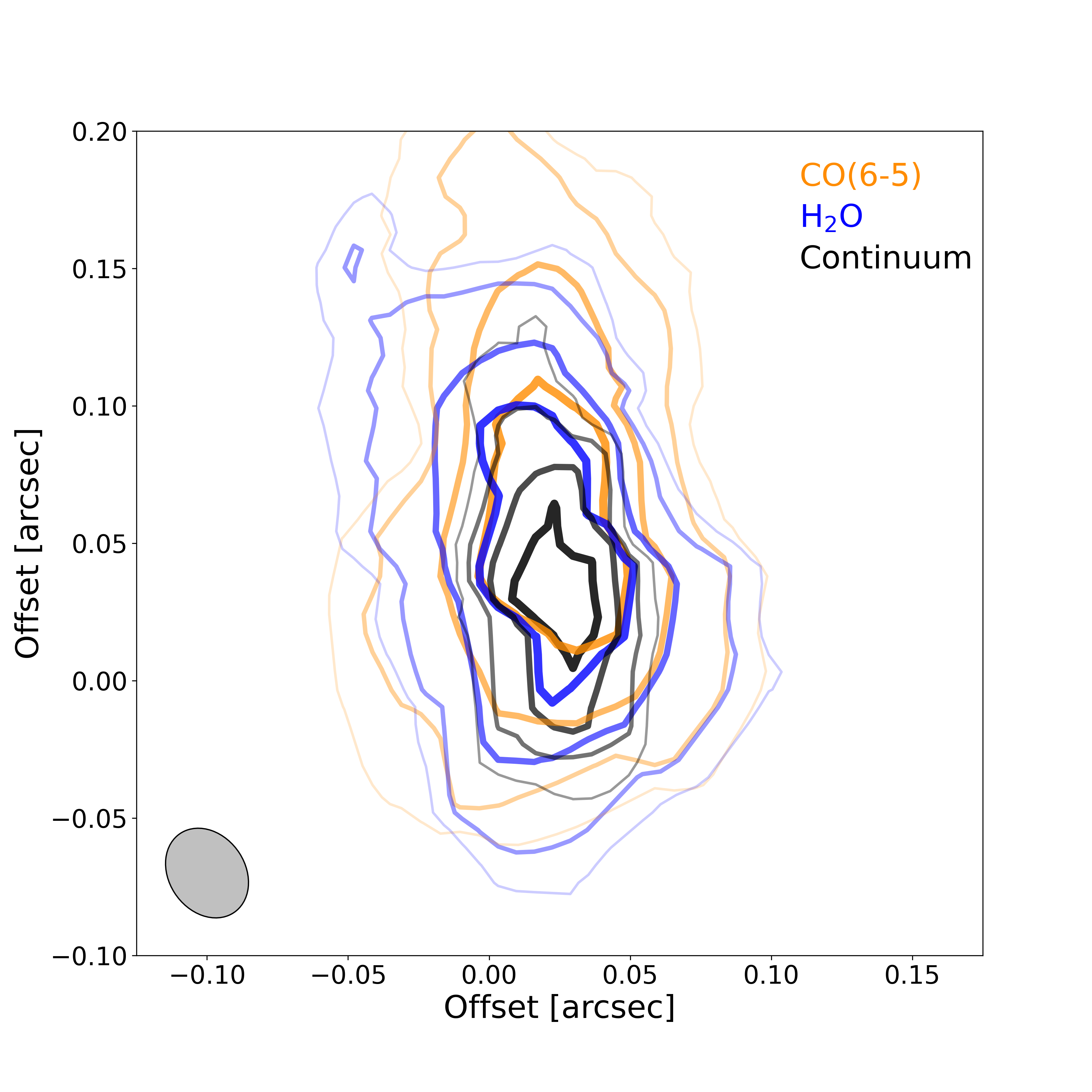}
    \includegraphics[width = 1.0\linewidth]{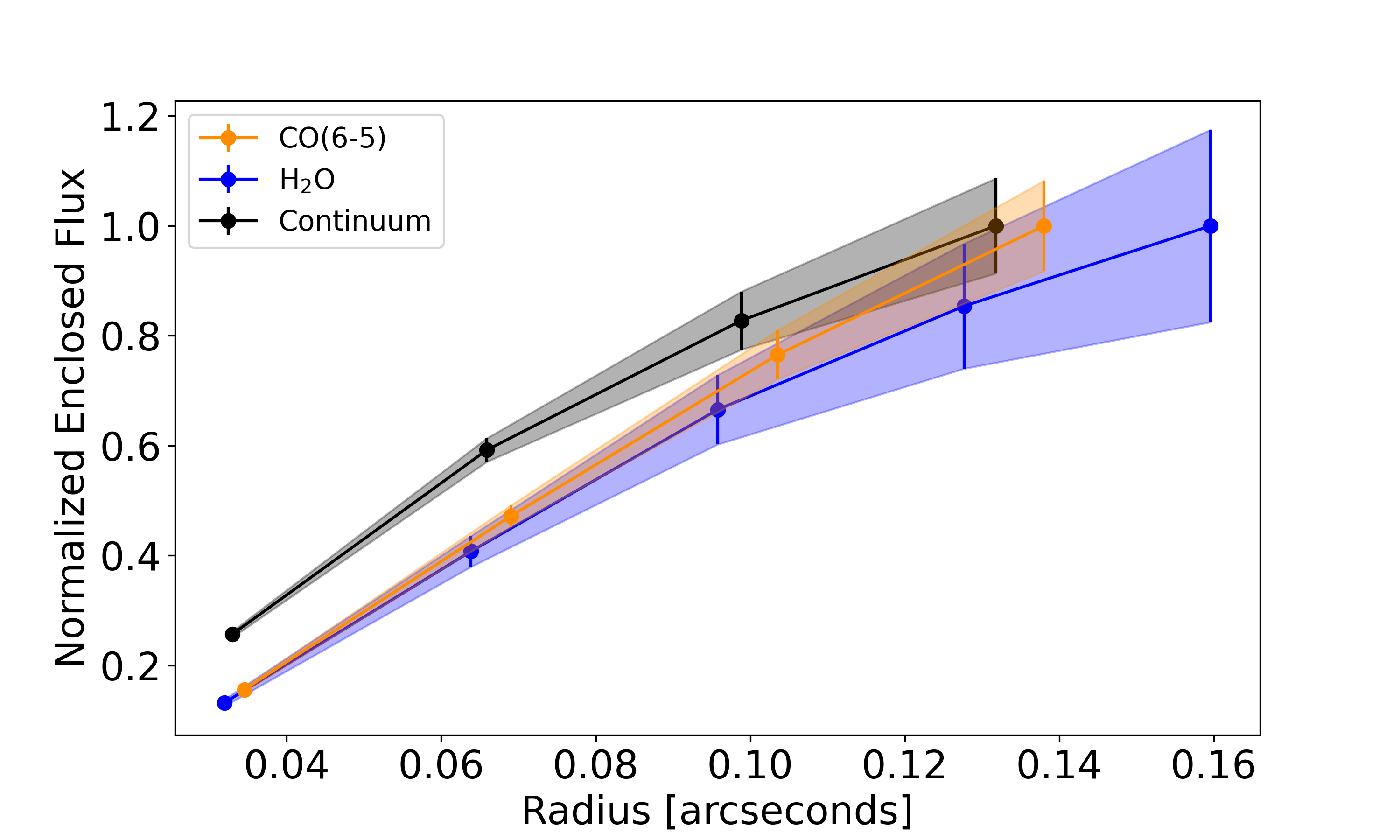}
    \caption{Top: Source plane contour plots of the dust continuum, CO(6--5), and \water emission in G09v1.97. The contours are at $30, 40, 60, 80, 100\%$ levels of the maximum value of the individual tracer. The approximate beam, assuming a static magnification factor across the image, is shown in the bottom left of the image. Bottom: Normalized enclosed flux in circular apertures increasing from the center of the CO(6--5), dust continuum, and \water emission out to 80\% of the total flux of the image. It is clear that the dust continuum emission is the most compact of the emission types, whereas the CO(6--5) and \water emission closely match in emission extent.}
    \label{fig:src_plane_contour}
\end{figure}

We investigate G09v1.97 in the source plane primarily using the non-parametric source-plane reconstruction images as described in Section \ref{subsec:cube_and_spec_reconstructions}. We do not include the \waterp emission in this analysis due to the significantly lower angular resolution, which does not allow for a robust comparison. We opt to exclude the parametric source models from this investigation due to the a priori source morphology assumptions inherent to these models. We show the spatial extent of the continuum emission, CO(6--5), and \water emission from the non-parametric source reconstruction of the emission (see Section \ref{subsubsec:nonparametric_lensmodeling}) in Fig. \ref{fig:src_plane_contour}. Rather than showing $\sigma$ level contours, we show the contours at percentage levels (30\%, 40\%, 60\%, 80\%, and 100\%) of the peak flux for the specific tracer with darker colors corresponding to higher percentages. We opt to show the extent in this manner due to the uncertainties in the lens modeling towards the edges of emission regions where the flux levels are lower. We primarily focus on the regions where the flux is at least 50\% of the maximum value for that emission type. In addition, we show the enclosed flux in circular apertures increasing as a function of the beam size out to 80\% of the total flux in the image. The noise in the image was sampled, similar to the sampling performed for the spectra (see Section \ref{sec:results}), corresponding to the size of the aperture. 

The dust continuum emission seems to be the most compact of the three tracers. We find that the distribution of the \water and the CO(6--5) are very similar, with CO(6--5) exhibiting a slightly larger extent. CO(6--5) and \water emission have previously been found to be correlated with regions of star formation within the galaxy \citep[][]{Greve14, Liu15, Yang13, Yang16, Omont13, Jarugula21, Pensabene22}, and therefore, their extents should be similar \citep[e.g.,][]{Quinatoa24}. However, the \water emission is expected to be produced in part by infrared pumping \citep[e.g.,][]{GA10, GA14, GA22} and is correlated with warm dust \citep[T$_{\rm dust} \approx 40-70$\,K ][]{Liu15}. Therefore, it is interesting that the continuum appears more compact than the \water emission. However, a significant fraction of the \water emission can originate from collisional excitation \citep[e.g.,][]{Yang13, Yang19, GA22, Liu17}. In addition, even assuming that the bulk of the dust and \water are physically distributed in the same region of the galaxy, the scale of the dust emission will appear smaller due to a temperature gradient. The larger spatial extents of the molecular gas compared to that of the dust continuum emission has previously been seen in studies of $z \approx 2-3$ SMGs \citep[e.g.,][]{Hai12} with the proposed explanation of temperature and optical depth gradients \citep[e.g.,][]{CalistroRivera18}, which could indeed also be the case for G09v1.97. Overall, the distribution of the CO(6--5) emission closely matches that of the \water while the continuum appears more compact. The \water emission map in this paper is the highest angular resolution map to date at such a high redshift. We further note that the exact distribution of the different emission types is, in part, governed by the regularization coefficient described in Section \ref{subsubsec:nonparametric_lensmodeling}, particularly with regard to the apparent smoothness of emission. 

\subsubsection{Lensing comparison to \citet{Yang19}} \label{subsubsec:lensing_comparison}

Here, we further investigate the discrepancies between the lens modeling results between our analysis and \citet{Yang19}. As mentioned previously, \citet{Yang19} found that the best-fit lens model of G09v1.97 required two parametric S\'ersic sources. The authors postulated that the two peaks in the CO(6--5) (and other molecular line emission) corresponded to two different galaxies that were undergoing a merger or interaction. In this work, a single source was sufficient to produce successful lens modeling results for both the parametric and non-parametric source modeling performed. Of particular interest are the non-parametric models, which do not require morphological assumptions, but these models also do not provide any indication of two sources in the source plane (see Fig. \ref{fig:nonparametric_lens_modeling_images}). 

\begin{figure*}[h]
    \centering
    \includegraphics[width = 1.0\linewidth]{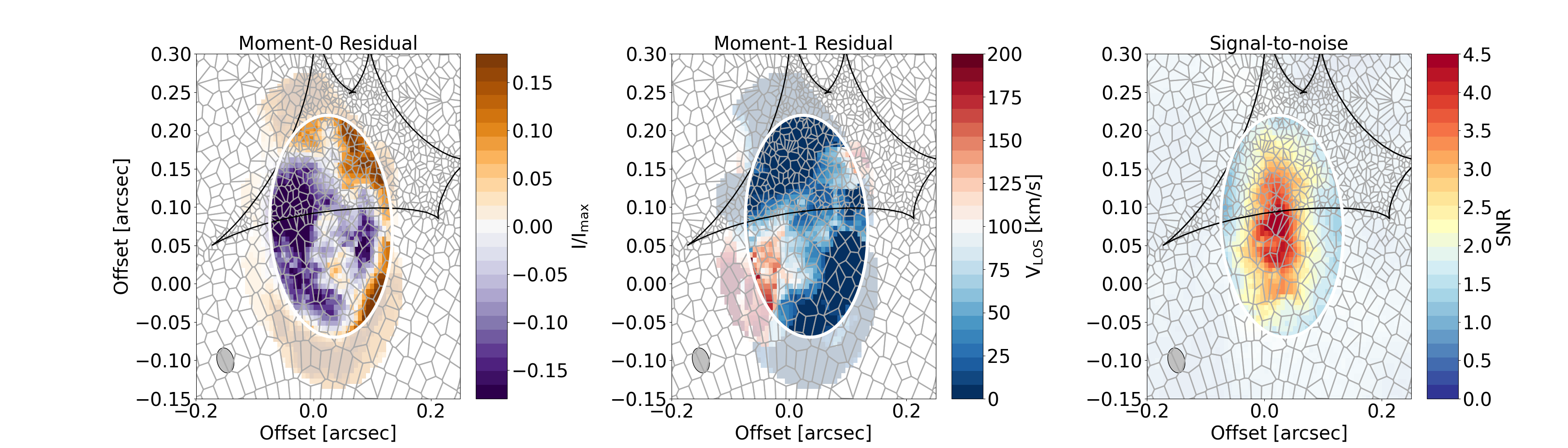}
    \caption{Residual moment-0 and moment-1 maps from {\sc $^{\rm 3D}$Barolo} fitting and the SNR map of the emission using the source plane error map (i.e., source plane intensity/moment-0 divided by the source plane intensity error map) as described in Section \ref{subsec:cube_and_spec_reconstructions}. Note that the first two panels are a zoomed-in version of the same images as in Fig. \ref{fig:barolo_modeling} rows 1 and 2 column 3 but with a different color scale. The grey grid shows the Voronoi pixels prior to interpolation to the square pixel grid, as described in Section \ref{subsec:cube_and_spec_reconstructions}. The black line shows the caustic line and the white ellipse shows the region used for {\sc $^{\rm 3D}$Barolo} fitting (i.e., the final ring). As in Fig. \ref{fig:barolo_modeling}, emission outside the fitting region are semi-transparent. The approximate beam, assuming a static magnification factor across the image, is shown in the bottom left of the image.}
    \label{fig:mom1_residual}
\end{figure*}

In order to better compare our results to those of \citet{Yang19} for the CO(6--5) emission, we performed non-parametric source reconstructions on the two peaks seen in the CO(6--5) spectrum. We used the two bins described in Section \ref{subsubsec:spec_line_shapes} covering velocities between $-300$ to 125\,km/s and 125 to 500\,km/s. These ranges correspond approximately to the blue and red components described in \citet{Yang19}\footnote{For simplicity here we assume that the two components `Rb' and `Rr' in \citet{Yang19} correspond to a single red component.}. We used the same procedure as described in Section \ref{subsubsec:nonparametric_lensmodeling} to perform the non-parametric lensing. 

We find that, when the spectrum is broken into these two bins, there does indeed appear to be a southern component corresponding to the blue peak of the CO(6--5) emission and a northern component corresponding to the red peak of the CO(6--5) emission. However, unlike in \citet{Yang19}, these two components are co-spatial around the caustic line, meaning that they do not fully appear as separate sources. We show the images, models, and residuals for all modeled emission in Fig. \ref{fig:nonparametric_red_blue_bins}. We suggest that the primary reason for the discrepancy between our modeling results and those from \citet{Yang19} is due to the velocity gradient seen in the de-magnified CO(6--5) source plane emission cube channel maps (Fig. \ref{fig:src_plane_channel_maps}). Without the additional per-channel information available in the source plane, the results of the two-bin model could easily be interpreted as two separate sources. This highlights the importance of creating source plane emission cubes during the lens modeling process.

\subsection{G09v1.97 - merger or disk?} \label{subsec:G09_merger_or_disk}

Studies in recent years have extended the redshift range of kinematic fitting significantly \citep[e.g.,][]{Smit18, Rizzo20, Fujimoto24, Rowland24}. This has led to further progress in understanding the kinematics of high-redshift massive galaxies. With regard to G09v1.97 and other DSFG/SMG galaxies, one of the primary motivations of kinematic studies is establishing whether DSFG/SMGs are largely merging or post-merger galaxies as they are often likened to ultra-luminous infrared galaxies \citep[ULIRGs][]{Sanders96} in the local universe due to their many shared properties \citep[e.g.,][]{Tacconi08, Engel10, Riechers11, Bothwell13}, which are primarily either undergoing or have recently undergone mergers \citep[e.g.,][]{Kim02, Veilleux02}. However, it has been shown that accurately distinguishing between mergers and disk galaxies at high redshift is very difficult, primarily for reasons relating to angular resolution and sensitivity challenges \citep[][]{Rizzo22, Peng23, Cathey24}. Although observations of gravitationally lensed galaxies typically have higher angular resolution and sensitivity due to the magnification effects of the lensing phenomenon, they also provide challenges in accurately determining the source plane properties of the lensed galaxy \citep[e.g.,][]{Dye14, Yang19, Rybak20}. For example, \citet{Rizzo22} simulated galaxies at high-, mid-, and low-angular resolutions and demonstrated the difficulties in accurately distinguishing between disks versus mergers. Below, we discuss the characteristics of G09v1.97 and attempt to classify the source as either a merger or a disk. Further, we examine some of the limitations caused by the gravitational lens modeling.

\begin{SCfigure*}[0.8][htbp] 
    \includegraphics[width=0.7\textwidth]{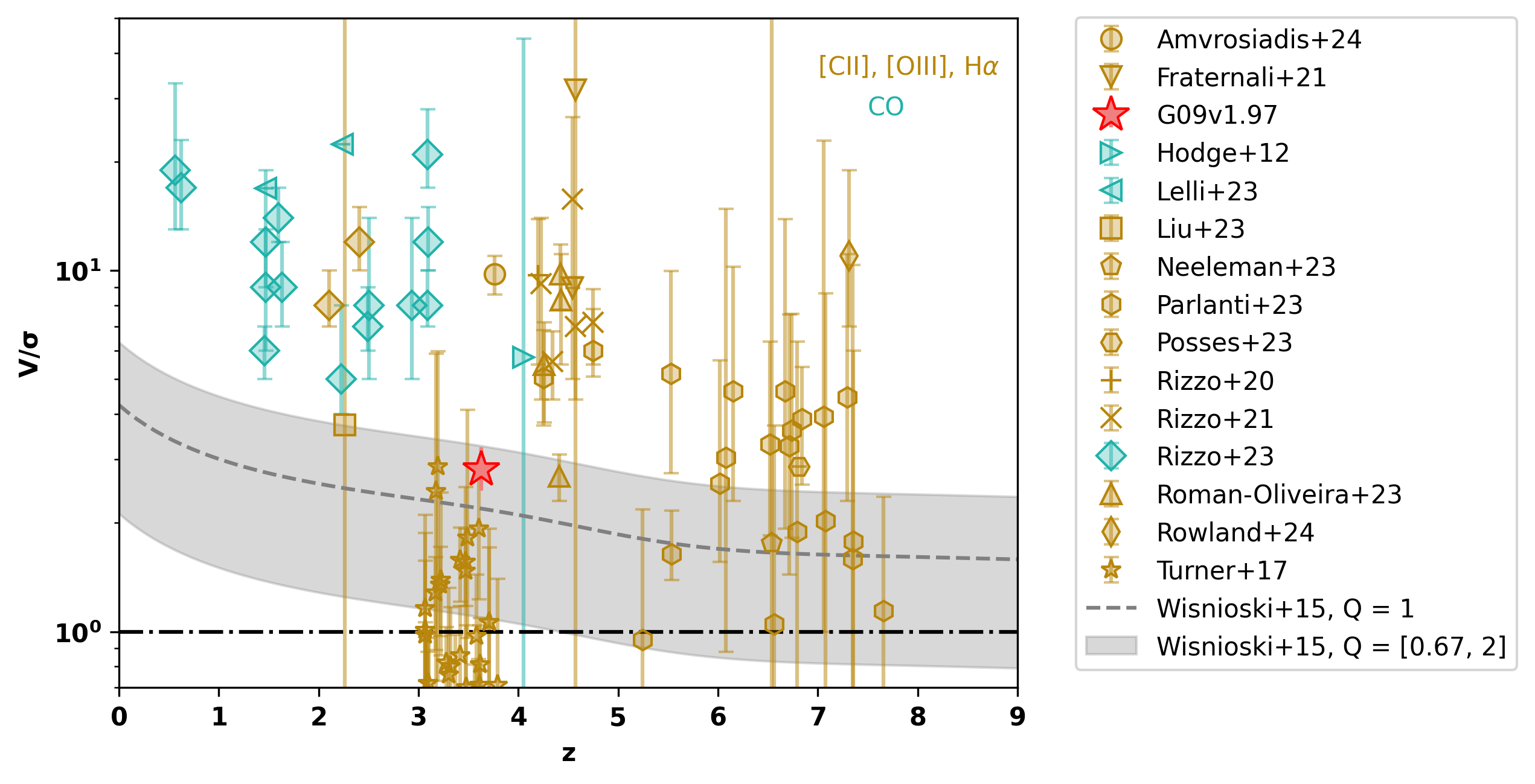}
    \caption{Evolution of $\rm V/\sigma$ with redshift for G09v1.97 (shown by the pink star) and literature sources. The literature sample is taken from \citet{Amvrosiadis24}, \citet{Fernanda23}, \citet{Hodge12}, \citet{Fraternali21}, \citet{Lelli23}, \citet{Liu23}, \citet{Neeleman23}, \citet{Parlanti23}, \citet{Posses23}, \citet{Rizzo20}, \citet{Rizzo21}, \citet{Rizzo23}, \citet{Rowland24}, \citet{Turner17}. For the literature sample, we show atomic tracers (in this case [C\,{\sc ii}], [O\,{\sc iii}], H$\alpha$ emission) in yellow and molecular gas tracers (in this case, different CO transitions) in blue. We show the semi-analytic model from \citet{Wisnioski15} in gray with Q values ranging between 0.67 to 2 and the fiducial model of Q = 1 in the dashed gray line. }
    \label{fig:v_over_sigma}
\end{SCfigure*}

We find $\rm V_{max}/\bar{\sigma} = 2.8 \pm 0.4$ for G09v1.97. This suggests that the source is dominated by ordered motion rather than turbulence. We compare the $\rm V_{max}/\sigma$ value found for G09v1.97 to other galaxies across a wide redshift range in Fig. \ref{fig:v_over_sigma}. We find that G09v1.97 has a lower $\rm V/\sigma$ value than has been found for other sources at comparable redshifts \citep[e.g.,][]{Hodge12, Rizzo23, Amvrosiadis24} and instead lies closer to the turbulent disks found at comparable and higher redshifts \citep[e.g.,][]{Turner17, Parlanti23}. However, there are a number of important caveats to note when using this comparison sample. The first is the choice of a gas tracer. CO(6--5) emission traces warmer and denser gas than \cii \citep{Cailli13}. The samples that use CO emission are limited to lower-$J$ transitions and, therefore, trace colder and more diffuse gas than the CO(6--5) emission used in this work. Secondly, the $\rm V/\sigma$ values used in different studies are calculated differently. Some studies use $\rm V_{max}/\bar{\sigma}$, as is used in this paper, while others use rotational velocities extracted at, for example, $\rm r_e$. Thirdly, there is a wide variety of data quality used for calculating the $\rm V/\sigma$ values plotted in the figure. Finally, $\rm H_{\alpha}$ is known to exhibit systematic differences in $\rm V/\sigma$ compared to those found for \cii and CO emission in the same galaxy \citep{Rizzo24}. However, we argue this sample still provides a relevant and meaningful basis for comparison. 

\citet{Rizzo22} present a method called {\sc PVsplit} that shows promise as a tool able to separate between disks and mergers. This method relies on three parameters, $P_{\rm major}$, $P_{\mathrm V}$, and $P_{\mathrm R}$, obtained from the PV major axis diagram. The {\sc PVsplit} method has been further investigated by \citet{Fernanda23} with a sample of mergers and disk galaxies from ALMA mock observations of SERRA simulated \citep{pallottini2022} galaxies together with a random selection of 9 local WHISP \citep{vanderhulst2001} survey disk galaxies. From this \citet{Fernanda23} suggested a plane dividing disks from mergers following the equation: $-0.63 P_{\rm major} - 0.27 P_{\mathrm V} + 2.78 P_{\mathrm R} -2.72 = 0$. 
{\sc PVsplit} is a promising method, but has only been tested on a very limited sample without galaxies with known outflows. Nonetheless, we calculate the {\sc PVsplit} parameters for G09v1.97 and find: $P_{\rm major} = -1.56$, $P_{\mathrm V} = -0.33$, and $P_{\mathrm R} = 0.50$. Using the dividing line suggested by \citet{Fernanda23}, this places G09v1.97 on the side of mergers, albeit close to the dividing plane. The distance between G09v1.97 and the dividing plane is very small, only 0.08. This suggests that G09v1.97 may be a merger, but it may also be one of the sources subject to scattering around the dividing plane or this may be the region of the $P_{\rm major} - P_{\mathrm V} - P_{\mathrm R}$ parameter space where disks with outflows, or late stage minor mergers, will populate when they have been placed into the {\sc PVsplit} scheme as the tool is tested on additional sources.

Using the classification scheme from \citet{Kohandel24} for \cii emission, G09v1.97 would be classified as a warm rotating disk. However, this classification scheme was developed for \cii and H$\alpha$ emission, and therefore, extrapolating to CO(6--5) emission could skew the interpretation of the $\rm V_{max}/\sigma$ value. Both \cii and CO(6--5) emission are thought to trace molecular gas, with \cii likely tracing more extended emission \citep[e.g.,][]{Cailli13}. 

It could be the case that the CO(6--5) emission has a higher dispersion value ($\sigma$) value since it is tracer warmer and therefore more turbulent gas, and thereby a lower $\rm V_{max}/\sigma$ value (under the assumption that we are probing $\rm V_{max}$ with the current data). On the other hand, we are likely not tracing the $\rm V_{max}$ for the entire galaxy with such a warm gas tracer, and therefore even if the dispersion value turns out to be higher in CO(6--5) than other, colder gas tracers, the overall   $\rm V_{max}/\sigma$ ratio may remain relatively unchanged. Additionally, observations of colder gas tracers that probe further into the disk may exhibit a smaller ratio due to larger $\rm V_{max}$ values and a lower average dispersion. To verify which of these scenarios is occurring, a kinematic analysis would need to be performed on, for example, very high angular resolution observations of \cii emission in G09v1.97. 

We find that G09v1.97 also follows the classification scheme for a rotating disk used by \citet{Rizzo22}. In this scheme, to be classified as a rotating disk, the galaxy must have $V/\sigma > 1.8$ to maintain sufficient rotational support over the turbulence \citep[e.g.,][]{Wisnioski15, Wisnioski19}. Additionally, the galaxy must exhibit a smooth velocity gradient in the moment-1 map. G09v1.97 fulfills both of these criteria. Similar criteria are commonly employed across a wide redshift range \citep[e.g.,][]{FA09, Wisnioski15, Smit18, Bakx20, Rizzo22}

We now turn to investigating the residuals found in the moment maps from our best-fit \barolo modeling. Fig. \ref{fig:mom1_residual} shows the zoomed-in moment-1 residual with a different colorscale and with the Voronoi grid along with the SNR map from the lens modeling as described in Section \ref{subsec:cube_and_spec_reconstructions}. We first note that the largest of these residuals in both the moment-0 and moment-1 maps are located outside of the region used for fitting in the \barolo models, meaning that the $\rm V_{max}/\bar{\sigma}$ ratio found for G09v1.97 is not affected by these residuals. Here we investigate different possibilities, primarily relating to the residuals inside this fitting ellipse where the significance of the emission is higher when considering the source plane error maps, to explain these residuals. We investigated the possibility that the residuals were caused by the interpolation from a Voronoi mesh onto a square pixel grid. We returned to the creation of the CO(6--5) source plane emission cube, described in detail in Section \ref{subsec:cube_and_spec_reconstructions}, and verify that the residuals are not caused by this interpolation through two methods. In the first, we used twice the number of pixels in the Voronoi mesh prior to interpolation to a square grid, and in the second, we used a rectangular mesh instead of a Voronoi mesh. In both cases, we found very similar residuals, suggesting that the interpolation from the Voronoi mesh to the square pixel grid was not the cause of the residuals. There does not appear to be a correlation between regions of high errors in the source plane error map (see Fig. \ref{fig:src_plane_errors}) and the strongest velocity residuals. We therefore conclude that there is no indication that the residuals result from a poorly fitting lens mass model. For the remainder of this paper, we consider only the residuals inside the region used for \barolo fitting, shown by the white ellipse in Fig. \ref{fig:mom1_residual}, due to the low SNR of regions outside this ellipse. 

Under the assumption that these features are not caused by the lens modeling, they could instead be indicative of non-circular motions in G09v1.97. First, we consider the significance of the moment-1 residuals. We calculated the significance of the velocity residuals by considering the moment-0 emission in regions where the moment-1 residuals exceed $125$\,km/s (within the fitting region used by {\sc $^{\rm 3D}$Barolo}) by taking the ratio of the moment-0 flux to total error as a proxy for their significance. We note that the independent resolution element, i.e. the assumed source plane beam area, is propagated into the measurement by calculating the total area of pixels with velocity $>125$\,km/s in the fitting region. These residuals are statistically significant ($\sim 6\sigma$), and we therefore turn to investigating their nature and discussing possible physical interpretations. 

One possibility to explain this emission is a bi-conical outflow approximately perpendicular to the major axis of the galaxy. In this scenario, the relative lack of a clear blue component could be explained by either interactions from this outflowing emission with the disk or simply an obscuration effect from the disk. We find no clear evidence of outflows in the CO(6--5) emission spectrum; however, the complex line profiles do not allow for a clear exclusion of this feature. Observations of specific outflow tracers (e.g., $\rm OH^+$) could help to establish if this explanation for the residual is correct. 

Another explanation could be that of an additional galaxy located behind G09v1.97 at spatial offsets of $>100$\,km/s \footnote{Note that this would be $>100$\,km/s away from the best-fit systemic velocity of $109$\,km/s, meaning $>220$\,km/s away from the location of the 0 velocity in the CO(6--5) spectrum.} However, we find no clear signs of merger activity that would be expected from such a nearby companion galaxy and would expect a lower $\rm V_{max}/\bar{\sigma}$ ratio in the case of an on-going merger or interaction. Previous studies have shown some evidence of flattening rotation curves with $\rm V/\sigma > 2$ while also exhibiting evidence of interactions/mergers \citep[e.g.,][]{Hai18}. In the case of G09v1.97, should a background galaxy be present, it would be expected that the residual intensity map would show elevated values at the same spatial extents. We do find a residual in the moment-0 residual map from \barolo (Fig. \ref{fig:mom1_residual}) that is slightly offset from the moment-1 residual, which could be interpreted as an additional source of emission. A similar but possibly more likely scenario is that of a tidal tail as a remnant of a merger or interaction. Given that G09v1.97 appears to be more consistent with a warmer rotating disk than a fully settled, cold, rotating disk, this could explain the relatively higher dispersion occurring in G09v1.97. This is further discussed in Section \ref{subsec:g09_as_warm_disk}. 

One final possibility for this residual is a bar, as has been suggested for similar residuals \citep[e.g.,][]{Amvrosiadis24}. This could explain the features in the major axis pv-diagram (Fig. \ref{fig:barolo_pvs}) and the bump in the $V_{\rm rot}$ and $\sigma$ at the innermost radii (Fig. \ref{fig:v_over_sigma}). However, the lack of a clear blue component in the moment-1 residual and the location of the moment-0 residual raise doubts about this possibility. 

Although these velocity residuals are statistically significant and indicative of non-circular motion in G09v1.97, their physical origin is not yet understood. Overall, we find that the results of the performed kinematic modeling are most consistent with G09v1.97 being a warm rotating disk, and suggest that the residual is likely some additional component to the galaxy, such as a tidal tail.

\subsection{Physical properties of the warm rotating disk G09v1.97} \label{subsec:g09_as_warm_disk}

G09v1.97 has been previously shown to be one of the most luminous dusty star-forming galaxies in the high-redshift universe with a total infrared luminosity of $L_{\mathrm{IR}} = (1.4 \pm 0.7) \times 10^{13} L_{\odot}$ \citep{Yang19}. Assuming a Chabrier initial mass function \citep[IMF;][]{Chabrier03} and the relation between SFR and $L_{\rm IR}$ as SFR $\sim 10^{-10} L_{\rm IR} [L_{\odot}]$, this implies a SFR of $1400 \pm 700$\,M$_{\odot}$\,yr$^{-1}$. Fitting the CO(6--5) emission in the source plane, we find a spatial extent of $0''.21 \pm 0''.03 \times 0''.12 \pm 0''.02$, corresponding to $1.5 \rm\, kpc \times 0.88\,kpc$ at $z = 3.63$. This implies a surface star formation rate density of $\Sigma_{\rm SFR} \approx 340$\,M$_{\odot}$\,yr$^{-1}$\,kpc${^{-2}}$. 

Following \citet{Wang13}, \citet{Venemans16}, and \citet{Yang17}, we calculate the dynamical mass of G09v1.97 assuming that the system is a rotating disk using the following,
\begin{equation}
    M_{\mathrm{dyn}} = (2.35 \times 10^5) (\frac{V_{\mathrm{c}}}{\mathrm{km\,s^{-1}}}) (\frac{\mathrm{r}}{\mathrm{kpc}}) [\rm M_{\odot}],
\end{equation}
where $\rm V_c$ is the rotational velocity at radius r. We consider the radius and rotational velocity in the outermost ring from {\sc $^{\rm 3D}$Barolo}, corresponding to $\rm V_c = 386$\,km/s and r = 1.07\,kpc, resulting in $M_{\mathrm{dyn}}= (3.7 \pm 0.1) \times 10^{10}$\,M$_{\odot}$ \footnote{We note that this value may be over-estimated by a factor $\sim1.5$, as described in \citet{Engel10}.}. This is in good agreement with the dynamical mass found in \citet{Yang17} for G09v1.97. 

We estimate the gas mass fraction by first calculating the gas mass from the CO(6--5) emission. We use the conversion ratio between the CO(6--5) and CO(1--0) emission based on single component large velocity gradient (LVG) modeling of the observed CO spectral line energy distribution (SLED) performed in \citet{Yang17}, wherein the ratio is very similar to LVG modeling of $z\approx2$ galaxies using two-component CO SLEDs \citep{Boogaard20}. This conversion assumes that the CO(6--5) and CO(1--0) emission come from the same spatial region, which is unlikely to be the case; however, it is the best approximation with the given data. We further note that, although conversions from higher-$J$ transitions to CO(1--0) are commonly performed in high-redshift studies, where obtaining CO(1--0) observations can be difficult, this conversion is inherently uncertain and therefore, the gas mass obtained here should be seen as an estimate. Assuming $\alpha_{\rm CO} = 0.8$ \citep[a typical value used for local ULIRGS][]{Downes98} and a helium contribution of 36\% \citep{Yang17}, we obtain a gas mass in the region of CO(6--5) emission of $M_{\mathrm{gas}} = (1.8 \pm 0.74) \times 10^{10}$M$_{\odot}$. We note that the value chosen for $\alpha_{\rm CO}$ can strongly impact the gas mass estimate \citep[see][for example]{Dye22} as this value can range from $\alpha_{\rm CO} = 0.8 - 4.0$, and we therefore incorporated a larger error on the gas mass (approximately a factor 4) to represent this range. This implies a depletion time (defined as $t_{\rm dep} = M_{\rm gas}/\mathrm{SFR}$) for G09v1.97 of $13 \pm 8$\,Myr, lower than the value found for G09v1.97 by \citet{Yang17} but within the error range. Similarly, this value is lower, but within errors, than the depletion time that has been found for other strongly lensed SMGs at similar redshifts \citep[e.g.,][]{Yang17}. This implies a gas mass fraction ($M_{\rm gas}$/$M_{\rm dyn}$) of $0.5 \pm 0.02$, lower than reported in \citet{Yang17} but still within an expected range for strongly lensed SMGs \citep[e.g.,][]{Yang17}. 

One possibility to explain such a high SFR is that the infrared luminosity could be contaminated by the presence of an AGN. However, \citet{Yang19} and \citet{Yang20} argue that there is no clear evidence for an obscured AGN at the center of G09v1.97; this is supported by G09v1.97’s similar line-to-IR luminosity ratio for both CO(6—5) and \water as well as the \textit{q}-parameter \citep{Condon92} as discussed in \citet{Yang16} and \citet{Yang20}. The surface brightness distribution of the CO(6--5) emission, as shown by, for example, the kinematic modeling, does not reveal an additional compact, bright, central component in the residual intensity map (Fig. \ref{fig:barolo_modeling}). Such a feature, if present, could have been indicative of AGN activity. In addition, we do not observe any excess emission appearing as a bright central source in either the \water or continuum maps. Hence, we find no evidence of an AGN in the observed frequencies. Should there be an AGN affecting the $L_{\rm IR}$ and thereby SFR, an alternative method to determine the SFR is through the use of the CO(6--5) emission given that mid-$J$ CO emission is thought to trace star-forming regions of the galaxy \citep[e.g.,][]{Greve14, Lu15}. Through this method, and assuming that the CO(6--5) and CO(7--6) emission are very similar, a reasonable estimate based on CO SLED modeling from \citet{Yang17}, we obtain $\rm SFR_{CO} \approx (2515 \pm 260)$\,M$_{\odot}$\,yr$^{-1}$ using the conversion between CO(7--6) and SFR given in \citet{Lu15}, implying a two times smaller depletion time. Thus, we return to the question of how such a high SFR could be triggered. 

Contrary to what was found by \citet{Yang19}, we find no clear evidence of merger activity in G09v1.97 and instead have demonstrated through kinematic modeling that G09v1.97 appears to be a warm rotating disk. However, we also find evidence in the residuals from {\sc $^{\rm 3D}$Barolo} modeling for an additional component such as a tidal tail or additional galaxy. In the former, G09v1.97, observed at its present epoch, is a post-merger coalescence galaxy that has relatively recently settled into a rotating disk. In the latter, given the close velocity separation between the additional component and G09v1.97, the additional component would be interpreted as an interaction. In both scenarios, the SFR of the galaxy is expected to increase due to the compression of the gas from ongoing or previous interactions. This is similar to results found by \citet{Kade24} and is consistent with the commonly accepted theoretical paradigm of massive galaxy evolution \citep[e.g.,][]{Hopkins08}.

\section{Conclusions} \label{sec:conclusions}
We report high-resolution ALMA observations of CO(6--5), \water($2_{11} - 2_{02}$), and the underlying dust continuum emission in the $z=3.63$ gravitationally lensed DSFG G09v1.97. We performed parametric and non-parametric source reconstructions of the different detected emission using the sophisticated, publicly available lens modeling code {\sc PyAutoLens} \citep{Nightingale21}. We performed source-plane kinematic modeling of the CO(6--5) emission using {\sc $^{\rm 3D}$Barolo}. Our conclusions are provided below.  

\begin{enumerate}
  \item We further improved the lensing model of G09v1.97 using the high-resolution dust continuum emission and applied this lens model to the line emission. The magnification factors from our lens model are in good agreement with those from \citet{Yang19}. 
  
  \item We used non-parametric pixelated modeling of the CO(6--5) emission to create a source-plane emission line cube. We find no evidence that the CO(6--5) emission requires two sources to successfully model G09v1.97; instead, we find a single source exhibiting a smooth velocity gradient across the CO(6--5) source-plane emission line cube. We suggest that the discrepancy between the results provided in \citet{Yang19} and those found in this work stems primarily from the velocity gradient found in the source plane. Without creating a de-magnified source plane emission cube, it is challenging to properly interpret the emission. 
  
  \item We compare the extent of the dust continuum, CO(6--5), and \water emission in the source plane. We find that the dust continuum emission is the most compact of the three emission types. Although \water emission is typically correlated with dust emission due to infrared pumping, it appears more extended than the dust continuum in this work. 

  \item We perform kinematic modeling on the source-plane CO(6--5) emission line cube using {\sc $^{\rm 3D}$Barolo} and find that the kinematics are well explained as a rotating disk with a maximum rotational velocity of $\rm V_{max} = 388^{+55}_{-36}$\,km/s and an average dispersion of $\bar{\sigma} = 138^{+7}_{-6}$\,km/s, implying $\rm V_{max} / \sigma = 2.8 \pm 0.4$. Hence, ordered rotation dominates over turbulence in G09v1.97.

  \item We compare the $\rm V/\sigma$ value found for G09v1.97 with a representative comparison sample between $1 < z < 8$ and find that G09v1.97 lies within predicted values from semi-analytic models but seems to exhibit relatively more turbulence than other disk galaxies at similar redshifts. According to a classification scheme developed from \citet{Kohandel24}, G09v1.97 should be classified as a warm disk. Alternatively, the in-development tool {\sc PVsplit} provides an indication that G09v1.97 is undergoing a merger event.

  \item We find evidence for non-circular motions in the velocity and dispersion residuals from the kinematic fitting. Similarly, we find evidence of a residual in the moment-0 map. These residuals cannot be explained by the lensing analysis and source plane cube reconstruction, and are statistically significant ($\sim 6\sigma$). We therefore suggest that these residuals represent an additional component, such as a biconical outflow, tidal tail, or additional background galaxy. 
  
  \item We calculate the dynamical mass of G09v1.97 using the rotational velocity in the final ring from {\sc $^{\rm 3D}$Barolo} as a proxy for the turnover radius. We find that, within 1\,kpc, the dynamical mass of G09v1.97 is $M_{\mathrm{dyn}}= (3.7 \pm 0.1) \times 10^{10}$\,M$_{\odot}$. Further, we calculate the gas mass and SFR from both infrared luminosity and CO emission; we find $M_{\mathrm{gas}} = (1.8 \pm 0.74) \times 10^{10}\, M_{\odot}$, $\rm SFR_{L_{IR}} = (1400 \pm 700)$\,M$_{\odot}$\,yr${-1}$, and $\rm SFR_{CO} \approx (2515 \pm 260)$\,M$_{\odot}$\,yr$^{-1}$. Together, these values imply a very low depletion time ($\rm 13\pm8 \, Myr$) in comparison to other galaxies at similar redshifts. 
  
\end{enumerate}

We suggest that G09v1.97 has either recently undergone or is currently undergoing an interaction or merger event, causing the residuals seen in the moment-1 emission from the {\sc $^{\rm 3D}$Barolo} modeling. Combined with the high SFR, and low depletion time, we suggest that the interaction triggered extreme star formation. In this scenario, G09v1.97 has recently settled into a turbulent, rotating disk that is rapidly consuming its gas and showing non-disk morphological features due to likely recent or on-going interactions or merger events.

Observations of a colder gas tracer, such as \cii, using similar or higher angular resolution observations could help probe further into the disk of G09v1.97 and provide additional clues to the on-going dynamical processes in the galaxy. Further, observations of specific outflow tracers, for example, OH$^{+}$, could assist in a more robust identification of the additional feature seen in the {\sc $^{\rm 3D}$Barolo} modeling should it be a bi-conical outflow or possible tidal tail. G09v1.97 remains a remarkable target for future observational campaigns aimed at understanding the ISM and dynamics of galaxies at this epoch. 

\begin{acknowledgements}

The authors thank the anonymous referee for their useful comments which improved the quality of the manuscript. The authors also thank Ian Smail and Hai Fu for useful comments that improved the manuscript. K.K. acknowledges support from the Nordic ALMA Regional Centre (ARC) node based at Onsala Space Observatory. The Nordic ARC node is funded through Swedish Research Council grant No. 2019-00208. M.Y. acknowledges support via research grants from the Knut and Alice Wallenberg Foundation. K.Kn. acknowledges support from the Swedish Research Council and the Knut and Alice Wallenberg Foundation. S.K. and C.Y. gratefully acknowledge funding from the European Research Council (ERC) under the European Union’s Horizon 2020 research and innovation programme (grant agreement No. 789410). SD acknowledges support from the Science and Technology Facilities Council (grant: ST/X000982/1). AC acknowledges support from NASA Astrophysics Data Program. E.I. gratefully acknowledges financial support from ANID - MILENIO - NCN2024\_112 and ANID FONDECYT Regular 1221846. M.J.M.~acknowledges the support of the National Science Centre, Poland through the SONATA BIS grant 2018/30/E/ST9/00208 and the Polish National Agency for Academic Exchange Bekker grant BPN/BEK/2022/1/00110. The National Radio Astronomy Observatory is a facility of the National Science Foundation operated under cooperative agreement by Associated Universities, Inc.

This paper makes use of the following ALMA data: ADS/JAO.ALMA\#2018.1.01710.S and ADS/JAO.ALMA\#2015.1.01320.S. ALMA is a partnership of ESO (representing its member states), NSF (USA) and NINS (Japan), together with NRC (Canada), NSTC and ASIAA (Taiwan), and KASI (Republic of Korea), in cooperation with the Republic of Chile. The Joint ALMA Observatory is operated by ESO, AUI/NRAO and NAOJ.

\end{acknowledgements}

\bibliographystyle{aa}
\bibliography{biblio}

\begin{appendix}

\onecolumn  

\section{Spectra}

\begin{SCfigure}[1.4][ht!] 
    \centering
    \includegraphics[width = 0.4\textwidth]{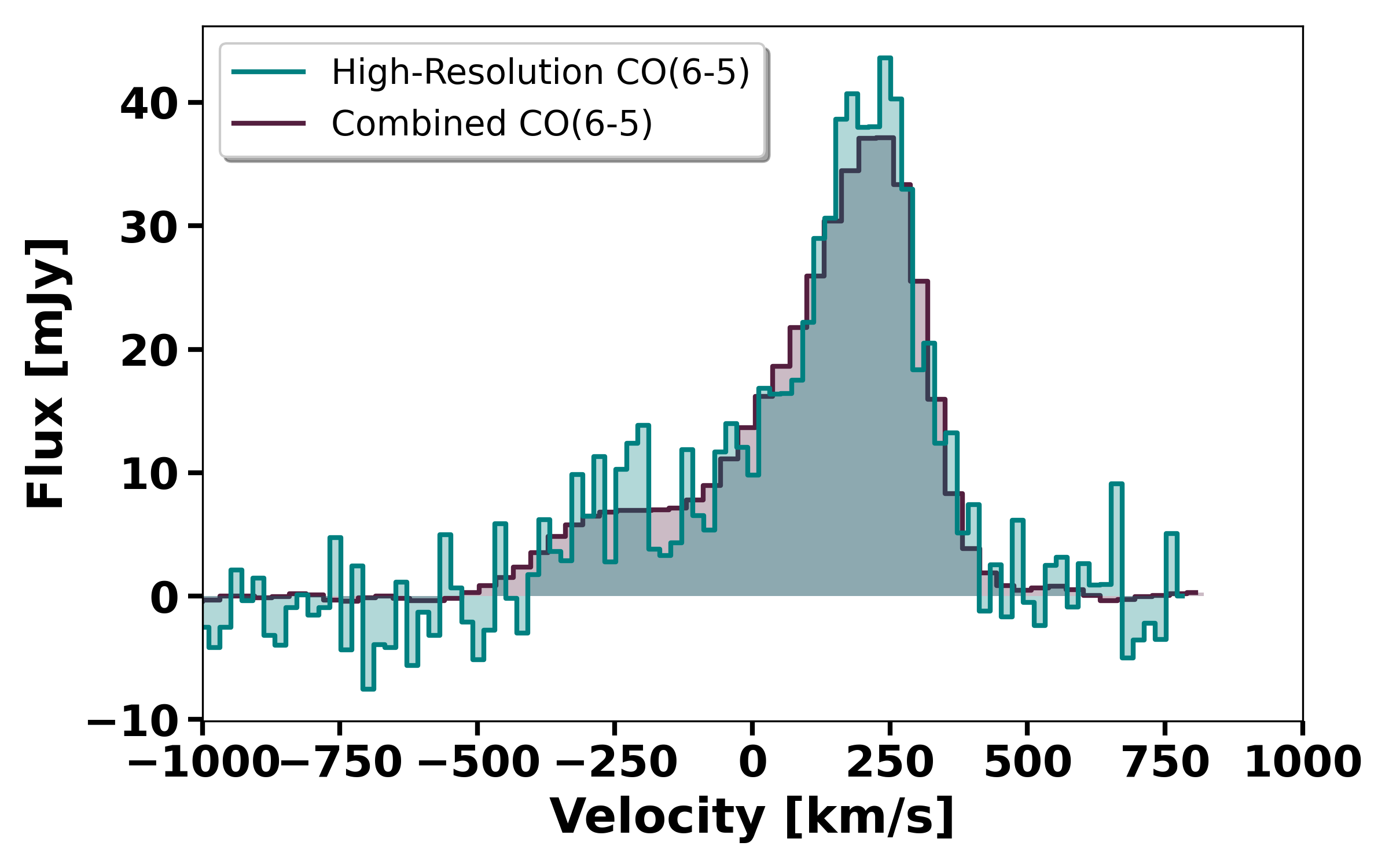}
    \caption{Comparison of the high-resolution CO(6--5) spectrum versus the spectrum from the combined dataset. Note that the spectral resolution between the two datasets was not the same. No significant difference is seen in the flux, suggesting that no emission has been filtered out in the high-resolution image.}
    \label{fig:co_combined_highres_spec_comparison}
\end{SCfigure}

\section{Lens modeling}

\begin{figure}[!ht]
    \centering
    \includegraphics[width = 0.19\textwidth]{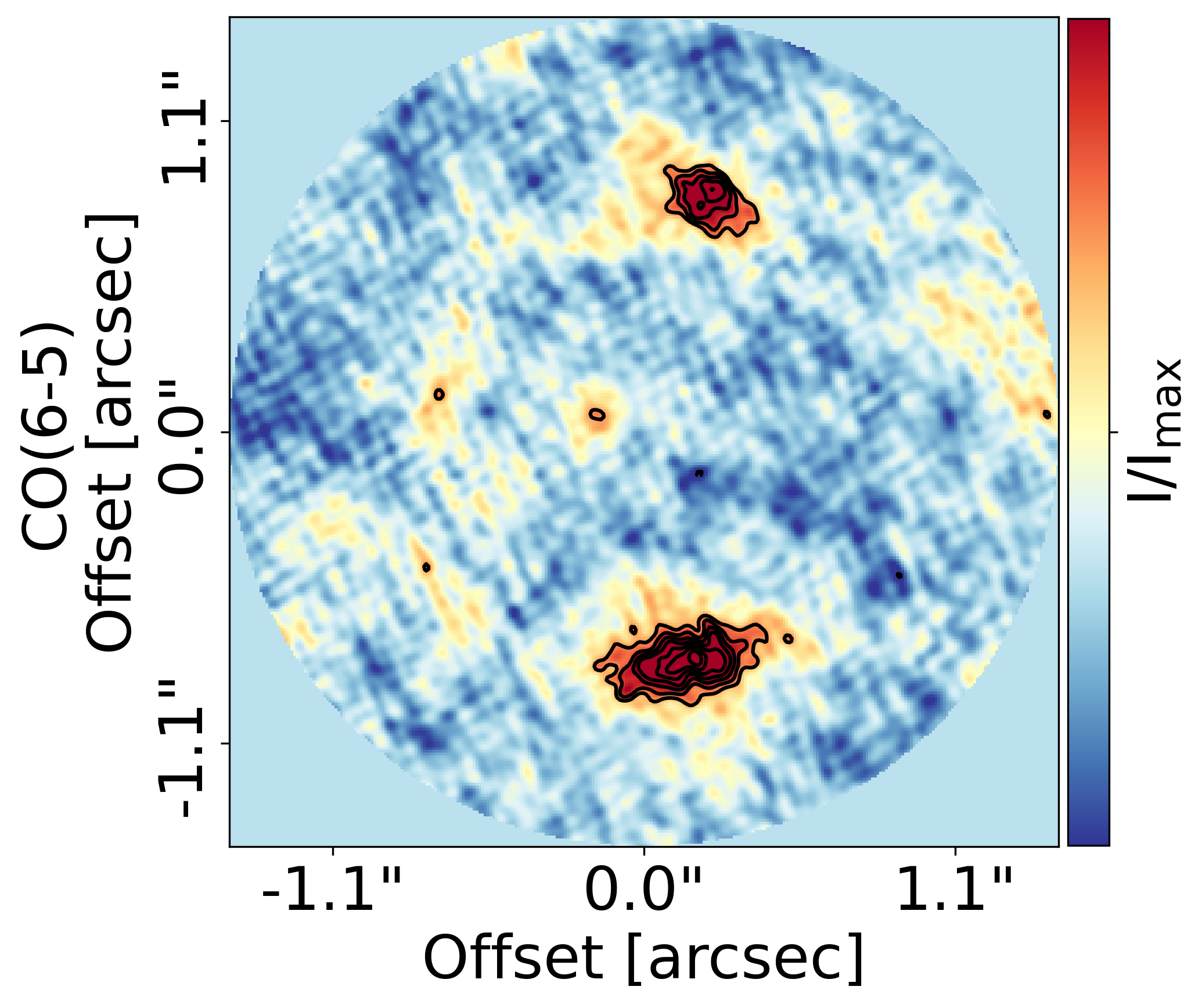}
    \includegraphics[width = 0.19\textwidth]{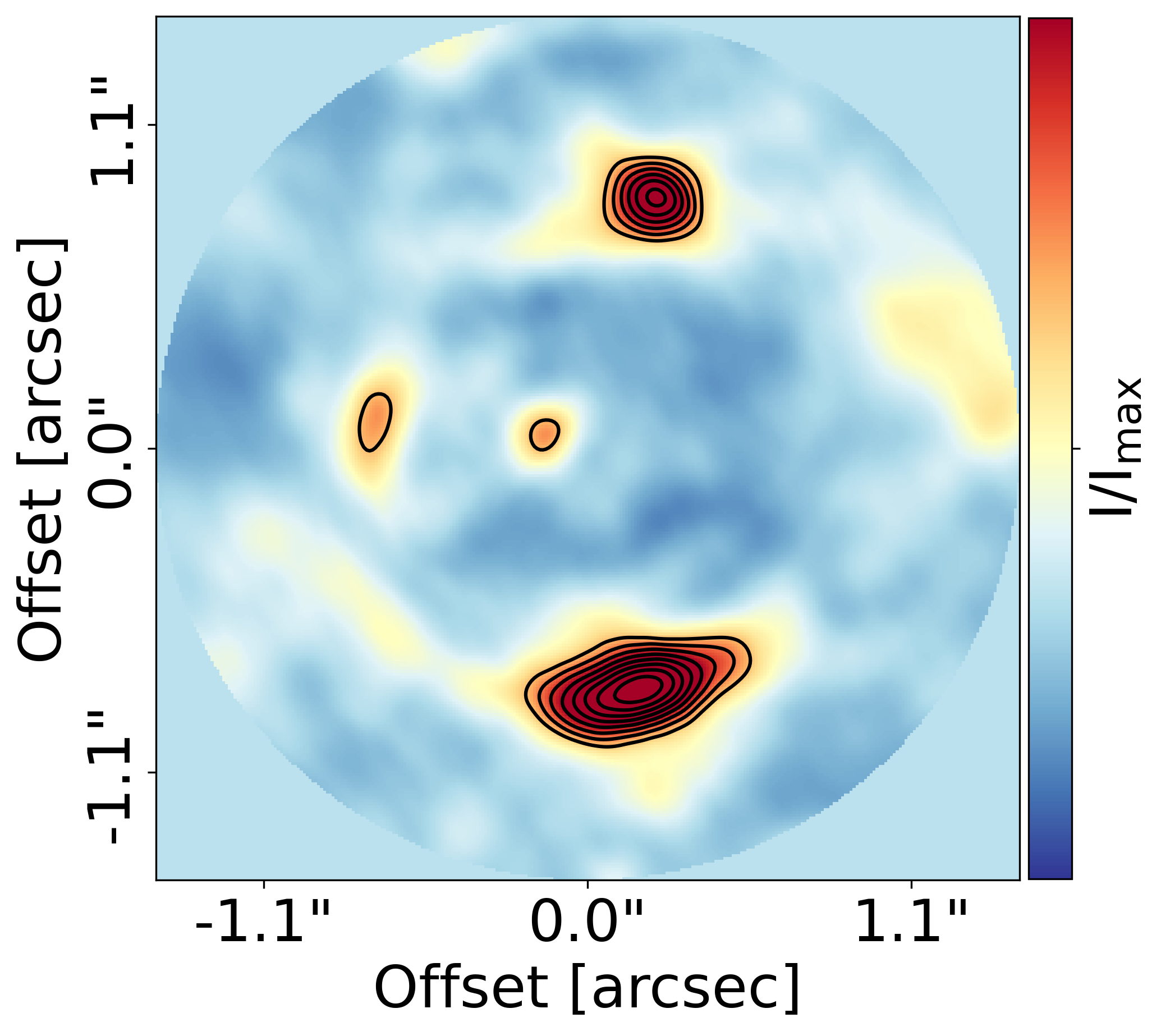}
    \includegraphics[width = 0.19\textwidth]{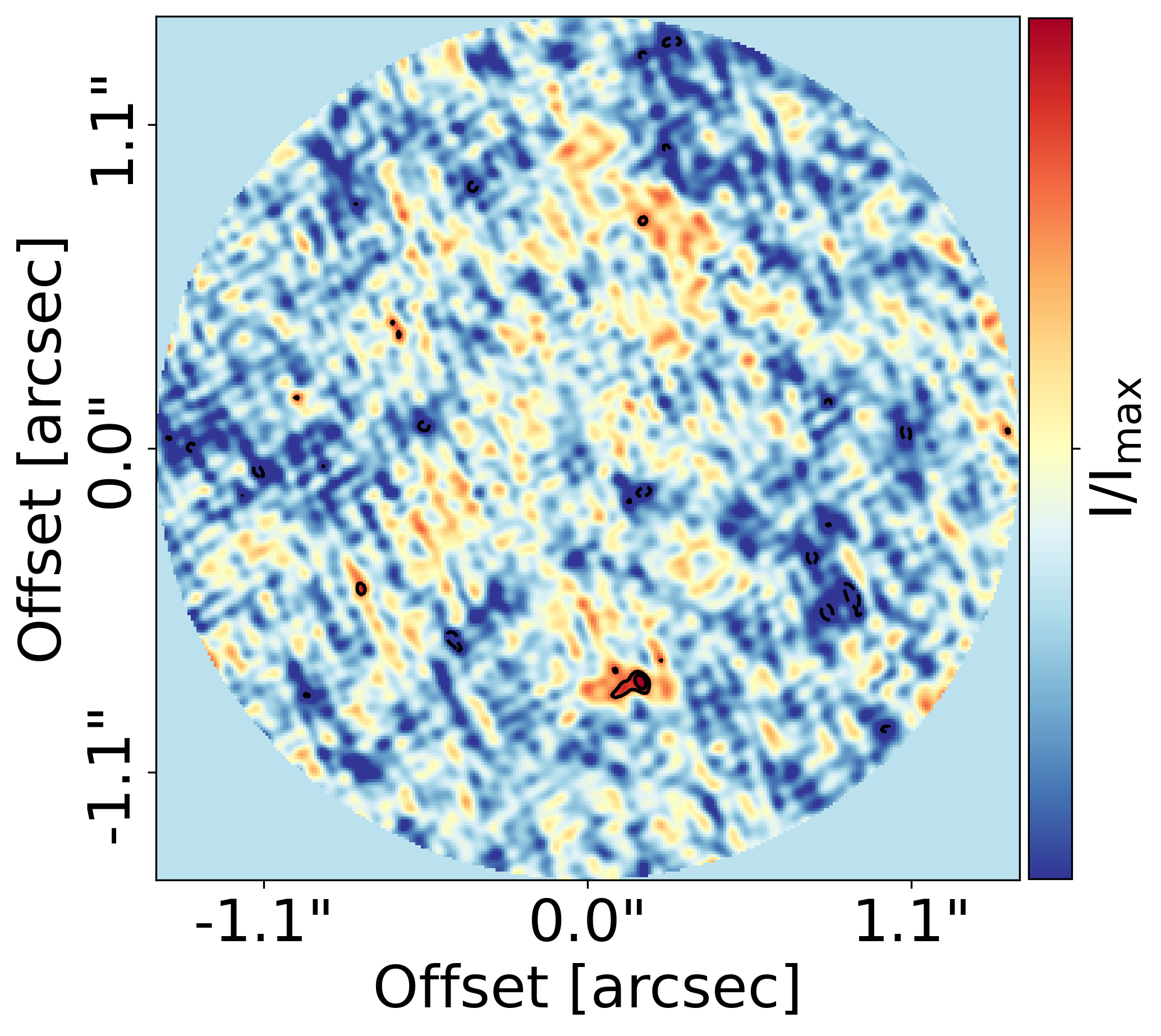}
    \includegraphics[width = 0.19\textwidth]{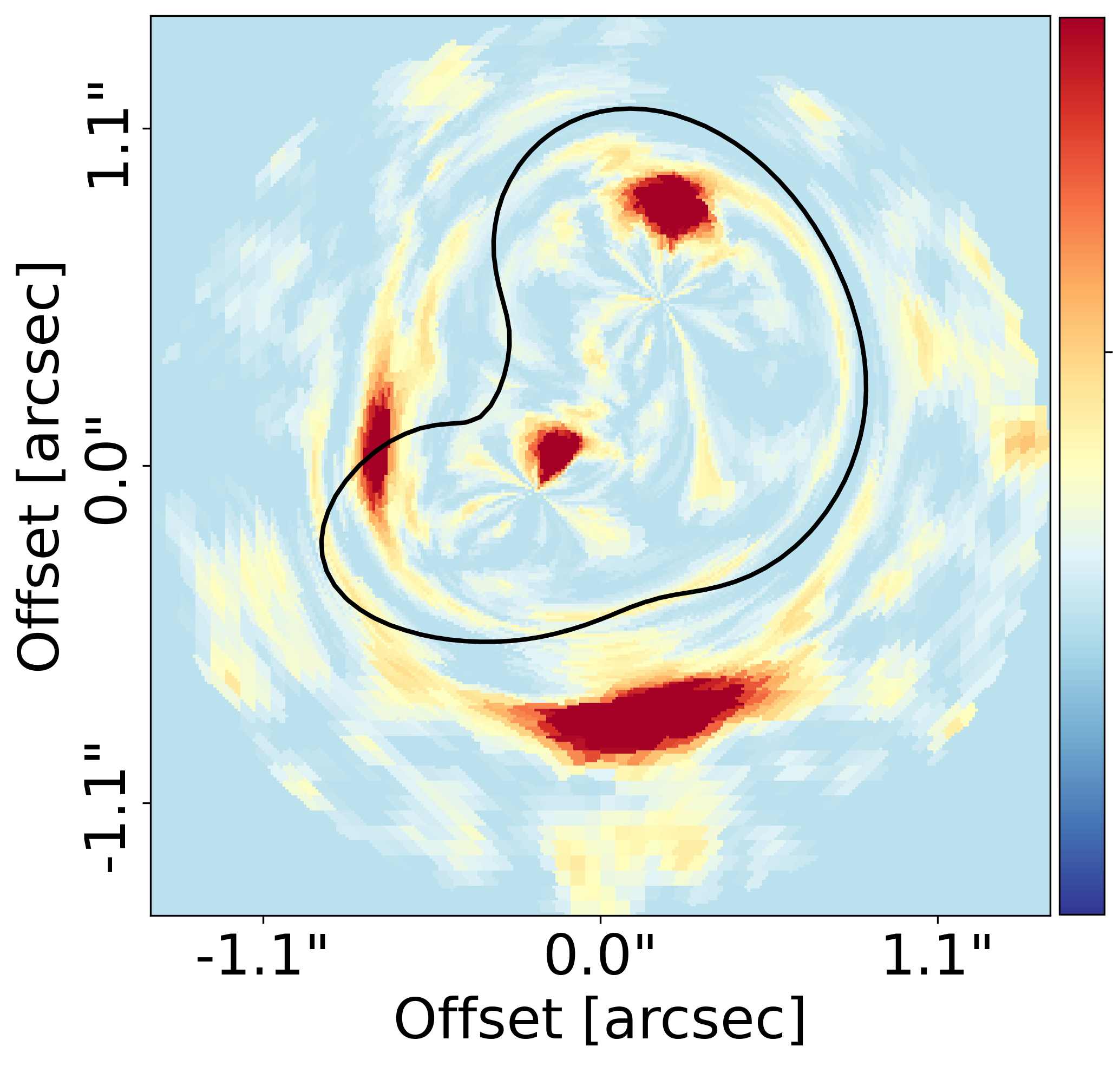}
    \includegraphics[width = 0.19\textwidth]{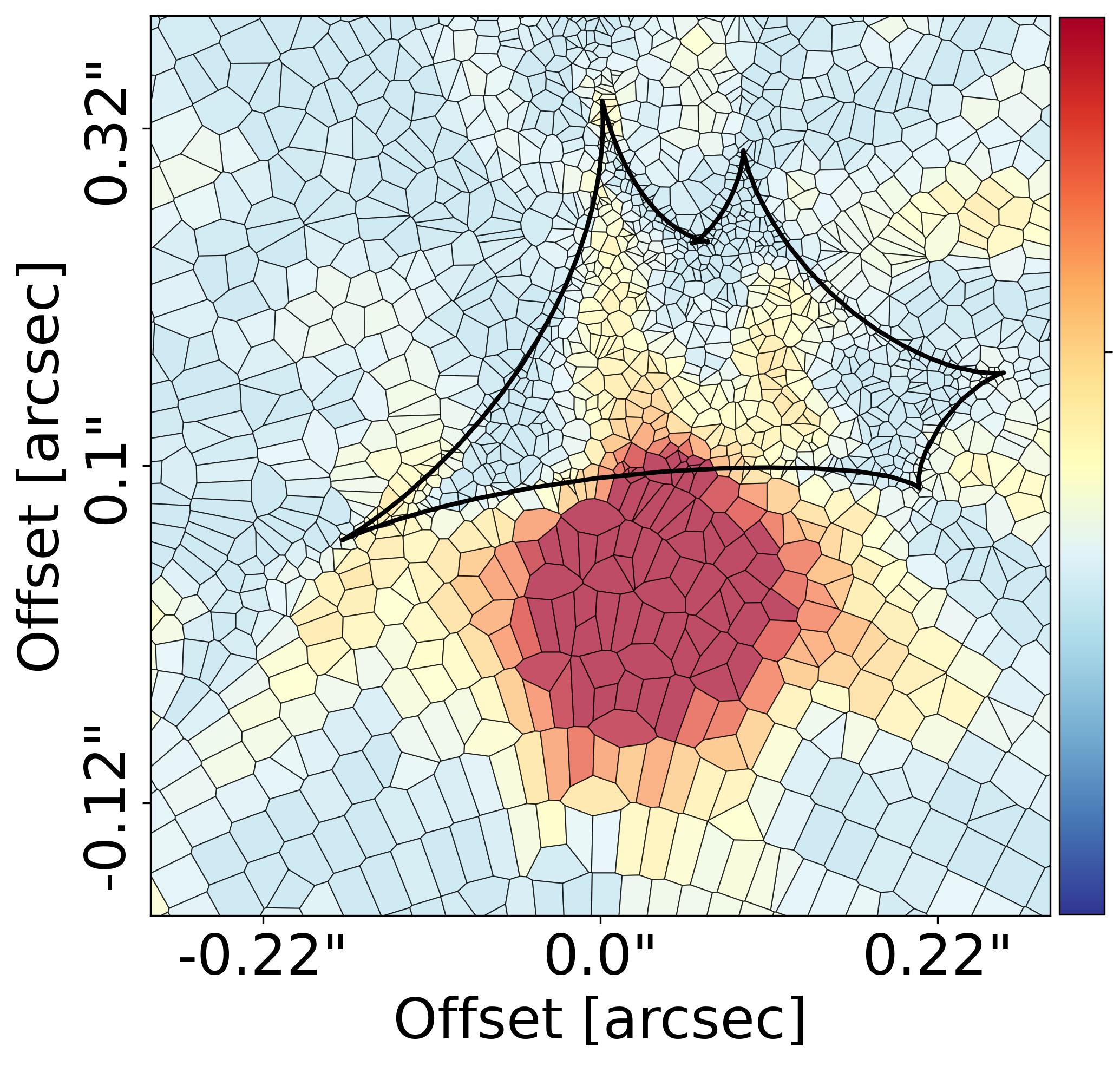}
    
    \includegraphics[width = 0.19\textwidth]{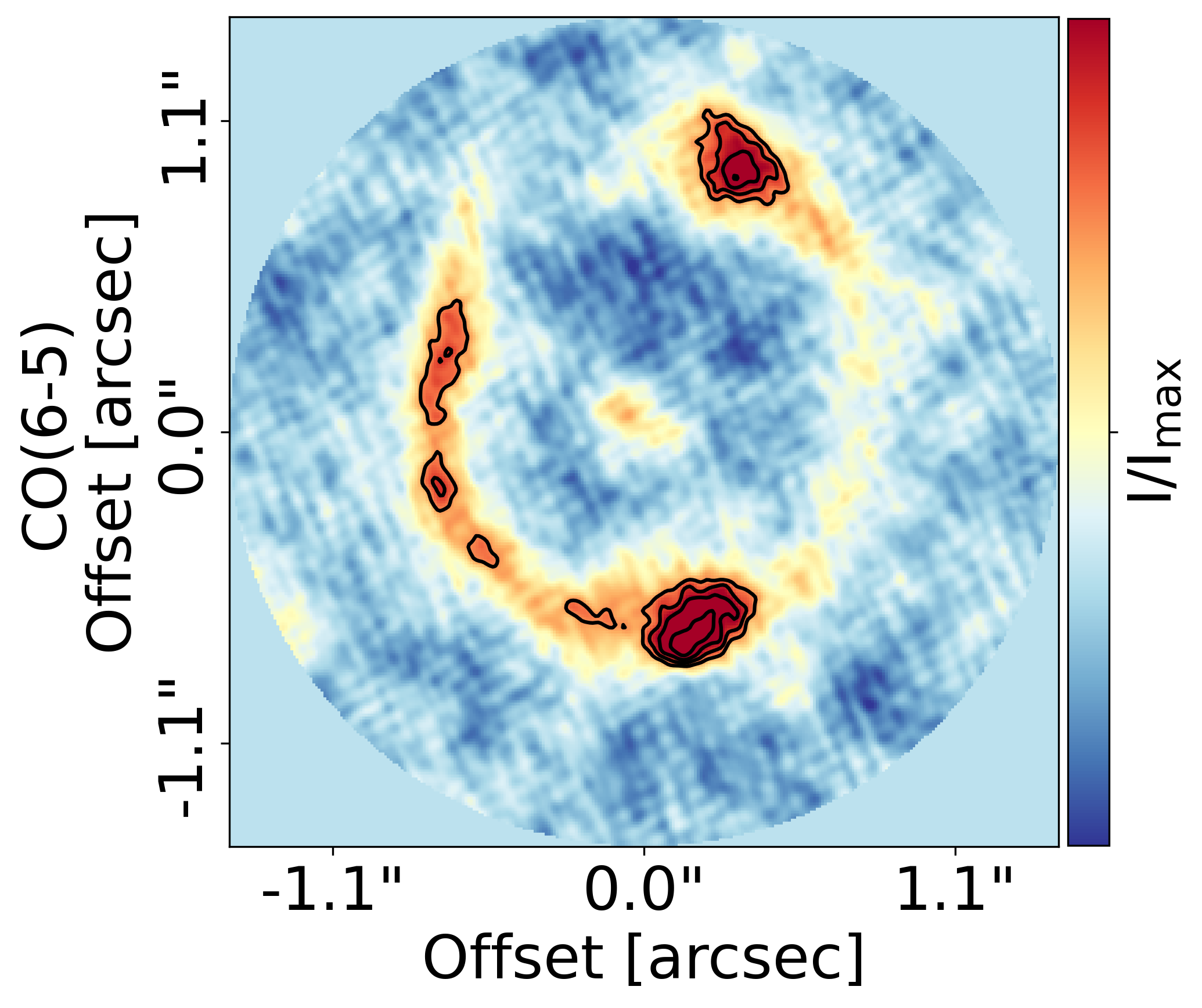}
    \includegraphics[width = 0.19\textwidth]{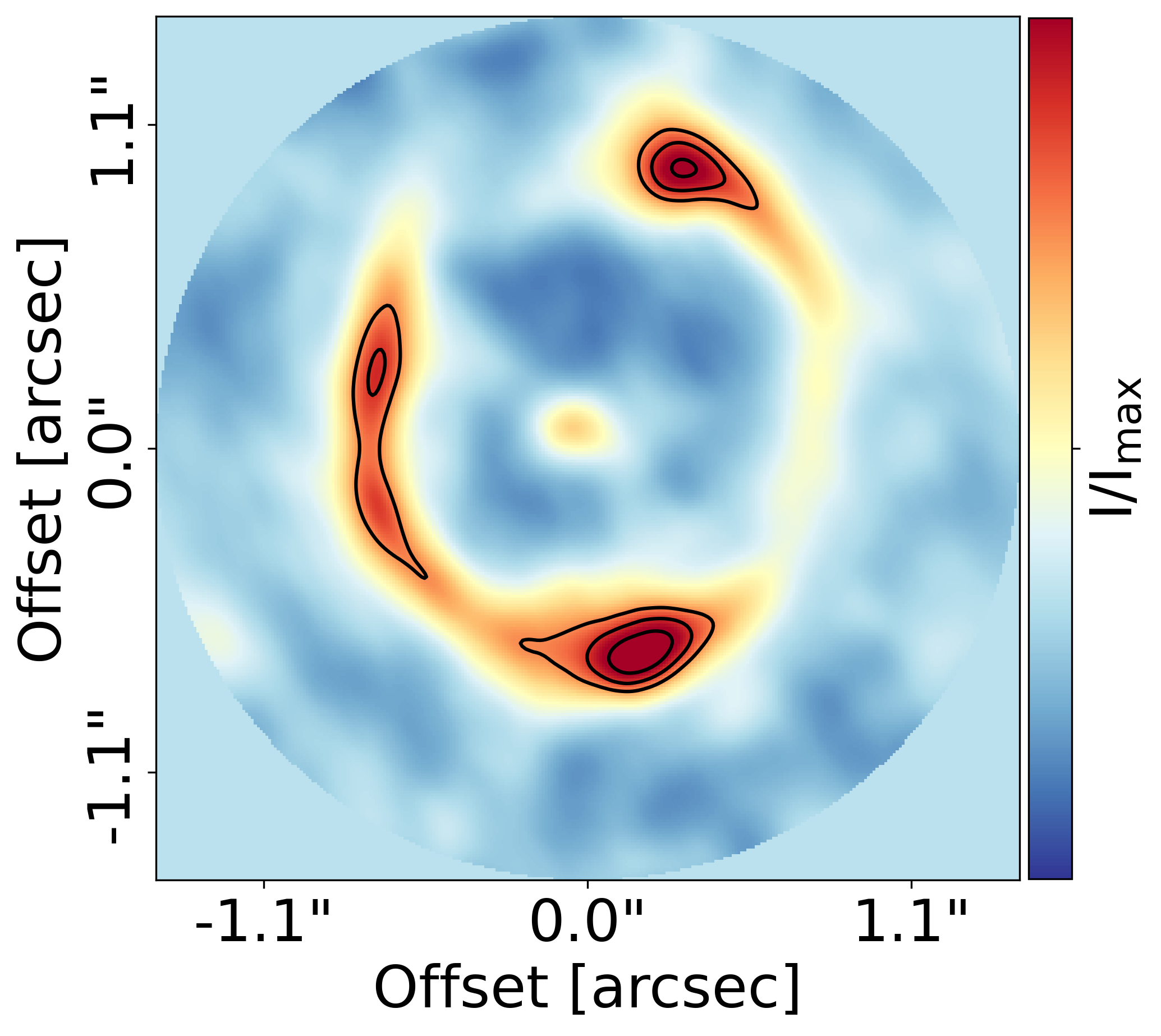}
    \includegraphics[width = 0.19\textwidth]{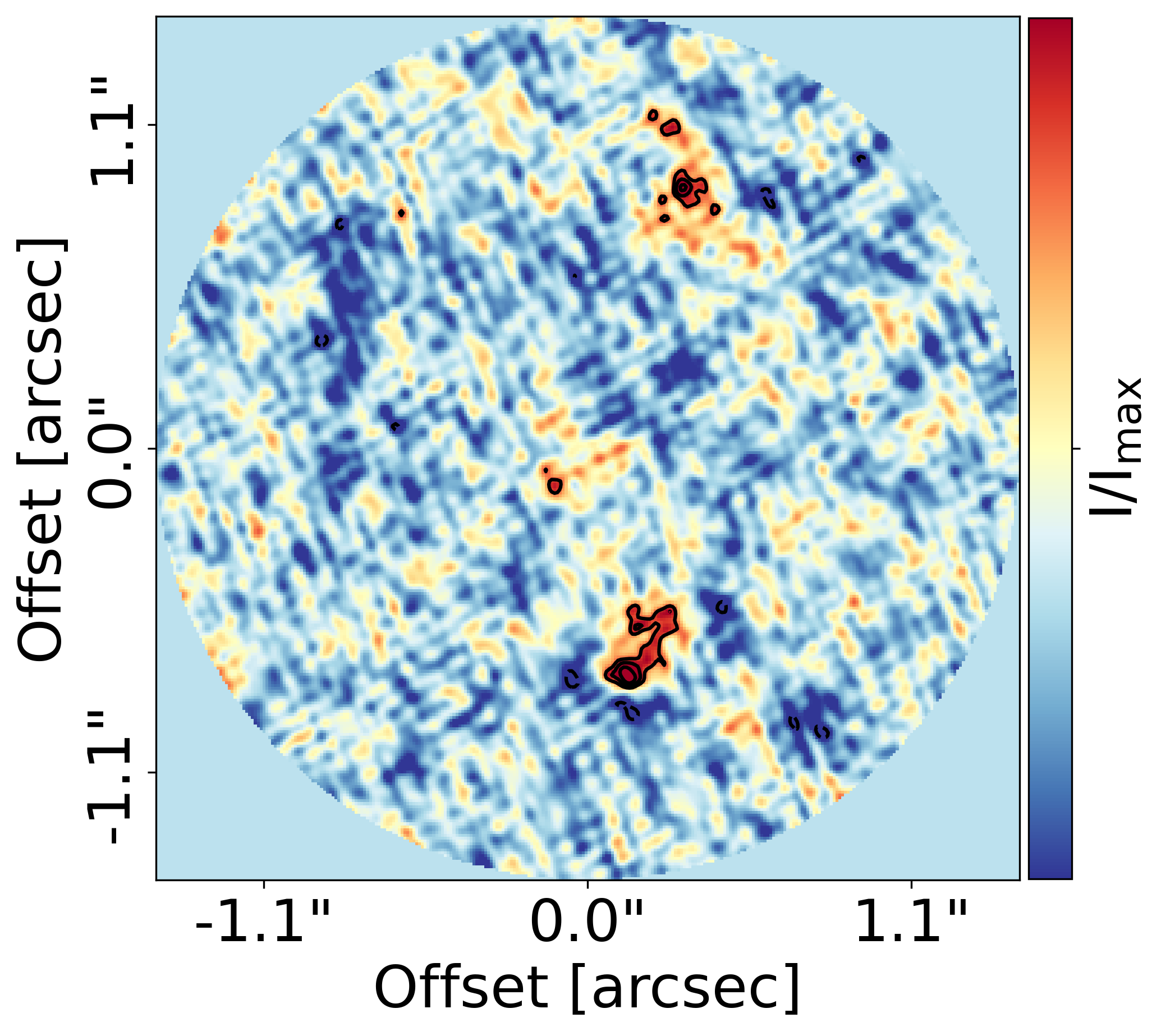}
    \includegraphics[width = 0.19\textwidth]{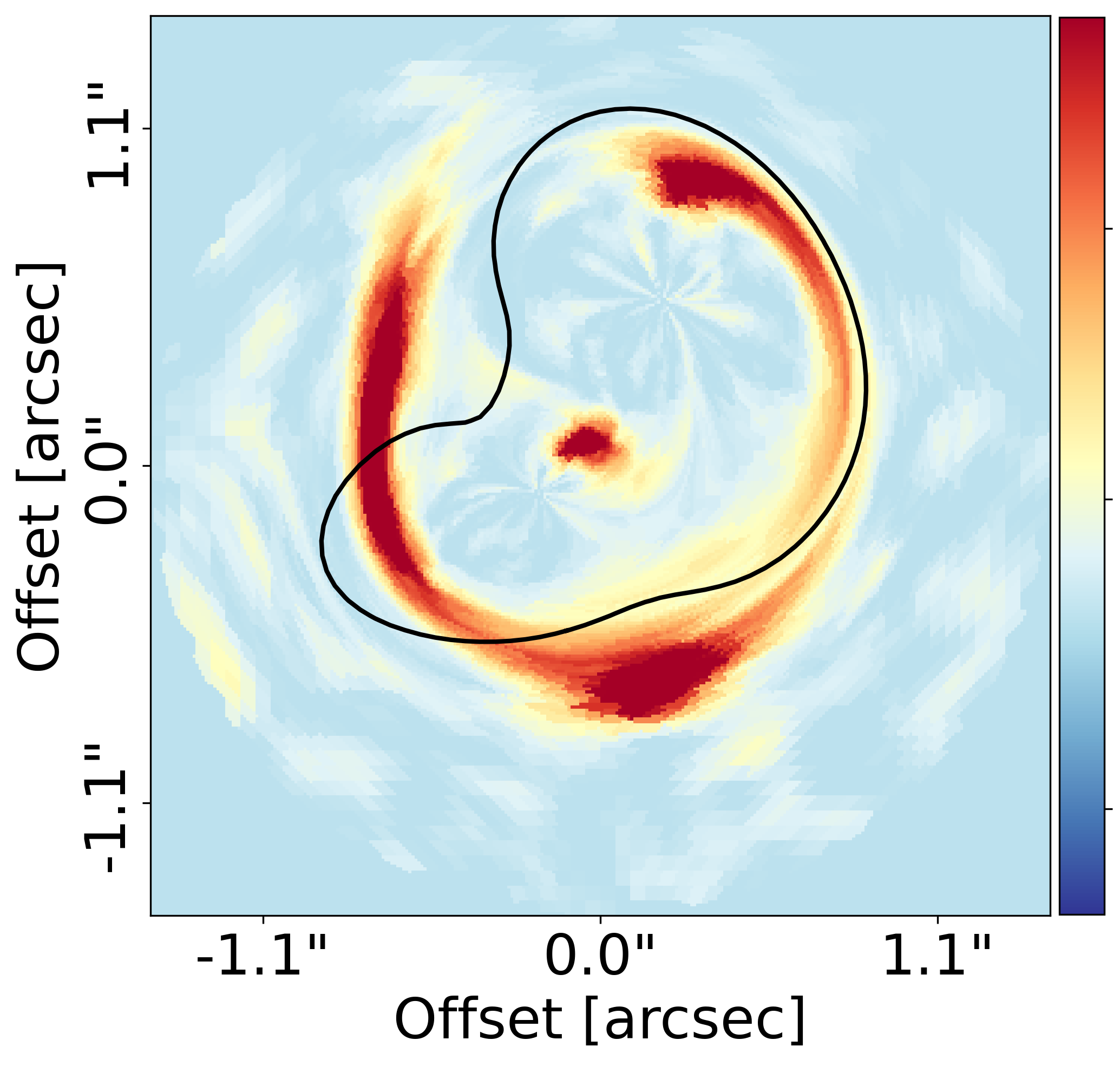}
    \includegraphics[width = 0.19\textwidth]{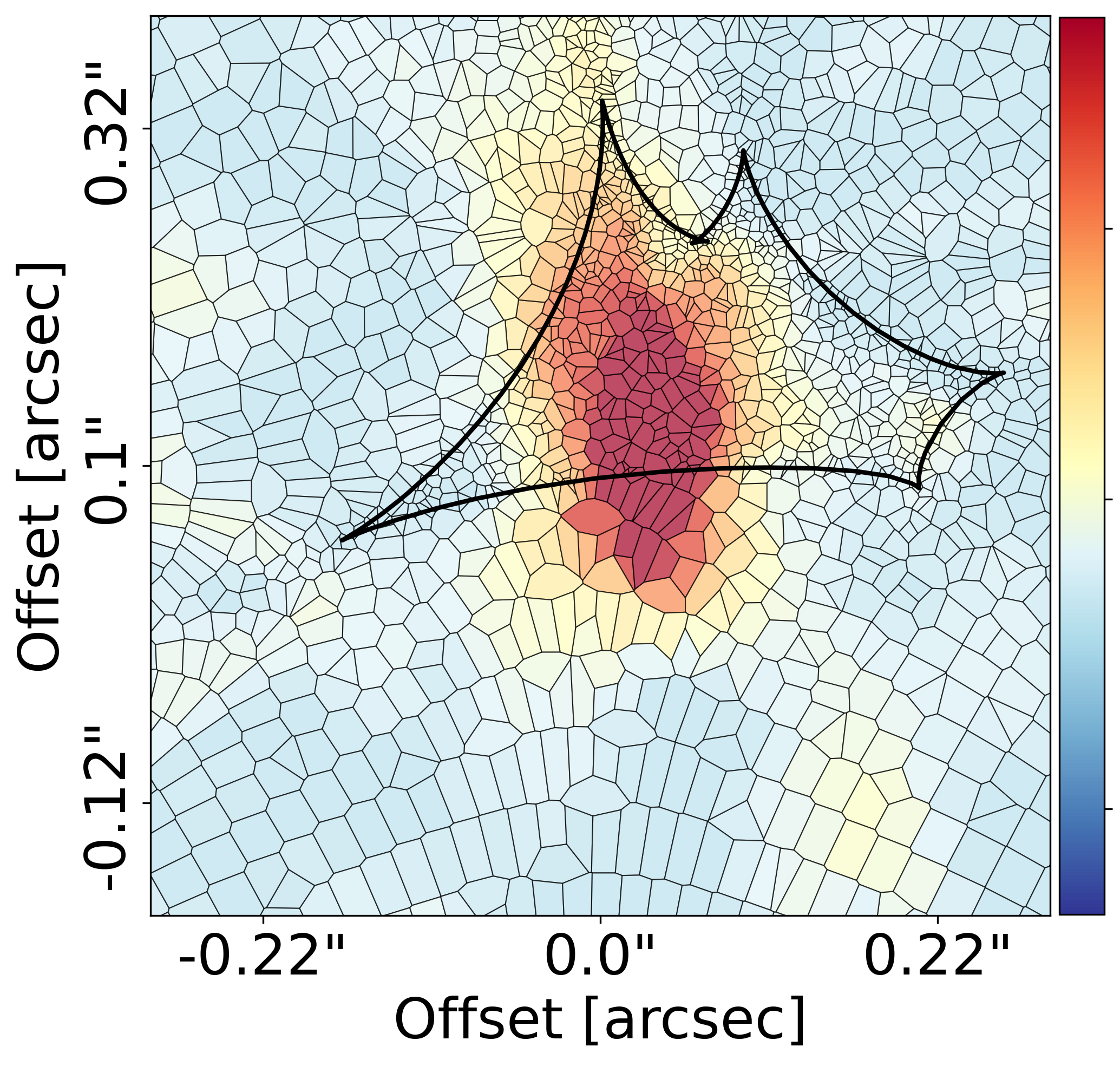}
    
    \caption{Non-parametric source modeling results for the blue and red bins of the CO(6--5) emission; first row: blue peak in the CO(6--5) spectrum, second row: red peak in the CO(6--5) spectrum. The first column shows the dirty image as produced by {\sc PyAutoLens} with contours shown at $3, 4, 5, 6, 7, 8, 9, 10\sigma$ levels. Note that this is not a cleaned image and structures may look slightly different than those shown in cleaned images. The second column shows the dirty model image as produced by {\sc PyAutoLens} with contours shown at $3, 4, 5, 6, 7, 8, 9, 10\sigma$ levels. The third column shows the dirty residual image produced by {\sc PyAutoLens} with contours shown at $-3, -2, 2, 3, 4, 5\sigma$ levels. The fourth column shows the image plane emission non-parametric model of the data produced by {\sc PyAutoLens}. The black line represents the critical line. The fifth column shows the source plane emission non-parametric model of the data produced by {\sc PyAutoLens}. The black line represents the caustic line. All images are centered around the ALMA phase center for each image. Note that the \waterp emission is of significantly lower angular resolution as the lensing model was created using the combined data, as described in Section \ref{sec:observation_details}.}
    \label{fig:nonparametric_red_blue_bins}
\end{figure}

\begin{figure}[!ht]
    \centering
    \includegraphics[width = 1.0\textwidth]{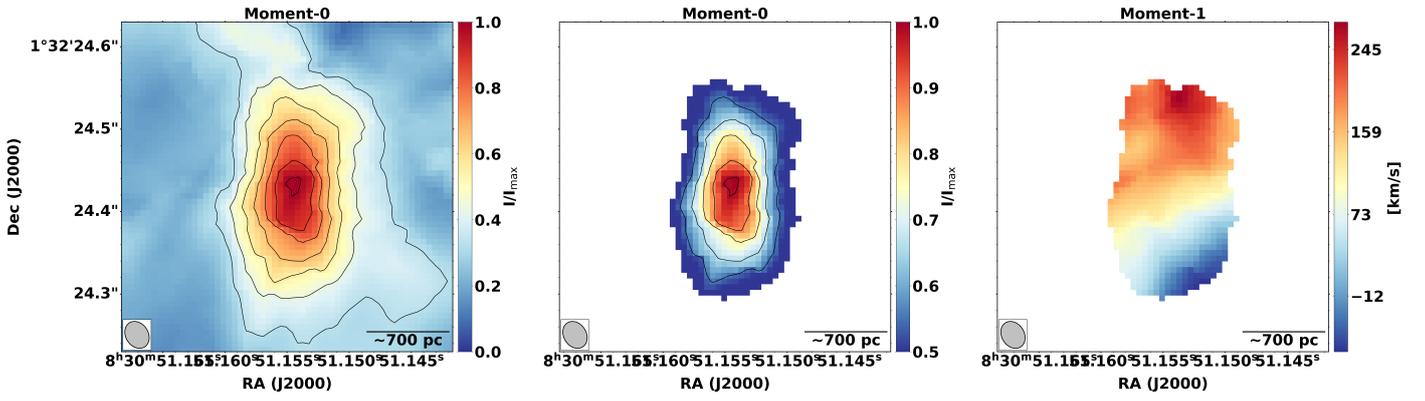}
    \caption{Source plane CO(6--5) masked and unmasked moment-0 and moment-1 maps created from the source plane emission cube described in Section \ref{subsec:cube_and_spec_reconstructions}. The first panel shows the full moment-0 map with contours shown at $-3, -2, 3, 4, 5, 6, 7, 8, 9, 10\sigma$ levels. The middle panel shows the moment-0 map masked to show only values above $4\sigma$ levels where the contours are shown at $5, 6, 7, 8, 9, 10\sigma$ levels. The right panel shows the moment-1 map masked to show only values above $4\sigma$ levels. The intensity values are in arbitrary units for both the unmasked and masked moment-0 maps as output from {\sc PyAutoLens}. The approximate beam, assuming a static magnification factor across the image, is shown in the bottom left of each image. A clear velocity gradient is seen in the moment-1 emission map.}
    \label{fig:src_plane_mom0_mom1}
\end{figure}

\clearpage

\begin{figure}[!ht]
    \centering
    \includegraphics[width = 0.7\textwidth]{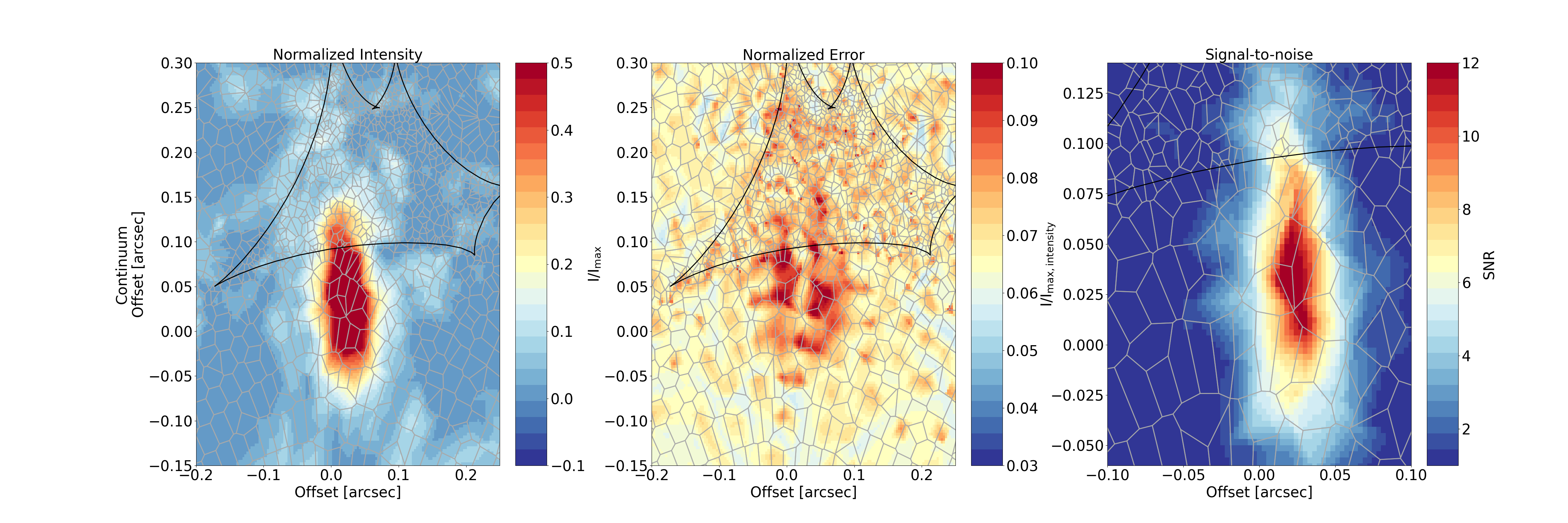}
    \includegraphics[width = 0.7\textwidth]{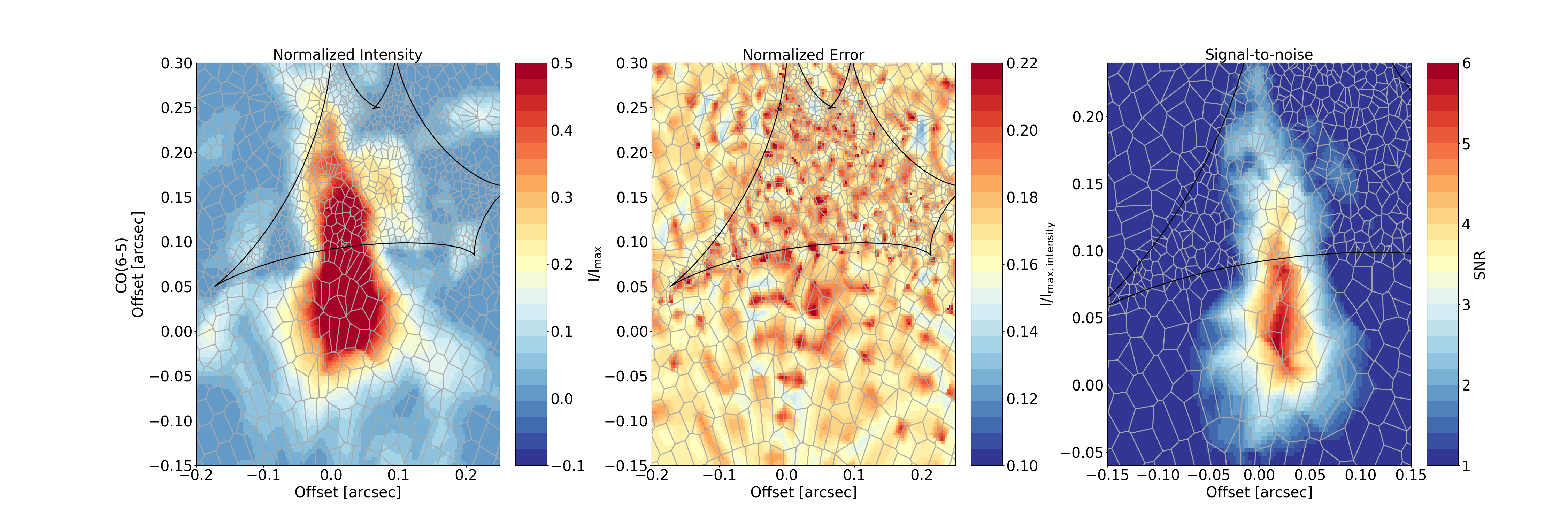}
    \includegraphics[width = 0.7\textwidth]{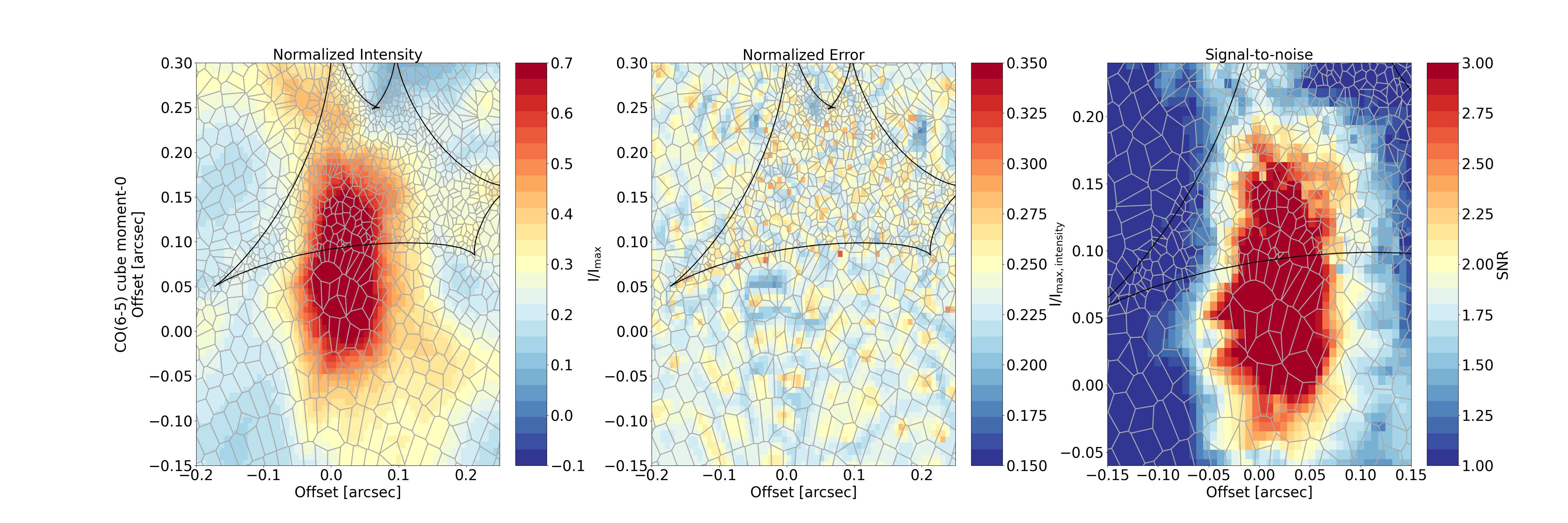}
    \includegraphics[width = 0.7\textwidth]{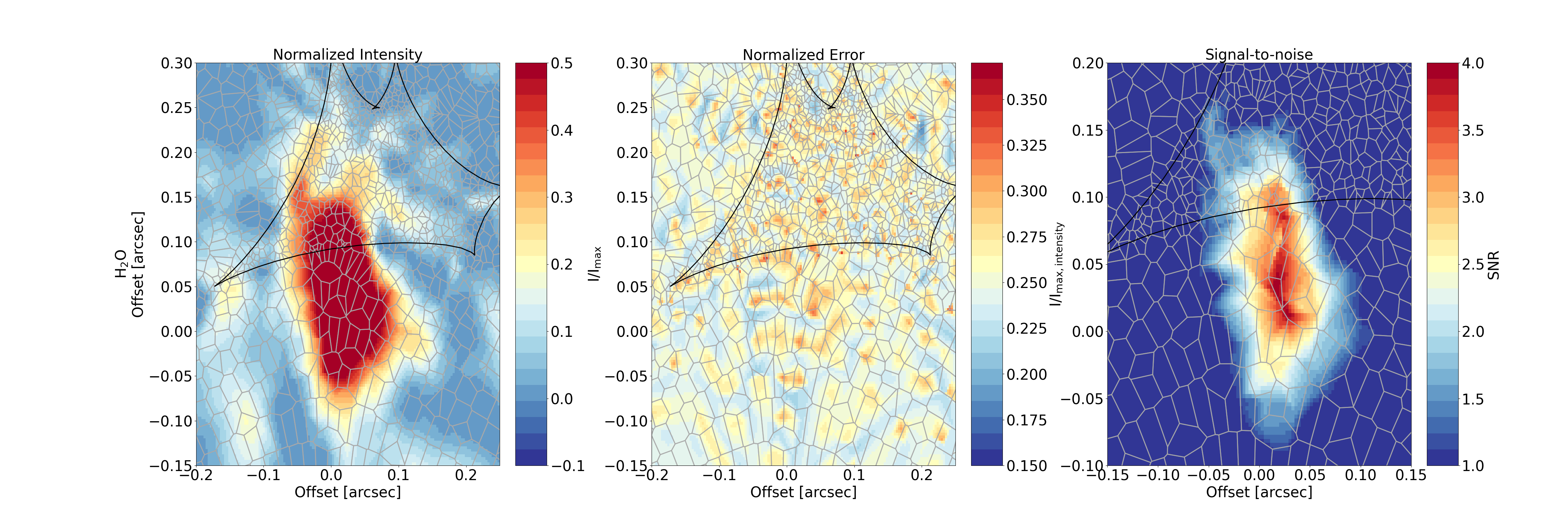}
    \includegraphics[width = 0.7\textwidth]{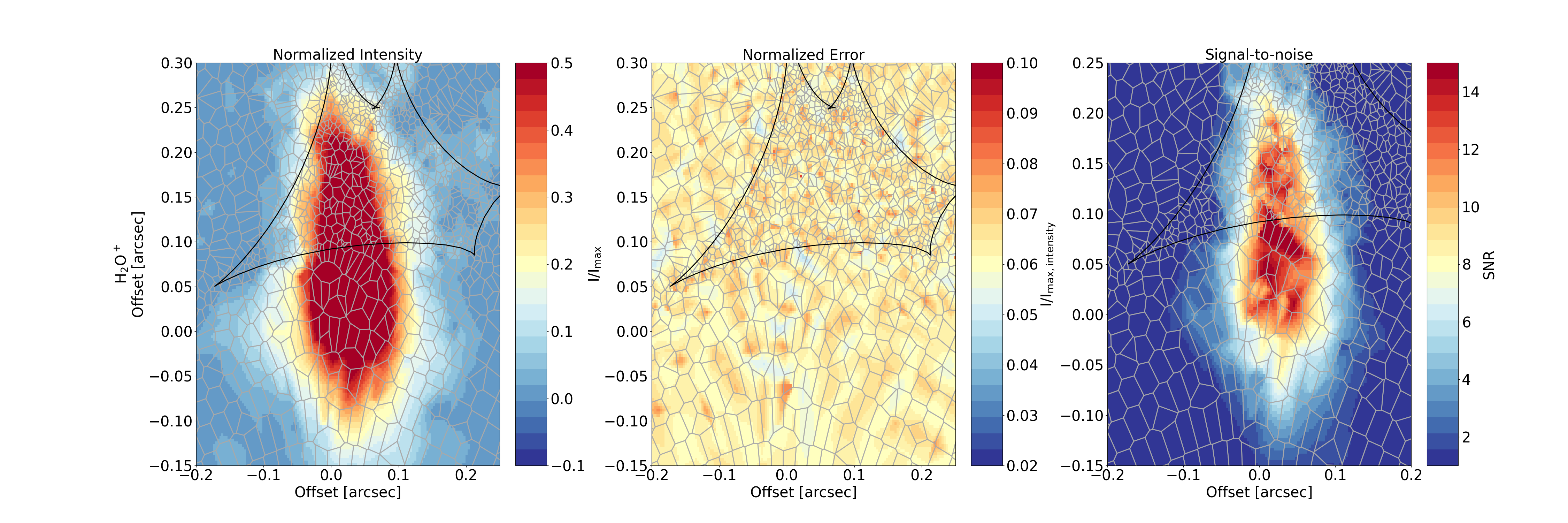}
    
    \caption{Source plane intensity (left panels), error map (middle panels), and SNR map (right panels) for the continuum (top), CO(6--5) (second row), CO(6--5) source plane emission cube (third row), \water (fourth row), and \waterp (bottom row) emission. The grey grid shows the Voronoi pixels prior to interpolation to the square pixel grid, as described in Section \ref{subsec:cube_and_spec_reconstructions}. The black line shows the caustic line. Note that the error maps have been normalized by the maximum value in the respective intensity map, and therefore the error maps can be seen as a percentage error for each pixel. For the CO(6--5) source plane cube the intensity map is the source plane moment-0 map and the intensity map is a moment-0 map of the error cube as described in Section \ref{subsec:cube_and_spec_reconstructions}. Note that the SNR map is zoomed in compared to the other two panels.}
    \label{fig:src_plane_errors}
\end{figure}

\clearpage

\section{Kinematic modeling}

\begin{SCfigure}[1.2][ht!] 
    \centering
    \includegraphics[width=0.5\textwidth]{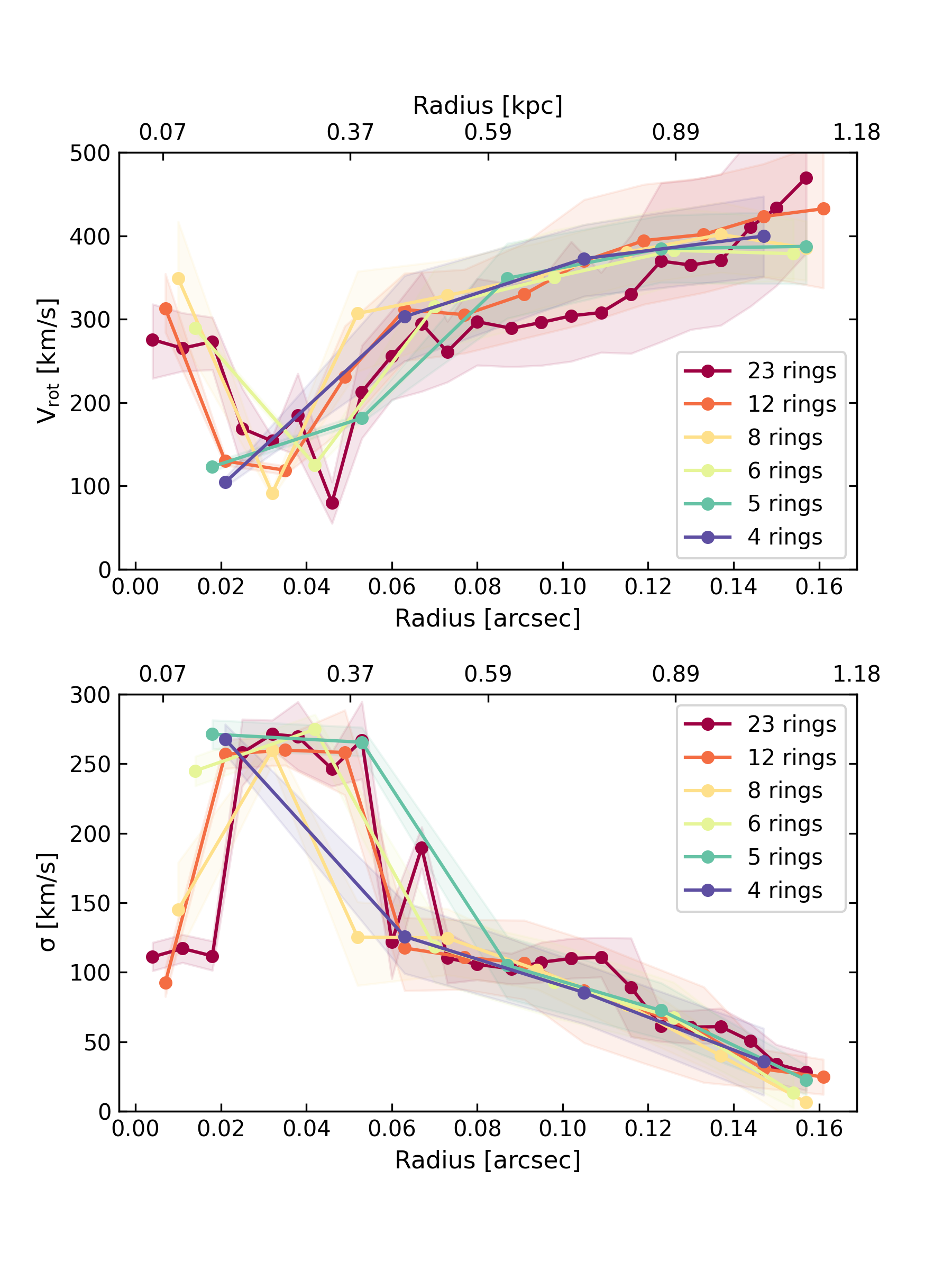}
    \caption{Rotational velocity and dispersion as a function of radius when varying the number of rings used in the \barolo kinematic modeling. The errors are represented by the shaded regions for both. }
    \label{fig:v_sig_radius_perterbbeam}
\end{SCfigure}

\end{appendix}

\end{document}